\documentclass[12pt]{article}%
\pdfoutput=1

\usepackage[normalem]{ulem}
\usepackage{cancel}
\usepackage{booktabs}
\usepackage[para]{threeparttable}
\usepackage{multirow}
\usepackage{subcaption}
\usepackage{rotating}
\usepackage{cite}
\usepackage[all]{xy}
\usepackage{amsmath,amssymb,amsthm,amsfonts,mathrsfs}
\theoremstyle{definition}

\usepackage{mathtools}
\usepackage{amsfonts}
\usepackage{amssymb}
\usepackage{authblk}
\usepackage{bbm}
\usepackage{bm}
\usepackage{caption}
\usepackage{epsfig}
\usepackage{epstopdf}
\usepackage{heck}
\usepackage[margin=1in]{geometry}
\usepackage{tabularx}
\usepackage[usenames,dvipsnames]{xcolor}
\usepackage{graphicx}
\numberwithin{equation}{section}
\usepackage{tikz}

\tikzstyle{startstop} = [rectangle, rounded corners, minimum width=3cm, minimum height=1cm,text centered, draw=black, fill=blue!10]
\tikzstyle{startstop} = [rectangle, rounded corners, minimum width=3cm, minimum height=1cm,text centered, draw=black, fill=blue!10]
\tikzstyle{arrow} = [thick,->,>=stealth]

\usetikzlibrary{decorations.pathreplacing}
\usetikzlibrary{calc}

\usepackage{hyperref}
\usepackage[nameinlink,capitalize]{cleveref}

\makeatletter
\newcommand*{\diff}{\@ifnextchar^{\DIFF}{\DIFF^{}}}
\def\DIFF^#1{\mathop{\text{\mathstrut d}}\nolimits^{#1}\gobblespace}
\newcommand*{\bigdiff}{\@ifnextchar^{\BIGDIFF}{\BIGDIFF^{}}}
\def\BIGDIFF^#1{\mathop{\text{\mathstrut \mathcal{D}}}\nolimits^{#1}\gobblespace}
\def\gobblespace{\futurelet\diffarg\opspace}
\def\opspace{%
	\let\DiffSpace\!%
	\ifx\diffarg(%
		\let\DiffSpace\relax
	\else
		\ifx\diffarg[%
			\let\DiffSpace\relax
		\else
			\ifx\diffarg\{%
				\let\DiffSpace\relax
			\fi
		\fi
	\fi
	\DiffSpace
}
\makeatother

\newcommand{\bbZ}{\mathbb{Z}}

\def\bbZ{\mathbb{Z}}
\def\fkg{\mathfrak{g}}

\usepackage{tikz}
\usetikzlibrary{arrows}
\usetikzlibrary{arrows.meta}
\usetikzlibrary{shapes.geometric,calc,arrows, positioning,shapes.misc,decorations.markings}
\tikzset{
  big arrow/.style={
    decoration={markings,mark=at position 1 with {\arrow[scale=2,#1]{>}}},
    postaction={decorate},
    shorten >=0.4pt},
  big arrow/.default=black}

\pgfdeclarelayer{edgelayer}
\pgfdeclarelayer{nodelayer}
\pgfsetlayers{edgelayer,nodelayer,main}
\tikzstyle{none}=[inner sep=0pt]

\tikzstyle{NodeCross}=[draw, shape=circle, cross out, inner sep=0pt, minimum size=6pt,line width=0.25mm]
\tikzstyle{Circle}=[draw, shape=circle, black, fill=black, inner sep=0pt, minimum size=6pt]
\tikzstyle{Star}=[draw, shape=star, fill=black, star points=6, inner sep=0pt, minimum size=8pt]

\tikzstyle{DashedLine}=[-, densely dashed, line width=0.25mm]
\tikzstyle{DottedLine}=[-, dotted, line width=0.25mm]
\tikzstyle{ThickLine}=[-, line width=0.25mm]
\tikzstyle{ArrowLineRight}=[-, -{Stealth[scale=1.75]}, line width=0.15mm, scale=5]
\tikzstyle{RedLine}=[-, draw={rgb,255: red,191; green,0; blue,0}, fill=none, line width=0.25mm]
\tikzstyle{DashedLineThin}=[-, densely dashed, line width=0.125mm, fill=none, draw=black]
\tikzstyle{ArrowLineRed}=[-, draw={rgb,255: red,191; green,0; blue,0}, -{Stealth[scale=1.75]}, line width=0.15mm, scale=5]

\newcommand{\bea}{\begin{eqnarray}}
\newcommand{\eea}{\end{eqnarray}}
\newcommand{\be}{\begin{equation}}
\newcommand{\ee}{\end{equation}}
\newcommand{\ba}{\begin{aligned}}
\newcommand{\ea}{\end{aligned}}
\newcommand{\bit}{\begin{itemize}}
\newcommand{\eit}{\end{itemize}}
\newcommand{\ben}{\begin{enumerate}}
\newcommand{\een}{\end{enumerate}}

\begin{document}

\date{May 2022}

\title{6d SCFTs, Center-Flavor Symmetries, and \\[4mm] Stiefel--Whitney Compactifications}

\institution{PENN}{\centerline{$^{1}$Department of Physics and Astronomy, University of Pennsylvania, Philadelphia, PA 19104, USA}}
\institution{DESY}{\centerline{$^{2}$Deutsches Elektronen-Synchrotron DESY, Notkestr.~85, 22607 Hamburg, Germany}}
\institution{CERN}{\centerline{${}^{3}$CERN Theory Department, CH-1211 Geneva, Switzerland}}
\institution{LEUVEN}{\centerline{${}^{4}$Instituut voor Theoretische Fysica, KU Leuven, Celestijnenlaan 200D, B-3001 Leuven, Belgium }}

\authors{
Jonathan J. Heckman\worksat{\PENN}\footnote{e-mail: \texttt{jheckman@sas.upenn.edu}},
Craig Lawrie\worksat{\DESY}\footnote{e-mail: \texttt{craig.lawrie1729@gmail.com}},
Ling Lin\worksat{\CERN}\footnote{e-mail: \texttt{ling.lin@cern.ch}},\\[4mm]
Hao Y. Zhang\worksat{\PENN}\footnote{e-mail: \texttt{zhangphy@sas.upenn.edu}}, and
Gianluca Zoccarato\worksat{\LEUVEN}\footnote{e-mail: \texttt{gianluca.zoccarato@kuleuven.be}}
}

\abstract{The center-flavor symmetry of a gauge theory specifies the global form of
consistent gauge and flavor bundle background field configurations.
For 6d gauge theories which arise from a tensor branch deformation of a
superconformal field theory (SCFT), we determine the global structure of such background field configurations, including possible
continuous Abelian symmetry and R-symmetry bundles. Proceeding to the conformal fixed point, this provides a prescription for
reading off the global form of the continuous factors of the zero-form symmetry, including possible non-trivial mixing between
flavor and R-symmetry. As an application, we show that this global structure leads to a large class of 4d $\mathcal{N} = 2$ SCFTs
obtained by compactifying on a $T^2$ in the presence of a topologically non-trivial flat flavor bundle characterized by an 't Hooft magnetic flux. The resulting ``Stiefel--Whitney twisted'' compactifications realize several new infinite families of 4d $\mathcal{N} = 2$ SCFTs, and also furnish a 6d origin for a number of recently discovered rank one and two 4d $\mathcal{N} = 2$ SCFTs.}

\begin{flushright}
  {\small \texttt{\hfill CERN-TH 2022-074}}\\[-1mm]
  {\small \texttt{\hfill DESY-22-069}}
\end{flushright}

\maketitle

\setcounter{tocdepth}{2}

\tableofcontents

\newpage

\section{Introduction}\label{sec:intro}

Since their discovery \cite{Witten:1995zh,Strominger:1995ac,Seiberg:1996qx},
6d SCFTs have been a fount of insight into the non-perturbative structure of quantum field theory in diverse dimensions.
In particular, knowledge of the six-dimensional theory and the compactification geometry can make hard-to-access non-perturbative features in lower-dimensional systems manifest. On the other hand, general arguments indicate that such theories cannot be realized via perturbations of a Gaussian fixed point, and so in this sense they are intrinsically strongly coupled \cite{Cordova:2016xhm}. This in turn complicates the construction and study of such theories.

A conjectural classification of all 6d SCFTs was proposed in \cite{Heckman:2013pva,Heckman:2015bfa} (see also \cite{DelZotto:2014hpa, Heckman:2014qba, Bhardwaj:2015xxa, Tachikawa:2015wka, Bhardwaj:2015oru, Bhardwaj:2018jgp, Heckman:2018pqx, Bhardwaj:2019hhd,Distler:2022yse}). The main idea in this classification program is to engineer such theories via F-theory backgrounds involving a non-compact elliptically fibered Calabi--Yau threefold with a canonical singularity. This has led to a vast class of new theories, and a remarkably simple unifying description of nearly all such theories on their partial branch as generalized quiver gauge theories. This perspective has been used to extract a number of calculable quantities from such systems, including, for example the anomaly polynomial \cite{Intriligator:2014eaa, Ohmori:2014pca, Ohmori:2014kda, Cordova:2018cvg}, as well as operator scaling dimensions of certain operator subsectors \cite{Heckman:2014qba, Bergman:2020bvi,Baume:2020ure, Heckman:2020otd, Baume:2022cot}. Compactification of such theories to four-dimensional systems also provides a systematic way to generate a broad class of 4d SCFTs with varying amounts of supersymmetry \cite{Gaiotto:2015usa,DelZotto:2015rca,Franco:2015jna,Coman:2015bqq,Garcia-Etxebarria:2016erx,Razamat:2016dpl,Baume:2021qho,Ohmori:2015pua,Ohmori:2015pia,Mekareeya:2017jgc,Mekareeya:2017sqh,Heckman:2016xdl,Bah:2017gph,Bourton:2017pee,Kim:2017toz,Apruzzi:2018oge,Razamat:2018zus,Kim:2018lfo,Razamat:2018gro,Kim:2018bpg,Razamat:2018gbu,Chen:2019njf,Pasquetti:2019hxf,Sela:2019nqa,Razamat:2019mdt,Razamat:2019ukg,Razamat:2020bix,Sabag:2020elc,Bourton:2020rfo,Nazzal:2021tiu,Kang:2021lic,Hwang:2021xyw,Kang:2021ccs,Bourton:2021das,Razamat:2022gpm}.\footnote{A recent overview of superconformal field theories in dimensions three to six is \cite{Argyres:2022mnu}.}

In general terms, global symmetries also play an important role in constraining correlation functions of local operators, and also figure into the analysis of higher symmetries \cite{Gaiotto:2014kfa}. This is no less true in 6d SCFTs, and also plays an important role in in the study of compactifications of such systems. As a recent example, \cite{Ohmori:2018ona} (see also \cite{Razamat:2016dpl}) demonstrated that starting from certain 6d $\mathcal{N} = (1,0)$ SCFTs, compactification on a $T^2$ in the presence of a topologically non-trivial but flat bundle associated with an 't Hooft magnetic flux can be used to generate a class of 4d $\mathcal{N} = 2$ SCFTs. In particular, this requires knowing not just the global symmetry \textit{algebra} of the 6d theory, but the actual \textit{group}.

Our aim in this paper will be to extract the continuous zero-form group symmetries of 6d SCFTs, and to use this in the construction of 4d $\mathcal{N} = 2$ SCFTs via Stiefel--Whitney twisted compactifications. Now, although the actual method of constructing such 6d SCFTs involves the geometry of the F-theory compactification, geometry can sometimes obscure some of the symmetries \cite{Bertolini:2015bwa}. These top down considerations can often be supplemented by various bottom up considerations, including Higgsing from theories with known flavor symmetry algebras \cite{Heckman:2016ssk,Heckman:2018pqx,Hassler:2019eso,Apruzzi:2020eqi}, and thus in many cases we know the continuous global symmetry algebra.
Consequently, we can specify a corresponding ``naive'' flavor symmetry $\widetilde{G}_{\text{flavor}}$, where all simple non-Abelian factors are simply connected, and there is no finite group action on any $U(1)$ factors. One can also supplement this by the R-symmetry $SU(2)_{R}$, which is difficult to track in the F-theory construction, but which must be present in any 6d SCFT.\footnote{Recall that in a supersymmetric theory, the flavor symmetry commutes with the supercharges, whereas the R-symmetry (by definition) does not.} Since we also have the gauge symmetry on the tensor branch, there is a corresponding ``naive'' group of continuous gauge and global zero-form symmetries:
\begin{equation}
\widetilde{G}_{\text{gauge-global}} \equiv \widetilde{G}_{\text{gauge}} \times \widetilde{G}_{\text{flavor}} \times SU(2)_{R}\,,
\end{equation}
where we have kept implicit the spacetime global symmetries of the field theory. This answer is ``naive'', in the sense that the matter content and effective strings (coupling to the tensor multiplet chiral two-forms) of the effective field theory may be neutral under some subgroup of the center of $\widetilde{G}_{\text{gauge-global}}$.
Consequently, the global form may end up being quotiented by a subgroup of the center $\mathcal{C} \subset \widetilde{G}_{\text{gauge-global}}$:
\begin{equation}
G_{\text{gauge-global}} = \widetilde{G}_{\text{gauge-global}} / \mathcal{C}\,.
\end{equation}
This quotient can also act on the spacetime symmetries since the supercharges transform as spacetime spinors and R-symmetry spinors. 
The combined action on the gauge and flavor symmetry is often referred to as a ``center-gauge-flavor symmetry'', generalizing the notion of ``center-flavor symmetry" \cite{Cohen:1983sd,Aharony:2016jvv,Benini:2017dus,Cherman:2017tey,Shimizu:2017asf,Gaiotto:2017tne,Tanizaki:2017qhf,Tanizaki:2017mtm,Cordova:2018acb,Yonekura:2019vyz,Hidaka:2019jtv,Cordova:2019uob,Dierigl:2020myk,Apruzzi:2021vcu,Apruzzi:2021mlh,DelZotto:2022joo,Hubner:2022kxr,Cvetic:2022imb}.
Of course, from the perspective of the 6d SCFT, the defining data only makes reference to the global symmetries, and the same quotient, suitably projected, realizes the continuous part of the global symmetry group
\begin{equation}
G_{\text{global}} = \widetilde{G}_{\text{global}} / \mathcal{C}_{\text{global}}\,,
\end{equation}
in the obvious notation. In the conformal limit, the possible action of $\mathcal{C}_{\text{global}}$ on these internal symmetries 
can also be accompanied by a quotient on the conformal group. As already implicitly mentioned, knowing the global form of the zero-form symmetry group has important implications for the existence and structure of higher-dimensional defects in the theory, informing possible higher symmetry structures.

One of our core tasks will be to present a general algorithm for extracting $G_{\text{gauge-global}}$ and $G_{\text{global}}$ of the tensor branch effective field theory. This also amounts (upon projecting onto the global symmetry factors) to a prediction for the global continuous zero-form group of the 6d SCFT.
Extracting this data directly from the corresponding F-/M- theory background geometry \cite{Cvetic:2022imb} was recently carried out for a number of 5d supersymmetric quantum field theories obtained from circle reduction of the tensor branch of a 6d SCFT, and the ``bottom up'' approach developed here agrees with the ``top down'' results obtained in \cite{Cvetic:2022imb}. From a bottom up perspective, we simply work on
on the tensor branch where we have access to the large symmetry transformations of the system, and the correlated response from transformations on the chiral two-forms of the effective field theory. Such transformations are in turn sensitive to the global topology of background gauge/global bundle configurations \cite{Apruzzi:2020zot}.\footnote{We note that, since this analysis relies solely on the effective field theory description on the tensor branch, it can also be carried out straightforwardly for theories constructed from frozen singularities \cite{Tachikawa:2015wka, Bhardwaj:2018jgp}.} This technique has been used previously to explore some examples of non-Abelian flavor symmetry in related systems \cite{Apruzzi:2020zot, Apruzzi:2021mlh},\footnote{In gravitational theories where there are no global symmetries, the same methods give constraints on the global form of gauge symmetries \cite{Apruzzi:2020zot,Cvetic:2020kuw}, which for supergravity models in high dimensions are found to agree with patterns in string compactifications \cite{Font:2020rsk,Font:2021uyw,Cvetic:2021sjm,Cvetic:2022uuu}.} but as far as we are aware, a systematic study of all possibilities was not previously undertaken. In particular, we also show how to incorporate continuous Abelian symmetries.

Moreover, our analysis also extends to the global form of the R-symmetry, and its possible mixing with the center-flavor symmetry. This is difficult to extract from established index computations in the much-studied and related case of 4d theories (as obtained by compactification on a $T^2$ with no background bundles switched on), since in many cases, only specific R-charge sectors are counted.\footnote{We note that the possibility of mixing with the center of the R-symmetry resolves some puzzles in various claimed global forms of the flavor symmetry for certain 4d $\mathcal{N} = 2$ SCFTs which have appeared in earlier work (e.g., compare \cite{Distler:2019eky} with \cite{Bhardwaj:2021ojs}), a point we comment on in more detail later on.} In some cases, however, alternative methods have been explored for extracting the chiral ring of the corresponding Higgs branch \cite{Ferlito:2017xdq, Hanany:2018uhm}, which implicitly also determines a mod $2$ constraint on the global form of the center symmetry mixing with the R-symmetry. In these cases, we find that our analysis agrees with these constraints.

To illustrate the utility of this approach, we show in a number of examples how to extract the symmetry groups $G_{\text{gauge-global}}$ and
$G_{\text{global}}$. One large class of examples includes M5-brane probes of an ADE singularity $\mathbb{C}^2/{\Gamma_\text{ADE}}$ as well as their Higgs branch deformations. These flows are captured by group-theoretic data associated with nilpotent and semi-simple deformations of the corresponding flavor symmetry algebras. Since the corresponding tensor branch descriptions for these theories are all known, we can use our method to extract the corresponding continuous symmetry \textit{group}, including contributions from Abelian symmetry factors and mixing with the R-symmetry. Similar considerations hold for the ``orbi-instanton theories'' obtained from Higgs branch deformations of
M5-branes probing an ADE singularity $\mathbb{C}^2 / \Gamma_\text{ADE}$ wrapped by an $E_8$ nine-brane. In this case, deformations of the $E_8$ flavor symmetry factor are captured by finite group homomorphisms $\Gamma_\text{ADE} \rightarrow E_8$.

Analyzing this class of examples, we observe that many breaking patterns wind up generating a trivial quotienting subgroup $\mathcal{C}$ for the global symmetry. This occurs simply because, in many cases, there is no common center for the simply connected non-Abelian symmetry group factors. A general rule of thumb for realizing a common center-gauge-flavor symmetry is that the group-theoretic data such as a nilpotent orbit or a finite group homomorphism must have a sufficient multiplicity so that there is a non-trivial finite group action on the deformation parameter itself. This analysis also makes it clear that the vast majority of examples with non-trivial gauge-flavor symmetry mixing on the tensor branch will necessarily involve A-type symmetry algebras, simply because the corresponding Lie groups exhibit a far broader class of possible center subgroups (e.g., $SU(N)$ has center $\mathbb{Z}_N$), when compared with their non-A-type counterparts.

Once the center-flavor symmetry of a 6d SCFT is known, one can utilize it to generate a large class of lower-dimensional theories via  compactification. To illustrate, we primarily focus on the case of compactification on a $T^2$ in the presence of topologically non-trivial background bundle configurations. The corresponding 't Hooft magnetic fluxes are characterized by holonomies which commute in $G_{\text{flavor}} = \widetilde{G}_{\text{flavor}} / \mathcal{C}$, but which would not commute in $\widetilde{G}_{\text{flavor}}$ (see \cite{Witten:1997bs, Borel:1999bx}). These have been referred to as Stiefel--Whitney twisted theories in \cite{Ohmori:2018ona}. This provides a systematic way to generate a large class of 4d $\mathcal{N} = 2$ SCFTs. In particular, up to a small number of outliers, we show that after including further Higgs branch and mass deformations, this generates the full list of known rank two 4d $\mathcal{N} = 2$ SCFTs given in \cite{Martone:2021ixp}. The list of theories we generate in this way also has some overlap with other top-down constructions such as those based on D3-brane probes of $\mathcal{N} = 2$ S-folds (i.e., non-perturbative generalizations of an orientifold plane in the presence of a stack of flavor seven-branes) \cite{Apruzzi:2020pmv,Giacomelli:2020jel,Heckman:2020svr,Giacomelli:2020gee,Bourget:2020mez}. While there is indeed some overlap in 4d with suggestive evidence via string duality, we also find that there are some cases of Stiefel--Whitney twisted compactifications which resist a simple interpretation in terms of S-folds, an issue we leave for future investigations.

The rest of this paper is organized as follows. We begin by giving a brief review of the tensor branch of a 6d SCFT, with a particular emphasis on topological terms. In Section \ref{sec:6d}, we study the global structure of the flavor symmetry group of 6d $(1,0)$ SCFTs using the tensor branch description. In particular, we extract the overall center-flavor symmetry, including Abelian factors, as well as non-trivial mixing with R-symmetry factors, illustrating with a number of examples. Section \ref{sec:e8CF} serves as an intermezzo between the 6d and 4d analysis; we extract the center-flavor symmetry for a large class of, so-called, orbi-instanton theories which we then use in the next section. In Section \ref{sec:sfolds} we turn to the resulting 4d $\mathcal{N} = 2$ SCFTs generated by Stiefel--Whitney twisted compactifications of such 6d SCFTs. This provides us with a large class of new theories, and we also comment on the similarities and differences with 4d $\mathcal{N} = 2$ S-fold constructions. In Section \ref{sec:Esfolds}, we briefly explore the DE-type generalizations of the A-type 6d and 4d SCFTs that were studied in Sections \ref{sec:e8CF} and \ref{sec:sfolds}. We present our conclusions and areas of future investigation in Section \ref{sec:conc}. In Appendix \ref{app:JUSTHEFLUBRO}, we 
determine the continuous symmetry group for the $\mathcal{N} = (2,0)$ theories and the E-string theories. In Appendix \ref{app:ranktwo}, we show how to generate nearly all known rank two 4d $\mathcal{N}=2$ SCFTs via twisted Stiefel--Whitney compactifications, and we provide a comparison with previously obtained results in the literature in Appendix \ref{app:lit}. Appendix \ref{app:nilp} studies the nilpotent deformations in Stiefel--Whitney twisted theories inherited from the nilpotent deformations of their 6d parent theory. Finally, in Appendix \ref{app:CM}, we explore the global form of the flavor symmetry group for nilpotent deformations of conformal matter theories.

\section{Tensor Branch of 6d SCFTs}\label{sec:REVIEW}

In this section, we present a brief review of the tensor branch of a 6d SCFT, with a particular emphasis on the topological interaction terms.
Recently, much progress has been made in constructing 6d SCFTs by recasting the construction of such theories in terms of non-compact elliptically-fibered Calabi--Yau threefolds $X \rightarrow B$. In this description, one starts with a collection of curves in the base $B$, and with it a corresponding elliptic fibration. We can reach a conformal fixed point if the collection of curves can simultaneously contract to zero size. This results in a canonical singularity in the elliptic threefold (possibly partially frozen), and is the most systematic known method for realizing such theories \cite{Heckman:2013pva, Heckman:2015bfa}.

The configuration of curves prior to collapse gives a geometric realization of the so-called ``tensor branch'' of the 6d SCFT. In this regime, we have a collection of tensor multiplets, the bosonic content of each one consisting of a real scalar and an anti-chiral two-form potential. This anti-chiral two-form couples to effective strings, with tension controlled by the vacuum expectation value (vev) of the scalar.
We can potentially have 7-branes wrapped over each curve, and this results in non-Abelian gauge symmetries on the tensor branch. Collisions of 7-branes result in matter, which can include weakly coupled hypermultiplets, as well as (if we do not go to the full tensor branch) strongly coupled generalizations known as 6d conformal matter \cite{DelZotto:2014hpa,Heckman:2014qba}.

Letting $A^{ij}$ denote the intersection pairing matrix for curves in the base, a concise way to denote the tensor branch configuration is in terms of a quiver-like graph, where each node, denoted as $\overset{\fkg_i}{n_i}$, encodes the $i$th gauge algebra $\fkg_i$, whose associated tensor has self-pairing $A^{ii} = -n_i$. In what follows, we shall allow for the possibility that the gauge algebra is trivial, i.e., $\mathfrak{g}_i = \emptyset$, in which case no decoration is necessary. On the tensor branch, the condition of 6d gauge anomaly cancellation is, up to a small number of corner cases, enough to characterize the matter content of the tensor branch theory, including the spectrum of hypermultiplets.\footnote{There are a small number of cases, such as $\mathfrak{su}_6$ with $n_i = 1$ where the hypermultiplet spectrum is not uniquely fixed by the gauge algebra and self-pairing.}

Now, the F-theory model directly specifies the gauge symmetry, as associated with 7-branes wrapped on compact curves of the base $\mathcal{B}$, and this splits up into a collection of simple non-Abelian gauge symmetries:
\begin{equation}
\mathfrak{g}_{\text{gauge}} = \underset{i}{\bigoplus} \mathfrak{g}_{i}.
\end{equation}
Each factor here is a simple Lie algebra. Moreover, there are no gauged Abelian $\mathfrak{u}(1)$ factors, as follows directly from the structure of the local F-theory models \cite{Heckman:2013pva}. Turning next to the flavor symmetries of the 6d SCFT, the tensor branch description typically provides a good first approximation of the flavor symmetries of the 6d SCFT. For example, the hypermultiplets of the effective field theory often rotate under a global symmetry, and this persists at the conformal fixed point. In some cases, certain candidate flavor symmetries only become apparent once we approach the fixed point. A classic example of this phenomena is the E-string theory, namely the theory of an M5-brane probing an $E_8$ nine-brane. From the perspective of \cite{DelZotto:2014hpa,Heckman:2014qba}, the E-string, as well as other sub-configurations of matter fields can be viewed as the tensor branch of a generalized type of matter where the flavor symmetry is manifest, namely ``conformal matter''. All of this is to say that there is by now a general algorithm to read off the candidate non-Abelian flavor symmetry through a combination of the top-down F-theory geometry, and additional strong coupling enhancements (see, e.g., \cite{DelZotto:2014hpa,Heckman:2014qba,Heckman:2015bfa,Heckman:2016ssk} for some examples of such analyses). There is also a general algorithm for reading off candidate $\mathfrak{u}(1)$ symmetries which are free from mixed gauge symmetry/$\mathfrak{u}(1)$ anomalies, so-called ABJ anomalies \cite{Lee:2018ihr,Apruzzi:2020eqi}. Putting all of this together, the flavor symmetry algebra is of the general form:
\begin{equation}
\mathfrak{g}_{\text{flavor}} = \underset{a}{\bigoplus} \mathfrak{g}_{a} \oplus \underset{f}{\bigoplus} \mathfrak{u}(1)_{f},
\end{equation}
where each factor $\mathfrak{g}_{a}$ refers to a simple non-Abelian Lie algebra, and we have also included possible continuous Abelian symmetry factors. As a general point of notation, we shall distinguish the non-Abelian gauge and flavor symmetry algebras by the respective indices $i$ and $a$, while Abelian flavor symmetry algebras are indexed by $f$. Indeed, in the corresponding topological Green--Schwarz--Sagnotti--West terms, we will have couplings to both sorts of gauge bundle curvatures. Here, we have allowed for the possibility of various enhancements, as captured by working with conformal matter.
Finally, there is also the R-symmetry of the 6d SCFT, and this is also present on the tensor branch since it is unbroken. This provides an additional $\mathfrak{su}(2)_{R}$ global symmetry algebra.
Putting all of this together, the continuous global symmetry of the system is:
\begin{equation}
\mathfrak{g}_\text{gauge-global} = \mathfrak{g}_{\text{gauge}} \oplus \mathfrak{g}_{\text{flavor}} \oplus \mathfrak{su}(2)_R,
\end{equation}
where we have left implicit the spacetime symmetries. We again stress that in many cases, we can deduce the corresponding symmetry algebra from earlier work, so the main task reduces to determining the symmetry group, rather than just the algebra.

To accomplish this, we will need to know more about the topological sector of the theory. Much as in \cite{Apruzzi:2020zot}, we mainly claim that it suffices to study the topological terms of the tensor branch theory. Some of such terms are necessary for the theory to be free of gauge symmetry anomalies in the first place, while other terms inform us of global symmetry anomalies. All of these are captured by couplings between the anti-chiral two-forms and the Chern character of the non-Abelian gauge field strengths, as required to satisfy 6d anomaly cancellation via the Green--Schwarz--Sagnotti--West mechanism \cite{Green:1984bx, Sagnotti:1992qw}. Including background field strengths from global symmetries,
we get a set of topological couplings:
\begin{equation}\label{eq:green-schwarz-coupling}
    2 \pi \int_{{\cal M}_6} \Theta_i \wedge I^i \, ,
\end{equation}
where $\Theta_{i}$ refer to the anti-chiral two-forms of the $i$th tensor multiplet (denoted as $t_i$), and the $I^{i}$ are a collection of
four-forms:
\begin{align}
\begin{split}
    I^i = & - \sum_j A^{ij} \, c_2(F_j) - \sum_a B^{ia} \, c_2(F_a) + \sum_{f,f'} C^{i;f,f'} \frac{c_1(F_f) \wedge c_1(F_f')}{2} \\
          & + y^i c_2(R) - (2+A^{ii}) \tfrac14 p_1(T)\, .
\end{split}
\end{align}
Here, the $F_j$ refers to the gauge field strengths, the $F_a$ are the field strengths for the non-Abelian flavor symmetry factors, and the $F_f$ are the field strengths for Abelian symmetry factors, all of which we have expressed in terms of the corresponding Chern characters.\footnote{In our conventions, $\frac{1}{4}\text{Tr} F^2 = c_2(F)$ and $c_1(F) = \sqrt{-1} F$.} On the second line, we have also included the contribution from the R-symmetry, $c_2(R)$, as well as the first Pontryagin class of the spacetime tangent bundle.
Turning next to the coefficients appearing in $I^{i}$, the matrix $A^{ij}$ is, in our conventions, negative definite, and encodes the Dirac pairing for the effective strings, while the $B^{ia}$ are coefficients determined by the cancellation of all gauge-flavor anomalies, i.e., terms proportional to $\text{Tr}(F_i^2) \text{Tr}(F_a^2)$ in the full anomaly polynomial \cite{Ohmori:2014pca,Ohmori:2014kda,Intriligator:2014eaa,Cordova:2020tij}
\begin{equation}
    I_8 = I_\text{1-loop} + I_\text{GS} = I_\text{1-loop} - \frac12 (A^{-1})_{ij} I^i I^j \,.
\end{equation}
As an additional comment, the only $\mathfrak{u}(1)$ symmetry factors we can include are those which are free from ABJ-anomalies, which are encoded in the coefficients of $\text{Tr}(F_i^3) F_f$-terms of the anomaly polynomial.
Such terms must vanish at 1-loop for any quantum mechanically unbroken flavor $\mathfrak{u}(1)$.
For 6d SCFTs on their tensor branch, one can determine all such flavor $\mathfrak{u}(1)$s from a bottom-up approach \cite{Apruzzi:2020eqi} (see also \cite{Lee:2018ihr}). We note that, as opposed to non-Abelian flavor symmetries, these $\mathfrak{u}(1)$s are sometimes geometrically delocalized.

The main tool at our disposal for determining the global form of $G_{\text{gauge-global}}$ will be to track the global bundle structure of background field configurations using large symmetry transformations. Via the Green--Schwarz--Sagnotti--West mechanism, we know that this will also involve a non-trivial transformation from the anti-chiral two-forms $\Theta^{i}$, and the combined effect must be such that the full set of topological contributions remains invariant. We now proceed to the determination of this global structure.

\section{Topology of Global Symmetry Group for 6d SCFTs}\label{sec:6d}

In this section, we determine the global structure of the symmetry groups for 6d ${\cal N} = (1,0)$ SCFTs, based on their tensor branch characterization as a weakly-coupled gauge theory. In what follows, we assume that the symmetry algebra $\mathfrak{g}_{\text{gauge-global}}$ has already been specified. There is a corresponding ``naive'' answer for the zero-form symmetry:
\begin{equation}
\widetilde{G}_{\text{gauge-global}} = \widetilde{G}_{\text{gauge}} \times \widetilde{G}_{\text{flavor}} \times SU(2)_R \,,
\end{equation}
namely, for each non-Abelian Lie algebra, we take the corresponding simply connected Lie group, and all Abelian factors simply lift to $U(1)$. As before, we leave the spacetime symmetries implicit. The answer is naive, in the sense that this analysis does not distinguish between symmetries acting on genuine local operators, and those which are only defined as the endpoints of line operators (see \cite{Bhardwaj:2021wif, Lee:2021crt, Cvetic:2022imb, DelZotto:2022joo}). Indeed,
on general grounds, we expect that the actual zero-form symmetry group is quotiented by a subgroup of the common center for these factors. We shall refer to this as the gauge-global center symmetry, writing it as:
\begin{equation}
G_{\text{gauge-global}} = \widetilde{G}_{\text{gauge-global}} / \mathcal{C} \,.
\end{equation}
This leaves us with a residual center which is present in the actual tensor branch theory. With this in hand, we also have a candidate global symmetry for the 6d SCFT, as given by projection onto just the global symmetries of this quotient. Note that the group quotient specified by
$\mathcal{C}$ has a canonical restriction to just the global symmetries. In the obvious notation, we then have:
\begin{equation}
G_{\text{global}} = \widetilde{G}_{\text{global}} / \mathcal{C}_{\text{global}} \,.
\end{equation}

Our aim will be to extract $G_{\text{global}}$ by determining the corresponding center symmetry group $\mathcal{C}$. The analysis of this proceeds in several stages. First of all, we must require that all matter fields, including weakly coupled hypermultiplets as well as generalizations such as E-strings and conformal matter are all neutral under $\mathcal{C}$. Additionally, precisely because the group of gauge transformations in the 6d tensor branch theory also requires an accompanying transformation of the chiral two-forms of the associated tensor multiplets, we must \textit{also} require that the corresponding effective strings are neutral under $\mathcal{C}$ (see, e.g.,\cite{Apruzzi:2020zot}). In practical terms, what this amounts to is analyzing the topological sector of the tensor branch theory, and the response of the effective action under large field transformations. This leads to a non-trivial correlation between candidate 0-form symmetry bundles, which will in turn allow us to read off $G_{\text{global}}$.

In the rest of this section, we spell out the steps for extracting $G_{\text{global}}$ directly from the tensor branch. First, we begin by tracking the mixed gauge-flavor center symmetry for non-Abelian symmetry factors. We then show how to incorporate continuous Abelian symmetry factors, and then turn to possible mixing with the R-symmetry factors. The specific case of the R-symmetry group is particularly subtle, since it can evade detection via other means such as superconformal index computations. In each step, we present some illustrative examples, which we revisit to exhibit the full global symmetry structure.

\subsection{Anomalies for Center--Flavor Symmetry}\label{sec:cfanom}

We begin by considering the core example, based on mixing between the center of the gauge groups and non-Abelian symmetry factors. For now, we therefore suppress the contributions from Abelian symmetry factors as well as the R-symmetry. For a non-Abelian flavor symmetry that rotates matter charged under a gauge symmetry, a non-trivial global flavor symmetry structure generally requires a center-twisted gauge bundle that compensates the twisted flavor bundle. There is a potential obstruction to turning on such gauge and flavor bundles, which can be quantified from the tensor branch data \cite{Apruzzi:2020zot} and the topological couplings in equation \eqref{eq:green-schwarz-coupling}.

Turning on a center-twisted bundle for a simple algebra $\fkg$ (flavor or gauge), with simply-connected group $\widetilde{G}$ and center $Z(\widetilde{G})$ now leads to a fractionalization of $c_2(F)$ \cite{Kapustin:2014gua,Gaiotto:2014kfa,Gaiotto:2017yup,Cordova:2019uob}:
\begin{align}
    \tfrac14 \text{Tr}(F^2) = c_2(F) \equiv -\alpha_{\fkg} \, w(F) \cup w(F) \mod \bbZ \, .
\end{align}
Here, $w(F) \in H^2({\cal M}_6, Z(\widetilde{G}))$ is the $Z(\widetilde{G})$-valued characteristic class (the generalized Stiefel--Whitney class, also called the Brauer class in the mathematical literature) measuring the obstruction to lift a $\widetilde{G} / Z(\widetilde{G})$-bundle to a $\widetilde{G}$ bundle, and $w \cup w \equiv w^2$ is a 4-cocycle with integer periods.\footnote{To be precise, $c_2(F) \equiv \alpha_\fkg {\cal P}(w) \mod \bbZ$, where ${\cal P}$ is the Pontryagin square operation. If $w \in H^2({\cal M}, \bbZ_n)$, then for $n$ odd, ${\cal P}(w) \equiv w \cup w \in H^4({\cal M}, \bbZ_n)$; for $n$ even, ${\cal P}(w) \in H^4({\cal M}, \bbZ_{2n})$ reduces to $w \cup w$ modulo $n$.}
The fractionalization is due to the factors $\alpha_{\mathfrak{g}}$, whose fractional values depend $\fkg$ (with non-trivial center $Z(\widetilde{G})$):
\begin{equation}
\begin{aligned}
    & \mathfrak{g} = \mathfrak{su}_n \,  (\mathbb{Z}_n): \quad &&\alpha_\mathfrak{g} = \tfrac{n-1}{2n} \, , \qquad && \mathfrak{g} = \mathfrak{sp}_n \, (\mathbb{Z}_2): \quad &&\alpha_\mathfrak{g} = \tfrac{n}{4} \, , \\
    & \mathfrak{g} = \mathfrak{e}_{6} \,  (\mathbb{Z}_3): \quad &&\alpha_\mathfrak{g} = \tfrac{2}{3} \, , \qquad && \mathfrak{g} = \mathfrak{e}_{7} \, (\mathbb{Z}_2): \quad &&\alpha_\mathfrak{g} = \tfrac34 \, , \\
    & \mathfrak{g} = \mathfrak{so}_{4n+2} \, (\bbZ_4) : \quad &&\alpha_\mathfrak{g} = \tfrac{2n+1}{8} \, , \qquad && \mathfrak{g} = \mathfrak{so}_{2n+1} \, (\bbZ_2) : \quad &&\alpha_\fkg = \tfrac12 \, .
\end{aligned}
\end{equation}
In the case $\mathfrak{g} = \mathfrak{so}_{4n}$ and $\widetilde{G} = Spin(4n)$
with center $\mathbb{Z}_2^{(1)} \times \mathbb{Z}_2^{(2)}$, there are two contributions,
\begin{align}
    c_2 \equiv - \big( \tfrac{n}{4} (w^{(1)} + w^{(2)})^2 + \tfrac12 w^{(1)} \cup w^{(2)} \big) \mod \bbZ \, ,
\end{align}
originating from the center background $w^{(i)}$ of $\mathbb{Z}^{(i)}$.
In general, each $\mathbb{Z}_{\ell_s}$ factor of the full center $\prod_{i} Z(\widetilde{G}_i) \times \prod_a Z(\widetilde{G}_a) = \prod_s \bbZ_{\ell_s}$ is accompanied by a background field $w_s$. Again, the $i$ index refers to the gauge groups and the $a$ index refers to the non-Abelian flavor groups.

Because of the topological couplings in equation  \eqref{eq:green-schwarz-coupling}, a general background $\tilde{w} = (w_1 , ..., w_s, ...)$ for the center $\prod_s \bbZ_{\ell_s}$\footnote{Indexing by $s$ the individual cyclic factors distinguishes the two $\bbZ_2$ factors for a factor of $\widetilde{G}_g \cong Spin(4k_g)$.} will lead to a fractional 4-cocycle coupling to the tensor $\Theta_i$,
\begin{align}\label{eq:center_obstruction}
\begin{split}
    & \sum_j A^{ij} \, c_2(F_j) + \sum_a B^{ia} c_2(F_a) =: \sum_g {\cal A}^{ig} c_2(F_g) \\
    \equiv & \, - \sum_{\fkg_g \neq \mathfrak{so}(4n)} {\cal A}^{ig} \, \alpha_{\fkg_g} w_g^2 - \sum_{\fkg_g = \mathfrak{so}(4n_g)} {\cal A}^{ig} \left( \frac{n_g}{4} (w^{(1)}_g + w^{(2)}_g)^2 + \frac12 w^{(1)}_g \cup w^{(2)}_g \right)  \mod \bbZ \, ,
\end{split}
\end{align}
where the index $g$ runs over both gauge and flavor factors, and ${\cal A}^{ig}$ is the combined matrix of the tensor pairings $A^{ij}$ and non-Abelian flavor coefficients $B^{ia}$.
Because of this fractionalization, the action transforms anomalously under a large gauge transformation of the $i$th two-form tensor $\Theta_i$ \cite{Apruzzi:2020zot}, which poses an obstruction to turning on the corresponding twisted bundles.

However, for subgroups $Z \subset \prod_s \mathbb{Z}_{\ell_s}$, for which the $w_s$ are related to each other, it may be possible that different fractional contributions cancel, so that equation \eqref{eq:center_obstruction} is an integer class.
Concretely, for a cyclic $\bbZ_{n_r}$ subgroup with generator $(k^{(r)}_1, k^{(r)}_2, ...) \in  \prod_s \mathbb{Z}_{\ell_s}$, the corresponding center background is parametrized by $\tilde{w}^{(r)} = (k_1^{(r)} w^{(r)}, k_2^{(r)} w^{(r)}, \cdots)$, for a single independent 2-cocycle $w^{(r)}$.
If the fractionalizations vanish for a linear combination $\tilde{w} = \sum_r \tilde{w}^{(r)} = (\sum_r k_1^{(r)} w^{(r)}, \sum_r k_2^{(r)} w^{(r)}, ...)$ with generic backgrounds $w^{(r)}$ for a subgroup $Z = \prod_r \bbZ_{n_r}$, the global structure of the symmetry group is:\footnote{A short comment on notation: we reserve $\mathcal{C}$ for the full quotienting subgroup, with $Z$ the quotient on just the non-Abelian symmetry factors.}
\begin{equation}
G_{\text{gauge-global}} = \frac{\prod_i \widetilde{G}_i \times \prod_a \widetilde{G}_a}{Z}.
\end{equation}

Note that the candidate subgroups $Z$ of interest are in general severely limited by requiring that the hypermultiplet spectrum of the tensor branch theory must transform trivially under it. Any subgroup $Z$ of the full center which rotates these states by a non-trivial phase $\varphi \in U(1)$ is explicitly broken, i.e., one cannot twist the bundles by $Z$, regardless of the anomaly above.

For a simple group $G$ with $Z(G) = \bbZ_n$, a center element $x \,( \text{mod } n\bbZ) \in \bbZ_n$ acts on an irreducible representation ${\bf R}$ by the phase $\varphi^x({\bf R}) := (\varphi({\bf R}))^x$, with the phase $\varphi({\bf R})$ for the generator $1 \in \bbZ_n$ computed as follows:\footnote{In general, any irrep ${\bf R}$ of $G$ defines an element $\varphi({\bf R}) \in \text{Hom}(Z(G), U(1)) = \widehat{Z(G)} \cong Z(G)$ of the Pontryagin-dual.
The phase $\varphi^x({\bf R}) \in U(1)$ is then just the image of $x$ under $\varphi({\bf R})$.
}
\begin{itemize}
    \item for $G = SU(n)$, and ${\bf R}$ having a Young-tableaux with $m$ boxes, then $\varphi({\bf R}) = e^{2\pi i \tfrac{m}{n}}$;

    \item for $G = Sp(n)$, $\varphi({\bf R} = \text{fund}) = -1$ and $\varphi({\bf R} = \text{anti-sym}) = 1$;

    \item for $G = Spin(2n+1)$, $\varphi({\bf R} = \text{vector}) = 1$ and $\varphi({\bf R} = \text{spinor}) = -1$;

    \item for $G = Spin(4n+2)$, $\varphi({\bf R} = \text{vector}) = -1$ and $\varphi({\bf R} = \text{spinor}) = i$;

    \item for $G = E_6$, $\varphi({\bf R} = \text{fund}) = e^{\frac{2\pi i}{3}}$;

    \item for $G = E_7$, $\varphi({\bf R} = \text{fund}) = -1$.
\end{itemize}
For $G = Spin(4n)$ with $Z(G) = \mathbb{Z}_2^{(1)} \times \mathbb{Z}_2^{(2)}$, the phases associated with the generator $(1,0)$ are $\varphi^{(1)}({\bf R} = \text{vector})  = \varphi^{(1)}({\bf R} = \text{spinor}) = -1$, $\varphi^{(1)}({\bf R} = \text{co-spinor}) = 1$, and those associated with $(0,1) \in Z(G)$ are $\varphi^{(2)}({\bf R} = \text{vector})  = \varphi^{(2)}({\bf R} = \text{co-spinor}) = -1$, $\varphi^{(1)}({\bf R} = \text{spinor}) = 1$.
For a general element $(x_1, x_2) \in Z(G)$, the phase is then $\varphi^{(x_1,x_2)} ({\bf R}) = (\varphi^{(1)}({\bf R}))^{x_1} \, (\varphi^{(2)}({\bf R}))^{x_2}$.

For the center of a semi-simple group $\prod_g G_g \ni x = (x_g)$, one can analogously compute the phase from acting on a representation ${\bf R} = \bigotimes_g {\bf R}_g \equiv ({\bf R}_1, {\bf R}_2,...)$ as $\varphi^x({\bf R}) = \prod_g \varphi^{x_g}_g({\bf R})$.
Hence, for $Z \subset Z(\prod_g G_g) = \prod_g Z(G_g)$ to leave \emph{all} hypermultiplets invariant, $\varphi^x({\bf R}) = 1$ for all $x \in Z$ and all representations ${\bf R}$ that appear.
If, in addition, the obstruction in equation \eqref{eq:center_obstruction} vanishes, we propose that it is consistent to turn on the corresponding center twist (see also \cite{Apruzzi:2021mlh}).
This includes in particular the examples studied in \cite{Ohmori:2018ona}, and also agrees with expectations from explicit geometric constructions, where one can show that excitations of BPS-strings are invariant under $Z$ \cite{Apruzzi:2020zot,Bhardwaj:2020phs,Apruzzi:2021mlh}.

\subsubsection{Examples}

We now turn to examples illustrating how we extract the non-Abelian flavor symmetries. Let us also note that recently in \cite{Cvetic:2022imb}, geometric methods were developed to directly extract the global symmetry group for 5d conformal matter, i.e., the circle reduction of 6d conformal matter. Our bottom up analysis agrees with the results found there.

\paragraph{Example 1:}
Consider the SCFT with tensor branch description:
\begin{align}\label{eq:A-type_quiver_with_suN}
    [\mathfrak{su}_N^{(L)} ] \, \overset{\mathfrak{su}^{(1)}_{N}}{2} \, \overset{\mathfrak{su}^{(2)}_{N}}{2} \, \cdots \, \overset{\mathfrak{su}^{(m-1)}_{N}}{2} \, \overset{\mathfrak{su}^{(m)}_{N}}{2} \, [\mathfrak{su}_N^{(R)}] \, ,
\end{align}
which consists of $m$ gauge factors $\mathfrak{su}^{(i)}_N$ and has two $\mathfrak{su}_N^{(a)}$ ($a = L, R$) flavor factors at each end of the quiver.
The hypermultiplet spectrum consists of bifundamentals between each adjacent factor of $SU(N)_L \times \prod_i SU(N)^{(i)} \times SU(N)_R$:
\begin{align}
    {\bf R}^{(1)} = ({\bf N}, \overline{\bf N}, {\bf 1}, {\bf 1}, \ldots) \, , \quad {\bf R}^{(2)} = ({\bf 1}, {\bf N}, \overline{\bf N}, {\bf 1}, \ldots) \, , \quad \cdots \,.
\end{align}
With the tensor pairing matrix $A^{ij}$ being the negative $SU(m+1)$ Cartan matrix, the anomalies proportional to $\text{Tr}(F_i^2) \text{Tr}(F_a^2)$ are cancelled by a Green--Schwarz term with $B^{iL} = \delta^{i,1}$ and $B^{iR} = \delta^{m,R}$.
So the relevant part of the GS-coupling in equation \eqref{eq:green-schwarz-coupling} is
\begin{align}
    \Theta_i \wedge \left( -c_2(F_{i-1}) + 2 c_2(F_{i}) - c_2(F_{i+1}) \right) ,
\end{align}
where $F_{0} := F_L$ and $F_{m+1} := F_R$.

It is easy to see that the hypermultiplet spectrum is invariant under the diagonal center $\bbZ_N$ with generator
\begin{align}
    (1,1,...) \in \bbZ^{(L)}_N \times \prod_i \bbZ^{(i)}_N \times \bbZ_N^{(R)} = Z(SU(N)_L \times \prod_i SU(N)^{(i)} \times SU(N)_R) \, .
\end{align}
Since for this generator, all $-c_2(F_i) \equiv \frac{N-1}{2N} w^2$ fractionalize equally, they cancel out for each tensor multiplet $t_i$.
Therefore, the non-Abelian symmetry group is $[SU(N)_L \times \prod_i SU(N)^{(i)} \times SU(N)_R] / \bbZ_N$, and the non-Abelian flavor symmetry of the SCFT is $[SU(N)_L \times SU(N)_R]/\bbZ_N$, which agrees with known results \cite{Bah:2017gph, Cvetic:2022imb}. As an additional comment, we note that this case also has an overall $\mathfrak{u}(1)$ flavor symmetry \cite{Bah:2017gph, Apruzzi:2020eqi}, so we will revisit it when we discuss Abelian symmetry factors.

\paragraph{Example 2:}
For $N\geq 5$, there is an SCFT with tensor branch description:
\begin{align}\label{eq:single_1-curve-example}
    \underset{[\#\bigwedge^2 = 1]}{\overset{\mathfrak{su}_{N}}{1}}\, [\mathfrak{su}_{N+8}] \, ,
\end{align}
with a bifundamental hypermultiplet ${\bf R}^{(1)} = ({\bf N}, \overline{\bf N + 8})$ under $\mathfrak{su}_N \oplus \mathfrak{su}_{N+8}$, and one anti-symmetric ${\bf R}^{(2)} = (\frac{{\bf N}({\bf N -1})}{\bf 2}, {\bf 1}) = (\bigwedge^2, {\bf 1})$ that is uncharged under the $\mathfrak{su}_{N+8}$ flavor.
The Green--Schwarz four-form for the single tensor $\Theta$ of self-pairing $-1$ contains
\begin{align}
    I \supset c_2(F_N) - c_2(F_{N+8}) \,,
\end{align}
which ensures the absence of any $\text{Tr}(F_N^2) \text{Tr}(F_{N+8}^2)$ anomaly.

Some basic arithmetic reveals that there can be at most a non-trivial $\bbZ_2 \subset \bbZ_N \times \bbZ_{N+8} = Z(SU(N) \times SU(N+8))$ that acts trivially on the hypermultiplets, and that this can only occur when $N$ is even. So, for $N$ odd, there is no center-flavor symmetry. Restricting to $N$ even, the $\mathbb{Z}_2$ subgroup is generated by the element $(\tfrac{N}{2}, \tfrac{N+8}{2}) \in \bbZ_N \times \bbZ_{N+8}$.
For this candidate subgroup, the center-flavor anomaly indeed vanishes:
\begin{align}
    c_2(F_N) - c_2(F_{N+8}) \equiv \big( \underbrace{ - \tfrac{N^2}{4} \tfrac{N-1}{2N} + \tfrac{(N+8)^2}{4} \tfrac{N+7}{2N+16}}_{=7+2N} \big) w^2 \mod \bbZ \,.
\end{align}
Therefore, the faithfully acting non-Abelian symmetry group for $N$ even is $[SU(N) \times SU(N+8)]/\bbZ_2$, and the non-Abelian flavor symmetry of the SCFT is $SU(N+8)/\bbZ_2$.

\paragraph{Example 3:}
Consider next the SCFT with tensor branch description
\begin{align}\label{eq:example_3_tensor_config}
    [\mathfrak{su}_3^{(L)}] \, \overset{\mathfrak{e}_6}{3} \, 1 \, \overset{\mathfrak{su}_2}{2} \, [\mathfrak{so}_7^{(R)}] \, ,
\end{align}
which has
\begin{equation}
    -A^{ij} = \begin{pmatrix} 3 & -1 & 0 \\ -1 & 1 & -1 \\ 0 & -1 & 2 \end{pmatrix} \,,
\end{equation}
and hypermultiplets in the representations
\begin{align}
    {\bf R}^{(1)} = (\overline{\bf 3}, {\bf 27}, {\bf 1}, {\bf 1}) \, , \quad {\bf R}^{(2)} = \tfrac12 ({\bf 1}, {\bf 1}, {\bf 2}, {\bf 8}) \, ,
\end{align}
under the symmetry factors $\mathfrak{su}_3 \oplus \mathfrak{e}_6 \oplus \mathfrak{su}_2 \oplus \mathfrak{so}_7$. In the above, the ``$\frac{1}{2}$'' denotes a half-hypermultiplet, with matter in the spinor representation of $\mathfrak{spin}_7 \simeq \mathfrak{so}_7$.
Note also that this theory contains an undecorated $-1$ curve, so it provides an example where a subalgebra of the
E-string theory flavor symmetry has been gauged.

Let us now turn to the global structure of the symmetry group. The naive answer is $\widetilde{G}_{\text{gauge-global}} = SU(3) \times E_6 \times SU(2) \times Spin(7)$. Observe that the matter fields are invariant under a $\bbZ_3 \times \bbZ_2$ subgroup of the full center, where the $\bbZ_3$ is the diagonal of $Z(SU(3) \times E_6) = \bbZ_3 \times \mathbb{Z}_3$, and the $\bbZ_2$ the diagonal of $Z(SU(2) \times Spin(7)) = \bbZ_2 \times \mathbb{Z}_2$. Consider next the Green--Schwarz coupling to the tensor $\Theta_2$ of the unpaired middle node. This is an E-string not touching the flavor factors at the ends of the quiver, we find:
\begin{align}
    \Theta_2 \wedge \left( -c_2(F_{\mathfrak{e}_6}) - c_2(F_{\mathfrak{su}_2}) \right) \equiv \Theta_2 \wedge \left(\tfrac23 w_{\bbZ_3}^2 + \tfrac14 w_{\bbZ_2}^2 \right) \mod \bbZ \, ,
\end{align}
which would induce an anomaly for the large gauge transformations of $\Theta_2$.

As explained in \cite{Apruzzi:2020zot,Bhardwaj:2020phs}, the inconsistency of turning on such a twisted background, despite the absence of non-invariant hypermultiplets, can be also attributed to the excitations of the E-string, which transform in $E_8$ representations.
By decomposing the adjoint under $\mathfrak{e}_8 \supset \mathfrak{e}_6 \oplus \mathfrak{su}_3 \supset \mathfrak{e}_6 \oplus \mathfrak{su}_2$,
\begin{align}
\begin{split}
    {\bf 248} & \rightarrow ({\bf 78, 1}) \oplus ({\bf 1, 8}) \oplus  ({\bf 27, 3})  \oplus (\overline{\bf 27}, \overline{\bf 3}) \\
    & \rightarrow ({\bf 78, 1}) \oplus ({\bf 1, 3}) \oplus ({\bf 1, 2})^{\oplus 2} \oplus (({\bf 27, 2}) \oplus ({\bf 27, 1}) + \text{c.c}) \oplus ({\bf 1,1}) \, ,
\end{split}\label{eq:branching_e8_to_u1}
\end{align}
we indeed find states (the fundamentals under $E_6$ and $SU(2)$, respectively), which break the $\bbZ_3$ and $\bbZ_2$ twists, respectively.
Therefore, the non-Abelian flavor group is $SU(3) \times Spin(7)$.
However, as we will see below, the two discrete twists can be compensated if we take into the account the existence of $U(1)$ flavor factors.

\subsection{Anomalies for Center Symmetries of Abelian Factors}\label{sec:ab}

In the previous subsection we primarily focused on the non-Abelian symmetry factors. In some cases, there can also be continuous Abelian symmetry factors, which in many cases are delocalized. The procedure for extracting the global form of the center-flavor symmetry is to start with the ``naive'' gauge-global symmetry $\widetilde{G}_{\text{gauge-flavor}}$, and to then determine large symmetry transformations compatible with the presence of these $U(1)$ symmetry factors. The common center $\mathcal{C} \subset \widetilde{G}_{\text{gauge-flavor}}$ then specifies the quotient $G_{\text{gauge-flavor}} = \widetilde{G}_{\text{gauge-flavor}} / \mathcal{C} $. Note that we will also need to determine the overall normalization of $\mathfrak{u}(1)$ charges, a point we turn to shortly.

The analysis of the global form again relies on the same sort of topological terms $\Theta_i \wedge I^{i}$ encountered in our analysis of non-Abelian flavor symmetries. In the present case with Abelian symmetries, we recall that this includes:
\begin{equation}\label{eq:green-schwarz-4form_with_U1s}
    I^i = - \sum_j A^{ij} c_2(F_j) - \sum_a B^{ia} c_2(F_a) + \sum_{f,f'} C^{i;f,f'} \frac{c_1(F_f) \wedge c_1(F_f')}{2} \, .
\end{equation}
If we now activate a discrete twist $Z_f \cong \bbZ_{n_f}$ of a $U(1)_f$ bundle, then $c_1(F_f) = i F_f$ acquires a fractional part, thus affecting the large gauge transformations of the $\Theta_i$s. From this, we see that the symmetry group takes the general form
\begin{align}
    \frac{[\prod_i \widetilde{G}^{\text{gauge}}_i \times \prod_a \widetilde{G}^{\text{flavor}}_a]/Z \times \prod_f U(1)_f}{\prod_f Z_{f}} \, ,
\end{align}
where as before, the $\widetilde{G}$s refer to simply connected non-Abelian factors. In particular, notice that the quotient by $Z$, which we obtained in the previous subsection, just involves the condition of neutrality under a restricted set of center-symmetry transformations associated with the \textit{non-Abelian} symmetry factors. There can, of course, be more general symmetry transformations which involve the \textit{Abelian} factors, and this is accounted for by the $Z_f$s.

Indeed, for each $\Theta_i$, the fractionalizations of $c_2(F_i)$, $c_2(F_a)$, and $c_1(F_f)$ from the twists $Z$ and $Z_{f}$ must cancel. Note in particular that the quotienting procedure worked out for the non-Abelian symmetry factor is not contaminated by the appearance of the $U(1)$ factors. Said differently, the quotienting group $Z$ may end up only being a subgroup of the full $\mathcal{C}$ used to reach $G_{\text{gauge-global}} = \widetilde{G}_{\text{gauge-global}} / \mathcal{C}$.

To figure out the global quotient by $Z_f$, we need to know the overall normalization of the matter fields under the Abelian symmetries. This is rather subtle, because for a $U(1)$ factor, rescaling the charges is always a possibility. Importantly, such rescaling effects do not end up affecting the global form of the quotienting procedure. To demonstrate this, we now turn to an analysis of charge normalization for Abelian factors, and then illustrate how this works for hypermultiplets and E-string theories.

\subsubsection{Abelian Charge Normalization}

To determine the overall charge normalization for Abelian symmetry factors, as well as the contribution from fractional Chern classes,
it is instructive to consider bundles with structure group $U(N) = [SU(N) \times U(1)_f]/\bbZ_N$.
One can express a $U(N)$ bundle in terms of an $SU(N)/\bbZ_N$ and a $U(1)_f$ bundle, with curvatures $F$ and $F_f$ correlated via:
\begin{align}\label{eq:c1_U(N)_bundle}
    c_1(F_f) \equiv \tfrac{1}{N} w \mod \bbZ \, ,
\end{align}
where $w$ is the (generalized) Stiefel--Whitney class of the $SU(N)/\bbZ_N$ bundle.
In the language of generalized symmetries, one can think of the two 1-form center symmetries of $SU(N)$ and $U(1)_f$ being correlated through a single 2-cocycle $w$.
Namely, the background gauge field $b_e^{(2)}$ of the 1-form symmetry of $U(1)_f$ (which is itself $U(1)$-valued), which imposes $\int_{\Sigma_2} (c_1(F_f) - b_e^{(2)}) \in \bbZ$ for any 2-cycle $\Sigma_2$, is tied to the value of the $\bbZ_N$ 1-form symmetry gauge field of $SU(N)$, which in turn fixes the Stiefel--Whitney class to $w$.
We can verify explicitly that the fractional parts of the $SU(N)/\bbZ_N$ bundle, $c_2(F) \equiv - \tfrac{N-1}{2N}w^2 \mod \bbZ$, and of the $U(1)_f$ bundle, $Nc_1(F_f)^2 \equiv \tfrac{1}{N} w^2 \mod \bbZ$, cancel in $c_2(U(N)) = c_2(F) + \tfrac{N(N-1)}{2} c_1(F_f)^2$, which is indeed an integer characteristic class.

Importantly, the relation in equation \eqref{eq:c1_U(N)_bundle} holds only in a normalization of the $\mathfrak{u}(1)$ generator $\hat{q}$ where the charges span $\bbZ$, i.e., the fundamental representation of $\mathfrak{u}(N)$ has charge 1 in this normalization, and representations that are singlets under $\mathfrak{su}(N) \subset \mathfrak{u}(N)$ have charges 0 mod $N$.
In this case, the trivially acting $\bbZ_N$ center is generated by $(-1, e^{2\pi i \hat{q}/N}) \in  \bbZ_N \times U(1) = Z(SU(N) \times U(1))$.

More generally, once we fix the $U(1)_f$ charges of all representations ${\bf R}_q$ of a group $[\prod_g \widetilde{G}_g \times U(1)_f]/Z_{f}$ (with ${\bf R}$ a representation of $\prod_g \widetilde{G}_g$) to span $\bbZ$, there is no ambiguity to specify the generator of $Z_{f}$ as
\begin{align}\label{eq:normalized_center_generator_with_u1}
    (k_1,k_2,...; e^{2\pi i \hat{q} \frac{u_f}{l_f}}) \in \prod_g Z(G_g) \times U(1)_f \,,
\end{align}
where parameters must satisfy $\varphi({\bf R})^{(k_1,k_2,...)} \, \exp(2\pi i q \tfrac{u_f}{l_f}) = 1$ for any representation ${\bf R}_q$ of $[\prod_g \widetilde{G}_g \times U(1)_f]/Z_{f}$.
Then, the corresponding twist of the symmetry bundle is in terms of a 2-cocycle $w$:
\begin{align}\label{eq:1-form_twists_normalized_with_u1}
    c_1(F_f) \equiv \frac{u_f}{l_f} w \mod \bbZ \, , \quad w(F_g) = k_g w \, ,
\end{align}
with the understanding that when $\widetilde{G}_g \cong Spin(4m_g)$, we have $k_g \equiv (k_g^{(1)}, k_g^{(2)})$, and $w(F_g) \equiv (w_g^{(1)}, w_g^{(2)})$. Note that it is $c_1(F_f) c_1(F_{f'})$ that enters the four-forms $I^i$, whose fractional part,
\begin{align}
    c_1(F_f) c_1(F_{f'}) \equiv \frac{u_f u_{f'}}{l_f l_{f'}} w \cup w' + \frac{u_f}{l_f} w \cup \chi' + \frac{u_{f'}}{l_{f'}} w' \cup \chi \mod \bbZ \, ,
\end{align}
may depend on the integral parts, $\chi$ and $\chi'$, of $c_1(F_f)$ and $c_1(F_{f'})$, respectively.

Now, since we are dealing with Abelian symmetry factors, we can in principle consider rescaling the charges of the states so that we only span a rescaled subgroup of $\mathbb{Z}$, e.g., $\mathbb{Z} \rightarrow \lambda \mathbb{Z}$. Doing so has no effect on the structure of the topological Green--Schwarz couplings.\footnote{In F-theory models there is often a ``geometrically preferred'' normalization where $SU(N)$-fundamentals have charge $\frac{1}{N} \mod \bbZ$ \cite{Cvetic:2017epq}.} Indeed, on general grounds, the fractionality of $C^{i;f,f'} c_1(F_f) c_1( F_{f'})$ does not depend on the normalization.\footnote{From the formulae for $C^{i;f,f'}$ we will discuss shortly, it can be seen explicitly that the effect $c_1(F_f) \rightarrow \tfrac{1}{\lambda} c_1(F_f)$ under a charge rescaling $q_f \rightarrow \lambda q_f$ is absorbed by $C^{i;f,f'}$.}
A convenient normalization convention for $c_1(F_f)$ is to first normalize $U(1)_f$ such that the corresponding charges span $\bbZ$ (and rescale $C^{i;f,f'}$ accordingly).
We can then determine the generator in equation \eqref{eq:normalized_center_generator_with_u1} of the candidate subgroup $Z_f$ that acts trivially on all states, from which the characteristic classes in equation \eqref{eq:1-form_twists_normalized_with_u1} follow.

\paragraph{\textbf{Hypermultiplets}}

To illustrate how this works, consider the case of weakly coupled hypermultiplets charged under some $U(1)$s. Indeed, on the full tensor branch, the only source of $\text{Tr}(F_i^2) F_f F_f'$-terms in the 1-loop anomaly polynomial are the hypermultiplets in representation ${\bf R}$ under $\prod_i \widetilde{G}^{\text{gauge}}_i \times \prod_a \widetilde{G}^{\text{flavor}}_a$ and with $U(1)$-charge vector $\vec{q}$:
\begin{align}
\begin{split}
    I_\text{hyper}({\bf R}_{\vec{q}}) \supset \frac{1}{24} \text{tr}_{{\bf R}_{\vec{q}}}({\cal F}^4) \supset & \sum_{f,f',i} \frac{h_i({\bf R})}{4} \text{Tr}(F_i^2) q_f q_{f'} \, F_f F_{f'} \\
    = & - \sum_{f,f',i} h_i({\bf R}) \, q_f \, q_{f'} \, c_2(F_i) \, c_1(F_f) \, c_1(F_{f'})\, ,
\end{split}
\end{align}
where the decomposition of the curvature ${\cal F}$ of the full symmetry bundle into those of the non-Abelian (gauge) part ($F_i$) and those of the $U(1)$s ($F_f = -i c_1(F_f)$) introduces the index of the representation $h_i({\bf R})$.\footnote{For ${\bf R} = ({\bf R}^{(1)}, {\bf R}^{(2)}, \cdots)$ an irrep of a semi-simple group $\prod_g G_g$, $h_i({\bf R}) = \prod_{g\neq i} \dim({\bf R}^{(g)}) \, h_{\fkg_i}({\bf R}^{(i)})$.
In our normalization of the trace, $h_{\mathfrak{su}_N}({\bf N}) = \frac12$. For values of other representations ${\bf R}$, see \cite{Heckman:2018jxk}, where these are denoted $h_{\bf R}$.} Much as in our analysis of non-Abelian gauge and flavor anomalies, requiring the Green--Schwarz contribution $I_\text{GS} = -\frac12 (A^{-1})_{ij} I^i I^j$ to cancel the above terms of the 1-loop anomaly polynomial uniquely fixes the coefficients $C^{i;f,f'}$ in the Green--Schwarz four-forms in equation \eqref{eq:green-schwarz-4form_with_U1s}:
\begin{align}\label{eq:U(1)_anomaly_coefficients}
    C^{i;f,f'} = \sum_{{\bf R}_{\vec q}} 2 h_i({\bf R}) q_f q_{f'} \, .
\end{align}
For F-theory models (in particular, those with compact internal geometries describing 6d supergravity), where $U(1)_f$ corresponds to a rational section $\sigma_f$ (more precisely, the Shioda-map of a rational section) of the elliptic fibration, the coefficient $C^{i;f,f'}$ is the geometric intersection number of the compact curve ${\cal C}_i$ carrying the gauge algebra $\fkg_i$ with the so-called height pairing divisor $\pi(\sigma_f, \sigma_{f'})$ \cite{Park:2011ji,Morrison:2012ei}. In some cases, this structure persists even in local models \cite{Lee:2018ihr, Apruzzi:2020eqi}.

\paragraph{\textbf{E-String Contributions}}

In most cases, the $U(1)$ symmetry only acts on weakly coupled hypermultiplets. The states from an E-string sector can also be charged under a $U(1)$ factor embedded in the $E_8$ flavor symmetry factor. We can also extract the charge normalization in this case, and thus track its contribution to the global structure of the symmetry.

Concretely, consider a maximal embedding $\mathfrak{e}_8 \supset \bigoplus_{\beta \geq -d} \mathfrak{h}_\beta \oplus \bigoplus_\gamma \mathfrak{u}(1)_\gamma$ with simple algebras $\mathfrak{h}_\beta$, of which the first $d \geq 1$ factors $\mathfrak{h}_{\beta <0}$ are gauged,\footnote{At most two simple factors can be gauged, thus $d \leq 2$ \cite{Heckman:2015bfa}.} i.e., paired with tensors $\Theta_{i_\epsilon}$ ($\epsilon \in \{ -1,...,-d\}$) having $A^{\hat{\imath}, i_\epsilon} = 1$.
This leaves the commutant, $\bigoplus_{\beta \geq 0} \mathfrak{h}_\beta \oplus \bigoplus_\gamma \mathfrak{u}(1)_\gamma$, as the flavor symmetry, which receives no 1-loop anomalies from hypermultiplets (hence, in particular, no ABJ anomaly for the $\mathfrak{u}(1)$ \cite{Apruzzi:2020eqi}).\footnote{For simplicity, we will only consider rank 1 E-strings. However, the generalization to rank $Q$ is straightforward with the results from \cite{Ohmori:2014pca,Ohmori:2014kda}.}
Nevertheless, besides those of other flavor factors with labels $(a,f,f')$ in equation \eqref{eq:green-schwarz-4form_with_U1s}, there is an E-string contribution to the Green--Schwarz four-form involving the flavor backgrounds $F_{\beta \geq 0}$ and $F_\gamma$, with \cite{Ohmori:2014kda,Ohmori:2014pca,Baume:2021qho},
\begin{align}
    B^{i \beta} = \delta^{i, \hat{\imath}} \, \ell_\beta \, , \quad C^{i;\gamma,\gamma'} = -\tfrac12 \delta^{i, \hat{\imath}} \, r_{\gamma,\gamma'} \,,
\end{align}
associated to the decomposition of the trace
\begin{align}\label{eq:decompose_trace}
    \text{Tr}(F_{\mathfrak{e}_8}^2) \rightarrow \sum_{\epsilon=-d}^{-1} \text{Tr}(F_{\mathfrak{h}_\epsilon}^2) + \sum_{\beta \geq 0} \ell_\beta \text{Tr}(F_{\beta}^2) + \sum_{\gamma,\gamma'} r_{\gamma,\gamma'} F_\gamma F_{\gamma'} = \sum_{\beta \geq -d} \ell_\beta \text{Tr}(F_{\beta}^2) + \sum_{\gamma,\gamma'} r_{\gamma,\gamma'} F_\gamma F_{\gamma'} \, .
\end{align}

The coefficients $\ell_\beta$ are the Dynkin indices of $\mathfrak{h}_\beta$ associated to the embedding $\bigoplus_{\beta \geq -d} \mathfrak{h}_\beta \oplus \bigoplus_\gamma \mathfrak{u}(1)_\gamma \subset \mathfrak{e}_8$, with those of the gauged subalgebras, $\mathfrak{h}_{\beta <0}$, necessarily being 1.\footnote{For the significance of Dynkin index one embeddings in F-theory, see \cite{Esole:2020tby}.}
To compute these coefficients, we can consider the decomposition of any representation,
\begin{align}
    {\bf R} \rightarrow \bigoplus_j ({\bf R}^{(j)}_{-d}, \cdots, {\bf R}^{(j)}_{\beta},\cdots)_{q^{(j)}_1,\cdots,q^{(j)}_\gamma,\cdots} \, ,
\end{align}
interpreted as a decomposition of a vector bundle
\begin{align}
    V = \bigoplus V^{(j)} \, , \quad \text{with} \quad V^{(j)} = \bigotimes_{\beta \geq -d} U^{(j)}_\beta \otimes \bigotimes_\gamma W^{(j)}_\gamma \, ,
\end{align}
where $U_\beta^{(j)}$ is an $\mathfrak{h}_\beta$-bundle in the representation ${\bf R}_\beta^{(j)}$, and $W^{(j)}_\gamma$ a $\mathfrak{u}(1)_\gamma$-bundle in the charge $q_\gamma^{(j)}$ representation.
Using the decompositions of the Chern character, $\text{ch}(A \otimes B) = \text{ch}(A) \, \text{ch}(B)$ and $\text{ch}(A \oplus B) = \text{ch}(A) + \text{ch}(B)$, we have (here $[\cdot]_2$ extracts the degree-2 component of the total Chern character)
\begin{equation}
        \text{Tr}(F_{\mathfrak{e}_8}^2)  = \tfrac{1}{h({\bf R})} \text{tr}_{\bf R}(F_{\mathfrak{e}_8}^2) = -\tfrac{2}{h({\bf R})} [\text{ch}(V)]_2 = -\tfrac{2}{h({\bf R})} \sum_j [\text{ch}(V^{(j)})]_2 \, ,
\end{equation}
with
\begin{equation}
\begin{aligned}
       [\text{ch}(V^{(j)})]_2 = &  \sum_\beta  [\text{ch}(U_\beta^{(j)})]_2 + \big( {\textstyle \prod_\beta} \text{rk}(U_\beta^{(j)}) \big) \left( \sum_{\gamma < \gamma'} [\text{ch}(W^{(j)}_\gamma)]_1 [\text{ch}(W^{(j)}_{\gamma'})]_1 + \sum_\gamma [\text{ch}(W_\gamma^{(j)})]_2 \right) \\
       = & -\sum_\beta \frac{h({\bf R}_\beta^{(j)})}{2} \text{Tr}(F_{\mathfrak{h}}^2) + \big( {\textstyle \prod_\beta} \dim({\bf R}_\beta^{(j)}) \big) \left( \sum_{\gamma < \gamma'} q^{(j)}_\gamma q^{(j)}_{\gamma'} F_\gamma F_{\gamma'} + \sum_\gamma \frac{(q^{(j)}_\gamma)^2}{2} F_\gamma^2 \right) \,.
      \end{aligned}
\end{equation}
We have thus derived the coefficients in equation \eqref{eq:decompose_trace} as
\begin{align}\label{eq:coefficients_trace_decomposition_with_u1s}
    \ell_\beta = \sum_j \frac{h({\bf R}_\beta^{(j)})}{h({\bf R})} \, , \quad r_{\gamma, \gamma'} = -\sum_j \frac{{\textstyle \prod_\beta} \dim({\bf R}_\beta^{(j)})}{h({\bf R})} q^{(j)}_\gamma q^{(j)}_{\gamma'} \, .
\end{align}
Note that the values of $\ell_\beta$ relevant to gaugings of $\mathfrak{h}_\beta \rightarrow \mathfrak{e}_8$ can be found in \cite{Baume:2021qho}.
This result does not depend on the chosen representation ${\bf R}$, as long as the decomposition is done with a fixed normalization for each $\mathfrak{u}(1)_\gamma$.

Candidate subgroups of $Z(\prod_\beta \widetilde{H}_\beta \times \prod_\gamma U(1)_\gamma)$ that can be used to twist the symmetry bundles must leave the representations resulting from decomposing the ${\bf 248}$ of $E_8$ invariant, as associated with the 
decomposition of the adjoint-valued moment map operator of the E-string theory. For this candidate subgroup, we can then verify whether the twist induces any anomaly for the large gauge transformation of the E-string tensor multiplet.

\subsubsection{Examples}

Having presented a general prescription for incorporating the contribution from continuous Abelian symmetries, we now turn to some explicit examples, focusing on the same class of examples already treated in the case of the non-Abelian flavor symmetries.
For illustrative purposes, we only consider the background field of the center-flavor symmetry involving the $U(1)$ flavor symmetry.
In all cases, it is straightforwardly verified that the fractionalizations also cancel when we turn on the previously studied center twists involving only the non-Abelian flavor factors.

\paragraph{Example 1:}
The example in equation \eqref{eq:A-type_quiver_with_suN} of a chain of $m$ $\mathfrak{su}_N$ gauge nodes provides a simple example with a $U(1)_f$ flavor symmetry:
\begin{align}
    [\mathfrak{su}_N^{(L)} ] \, \overset{\mathfrak{su}^{(1)}_{N}}{2} \, \overset{\mathfrak{su}^{(2)}_{N}}{2} \, \cdots \, \overset{\mathfrak{su}^{(m-1)}_{N}}{2} \, \overset{\mathfrak{su}^{(m)}_{N}}{2} \, [\mathfrak{su}_N^{(R)}] \, .
\end{align}
There is an overall $U(1)$ which is free from ABJ anomalies \cite{Apruzzi:2020eqi}. The $m+1$ bifundamental hypermultiplets:
\begin{align}\label{eq:hyper_spectrum_22-example}
    {\bf R}^{(1)} = ({\bf N}, \overline{\bf N}, {\bf 1}, {\bf 1}, \ldots)_1 \, , \quad {\bf R}^{(2)} = ({\bf 1}, {\bf N}, \overline{\bf N}, {\bf 1}, \ldots)_1 \, , \quad \ldots
\end{align}
have equal charge $q$, which we normalize to 1. There are 1-loop contributions to the $\text{Tr}(F_i^2) F_f^2$-terms that come from the anomaly polynomial of the hypermultiplets:
\begingroup
\allowdisplaybreaks
\begin{align}
    & I_\text{hyper}({\bf R}^{(1)}) \supset \frac14 N \, \text{Tr}_{\overline{\bf N}} (F_1^2) F_f^2 = -\frac{N}{2} c_2(F_1) c_1(F_f)^2 \, , \nonumber\\
    & \vdots \nonumber\\
    & I_\text{hyper}({\bf R}^{(i)}) \supset \frac14 (N \, \text{Tr}_{\bf N}(F_{i-1}^2) + N \, \text{Tr}_{\overline{\bf N}}(F_i^2) ) \, F_f^2 = -\frac{N}{2} (c_2(F_{i-1}) + c_2(F_{i})) \, c_1(F_f)^2 \, , \nonumber\\
    & \vdots \nonumber\\
    & I_\text{hyper}({\bf R}^{(m+1)}) \supset \frac14 N \, \text{Tr}_{\bf N}(F_m^2) \, F_f^2 = -\frac{N}{2} c_2(F_m) c_1(F_f)^2 \, , \nonumber\\
    \Longrightarrow \quad & I_\text{hypers} = \sum_{i=1}^{m+1} I_\text{hyper}({\bf R}^{(m)}) \supset -N \left( \sum_{i=1}^{m} c_2(F_i) \right) \, c_1(F_f)^2 \, .
\end{align}
\endgroup
Including the Abelian flavor backgrounds in the Green--Schwarz four-form,
\begin{align}
    I^i \supset \sum_{j=1}^m (-A^{ij})c_2(F_j) -  B^{i,L} c_2(F_L) - B^{i, R} c_2(F_R) + \tfrac12 C^{i;f,f} c_1(F_f)^2 \, ,
\end{align}
with $B^{i,L} = \delta^{i,1}$ and $B^{i,R} = \delta^{i,m}$, one can cancel the above $c_2(F_i) c_1(F_f)^2$ terms in the full anomaly polynomial $I_8 \supset I_\text{hypers} - \frac12 (A^{-1})_{ij} I^i I^j$, by fixing the coefficients $C^{i;f,f}$ to be (see equation \eqref{eq:U(1)_anomaly_coefficients})
\begin{align}
    C^{i,f,f} = 2N \, .
\end{align}
Then, the Green--Schwarz mechanism couples each tensor $\Theta_i$ to
\begin{align}\label{eq:GS-coupling_example-22}
\begin{split}
    I^i \supset -c_2(F_{i-1}) + 2 c_2(F_i) - c_2(F_{i+1})  + N c_1(F_f)^2 \, ,
\end{split}
\end{align}
again with the convention $F_{0}:= F_L$ and $F_{m+1} = F_R$ being the $\mathfrak{su}_N^{(L/R)}$ flavor backgrounds.

Before, we have seen that, with trivial $c_1(F_f)$, one finds that a $\bbZ_N$ twist is possible, leading to the non-Abelian group structure $[SU(N)^{(L)} \times \prod_i SU(N)^{(i)} \times SU(N)^{(R)}]/\bbZ_N$.
To extend the analysis to the Abelian flavor factor, we first note that the hypermultiplet charges in equation \eqref{eq:hyper_spectrum_22-example} are already properly normalized, in that the charges of all matter states span $\bbZ_N$.
The spectrum is invariant under the $\bbZ_N$ center-flavor symmetry generated by
\begin{align}
    (1,2,...,m+2; e^{\frac{2\pi i}{N} \hat{q}}) \in
    Z(SU(N)^{(L)} \times \prod_i SU(N)^{(i)} \times SU(N)^{(R)} \times U(1)) \, .
\end{align}
This means that the first Chern-class of the $U(1)_f/\bbZ_N$-bundle and the Stiefel--Whitney class of the $SU(N)/\bbZ_N$ bundles are correlated via a single 2-cocycle $w$ as
\begin{align}\label{eq:shifts_classes_example-22}
\begin{split}
    & c_1(F_f) \equiv \tfrac{1}{N} w \mod \bbZ \, , \quad w(F_i) = (i+1)w \quad (i = 0,..., m+1) \, , \\
    \Longrightarrow \quad & c_2(F_i) \equiv -(i+1)^2 \tfrac{N-1}{2N} w^2 \mod \bbZ \, , \\
    & c_1(F_f)^2 \equiv \tfrac{1}{N^2} w^2 + \tfrac{2}{N} w \cup \chi \mod \bbZ \, ,
\end{split}
\end{align}
with $\chi$ an integer 2-cocycle.
Plugging these into equation \eqref{eq:GS-coupling_example-22}, one straightforwardly verifies that the non-integer parts for the tensor couplings vanish:
\begin{align}\label{eq:u1-center-anomaly-cancellation_A-type_quiver}
    \begin{split}
        2c_2(F_i) -c_2(F_{i-1}) - c_2(F_{i+1}) + N c_1(F_f)^2 \equiv \big(\underbrace{(i^2 - 2(i+1)^2 + (i+2)^2) \tfrac{N-1}{2N} + \tfrac{1}{N}}_{=1} \big) w^2 \text{ mod } \bbZ \, .
    \end{split}
\end{align}
Hence, the structure group admits also a $\bbZ_N \cong Z_f$ quotient
\begin{equation}
    [SU(N)_L \times \prod_i SU(N)^{(i)} \times SU(N)_R \times U(1)_f]/Z_f \,.
\end{equation}
This matches the intuition from M-theory constructions \cite{Bah:2017gph}, from which one expects the flavor symmetry group of this SCFT to be $S[U(N)_L \times U(N)_R]/\bbZ_N$: the $\bbZ_N$ in this quotient is the center-flavor symmetry involving just the $SU(N) \subset U(N)$ parts, while the quotient $Z_f$ is encoded in $S[U(N) \times U(N)] \cong [SU(N) \times SU(N) \times U(1)]/\bbZ_N$.

\paragraph{Example 2:}
The theory with tensor branch description as in equation \eqref{eq:single_1-curve-example}:
\begin{align}
    \underset{[\#\bigwedge^2 = 1]}{\overset{\mathfrak{su}_{N}}{1}}\, [\mathfrak{su}_{N+8}] \, ,
\end{align}
also has a flavor $U(1)_f$ free of ABJ anomalies \cite{Apruzzi:2020eqi}, under which the hypermultiplets have the following charges:
\begin{align}
    ({\bf N}, \overline{\bf N+8}) : \quad q= N-4 \, , \quad (\textstyle{\bigwedge^2}, {\bf 1}): \quad q=-(N+8) \, .
\end{align}
By equation \eqref{eq:U(1)_anomaly_coefficients}, the Green--Schwarz mechanism couples, to the single tensor $\Theta$,
the four-form
\begin{align}
    I \supset c_2(F_N) - c_2(F_{N+8}) + \tfrac12 C^{i;f,f} c_1(F_f)^2 \, ,
\end{align}
with
\begin{align}
    \tfrac12 C^{i;f,f} = N(N-1)(N+8) \, .
\end{align}
The conditions for an element $(k_1, k_2, e^{\frac{2\pi i}{l} \hat{q}}) \in \bbZ_N \times \bbZ_{N+8} \times U(1)_f$ to act trivially on ${\bf R}$ are
\begin{align}
    \begin{aligned}
        {\bf R} &= ({\bf N}, \overline{\bf N+8}) : && \quad \tfrac{k_1}{N} - \tfrac{k_2}{N+8} + \tfrac{N-4}{l} \equiv 0 \mod \bbZ \, , \\
        {\bf R} &= (\textstyle{\bigwedge^2}, {\bf 1}): && \quad \tfrac{2k_1}{N} - \tfrac{N+8}{l} \equiv 0 \mod \bbZ \, .
    \end{aligned}
\end{align}
For odd $N$, it turns out that there is no such combined transformation leaving the hypermultiplets invariant.

For even $N$, there is always a trivially acting combination, but the general solution is cumbersome, so we will focus on an example with $N=6$.
In this case, the solution is $(k_1, k_2, l) = (3,9,14)$, so the putative quotient is a $\bbZ_{14} \simeq \mathbb{Z}_2 \times \mathbb{Z}_7$.
Notice that the charges of the hypermultiplets have a greatest common divisor of two, so, in order to be in the proper $U(1)$ normalization, we have to divide the charges by two, which means that the value of the coefficient $C^{i;f,f}$ is divided by four, $\frac12 C^{i;f,f} = 105$.
In addition, in this normalization the twist inside the $U(1)_f$ is by $e^{2\pi i/7}$.
Hence, the fractionalization of the Chern classes for this discrete twist is
\begin{align}
\begin{split}
    & c_2(F_N) \equiv -9 \times \tfrac{5}{12} w^2 \mod \bbZ \, , \quad c_2(F_{N+8}) \equiv -81 \times \tfrac{13}{28} w^2 \mod \bbZ \, , \\
    & c_1(F_f) \equiv \tfrac{1}{7} w \mod \bbZ \quad \Rightarrow \tfrac12 C^{i;f,f'} c_1(F_f)^2 \equiv 105( \tfrac{1}{49} w^2 + \tfrac{2}{7} w \cup \chi) \mod \bbZ \, ,
\end{split}
\end{align}
for an integer cocycle $\chi$.
Almost miraculously, the fractional parts cancel out in $I$, thus verifying that the full symmetry group is:
\begin{equation}
G_{\text{gauge-flavor}} = \frac{SU(6) \times SU(14) \times U(1)}{\bbZ_{14}}\, .
\end{equation}
The superconformal flavor symmetry group is then
\begin{align}
G_{\text{flavor}} = \frac{SU(14) \times U(1)}{\bbZ_{14}} \, .
\end{align}

\paragraph{Example 3:}
Finally, let us return to the example in equation \eqref{eq:example_3_tensor_config} with tensor branch configuration
\begin{align}
    [\mathfrak{su}_3^{(L)}] \, \overset{\mathfrak{e}_6}{3} \, \underset{[\mathfrak{u}(1)_f]}{1} \, \overset{\mathfrak{su}_2}{2} \, [\mathfrak{su}_3^{(R)}] \, ,
\end{align}
where we now include the Abelian flavor factor.
The $\mathfrak{u}(1)_f$ is the commutant of $\mathfrak{e}_6 \times \mathfrak{su}_2$ inside $\mathfrak{e}_8$, under which the representations resulting from the branching in equation \eqref{eq:branching_e8_to_u1} of the $E_8$-adjoint have charges
\begin{align}
    {\bf 248} \rightarrow
({\bf 78, 1})_0 \oplus ({\bf 1, 3})_0 \oplus ({\bf 1, 2})_3 \oplus ({\bf 1,2})_{-3} \oplus (({\bf 27, 2})_1 \oplus ({\bf 27, 1})_{-2} + \text{c.c}) \oplus ({\bf 1,1})_0 \, .
\end{align}
These are uncharged under the non-Abelian flavor factors at the end of the quiver. In turn, the hypermultiplets
\begin{align}
    {\bf R}^{(1)} = (\overline{\bf 3}, {\bf 27}, {\bf 1}, {\bf 1})_0 \, , \quad {\bf R}^{(2)} = \tfrac12 ({\bf 1}, {\bf 1}, {\bf 2},  {\bf 8})_0 \,,
\end{align}
are uncharged under $U(1)_f$.
From this we find that the $\bbZ_3 \times \bbZ_2 \cong \bbZ_6 \subset Z(SU(3) \times E_6 \times SU(2) \times Spin(7))$ considered previously, which leaves the hypermultiplets invariant but not the E-string states, can be compensated by a $U(1)_f$ twist, such that both sectors are invariant.
This combined $\bbZ_6$ has generator
\begin{align}\label{eq:teenchoiceawards2022}
    (2,2,1,1; e^{\frac{2\pi i \hat{q}}{6}}) \in \bbZ_3^2 \times \bbZ_2^2 \times U(1) \cong Z(SU(3)_L \times E_6 \times SU(2) \times Spin(7)_R \times U(1)_f) \, ,
\end{align}
with fractionalizations
\begin{align}
\begin{split}
    & c_2(F_{L}) \equiv -\tfrac43 w^2 \, , \quad c_2(F_{\mathfrak{e}_6}) \equiv -\tfrac83 w^2 \, , \quad c_2(F_{\mathfrak{su}_2}) \equiv -\tfrac14 w^2 \, , \quad c_2(F_{R}) \equiv -\tfrac12 w^2 \, , \\
    & c_1(F_f) \equiv \tfrac16 w \quad \Rightarrow \quad c_1(F_f)^2 \equiv \tfrac{1}{36} w^2 + \tfrac{1}{3} w \cup \chi \mod \bbZ \, .
\end{split}
\end{align}
From the above decomposition involving the $U(1)$ charges and equation \eqref{eq:coefficients_trace_decomposition_with_u1s}, we further find that
\begin{align}
    \text{Tr}(F_{\mathfrak{e}_8}^2) \rightarrow \text{Tr}(F_{\mathfrak{e}_6}^2) + \text{Tr}(F_{\mathfrak{su}_2}^2) - 12 F_f^2 \, ,
\end{align}
where the $\mathfrak{e}_6$ and $\mathfrak{su}_2$ are gauged on the left and right, respectively, of the E-string.
With the formulae from \cite{Baume:2021qho} applied to the hypermultiplets above, this gives the flavor anomaly coefficients
\begin{align}
    B^{iL} = 6\delta^{i,1}\, \quad B^{iR} = \delta^{i, 3}\, , \quad C^{i;f} = 6 \delta^{i,2} \, .
\end{align}
Together with the matrix
\begin{equation}
-A^{ij} = \begin{pmatrix} 3 & -1 & 0 \\ -1 & 1 & -1 \\ 0 & -1 & 2 \end{pmatrix}\,,
\end{equation}
we can now verify that the above $\bbZ_6$ twist does not induce any anomaly for the large gauge transformations of the tensors:
\begin{align}
    \begin{split}
        & \Theta_1 : \quad \eta^{1j} I_j^{(4)} \supset 3 c_2(F_{\mathfrak{e}_6}) - 6 c_2(F_L) \equiv (-8 + 8)w^2 \mod \bbZ \, , \\
        & \Theta_2 : \quad \eta^{2j} I_j^{(4)} \supset -c_2(F_{\mathfrak{e}_6}) - c_2(F_{\mathfrak{su}_2}) + 3c_1(F_f)^2 \equiv \big( \underbrace{ \tfrac83 + \tfrac14 + \tfrac{1}{12}}_{=3} \big) w^2 \mod \bbZ \, , \\
        & \Theta_3: \quad \eta^{3j} I_j^{(4)} \supset 2c_2(F_{\mathfrak{su}_2}) - c_2(F_R) \equiv \big(-\tfrac12 + \tfrac12 \big) w^2 \mod \bbZ \, .
    \end{split}
\end{align}
From this, we conclude that the tensor branch gauge theory has symmetry group:
\begin{equation}
G_{\text{gauge-flavor}} = \frac{SU(3) \times E_6 \times SU(2) \times Spin(7) \times U(1)}{\bbZ_6},
\end{equation}
where the group action is specified by equation \eqref{eq:teenchoiceawards2022}.
This also provides a prediction for the SCFT flavor symmetry:
\begin{equation}
G_\text{flavor} = \frac{SU(3) \times Spin(7) \times U(1)}{\bbZ_6}.
\end{equation}

\subsection{Center Twists and R-Symmetry}\label{sec:R-symmetry_twist}

In addition to the flavor symmetries, all 6d $\mathcal{N} = (1,0)$ SCFTs have an $\mathfrak{su}(2)_R$ symmetry.
This, of course, is an additional global symmetry which can in principle also mix with the center of the gauge group and flavor symmetry.
It is also worth noting that this R-symmetry is not directly manifest in the target space geometry of the corresponding F-theory models, but is realized geometrically in various M-theory constructions of 6d SCFTs.

Now, before getting to the case of center / R-symmetry mixing in 6d SCFTs, it is already instructive to note that even in the context of 4d theories, entertaining this possibility resolves some apparent puzzles, which as far as we are aware have not been previously addressed in the literature.\footnote{We thank J.~Distler for helpful correspondence.} For example, in the context of 4d $\mathcal{N} = 2$ SCFTs, the $E_6$ Minahan--Nemeschansky was argued to have a non-Abelian $E_6 / \mathbb{Z}_3$ global symmetry \cite{Bhardwaj:2021ojs}, which is also in accord with some superconformal index computations \cite{Gadde:2010te}. On the other hand, a direct analysis of BPS states would appear to detect states in the $\mathbf{27}$ of $E_6$ \cite{Distler:2019eky}. The natural resolution of this puzzle is that the center of the $E_6$ flavor symmetry mixes with the $U(1)_R$ symmetry of an $\mathcal{N} = 2$ SCFT, namely we have the global structure $[E_{6} \times U(1)_R ] / \mathbb{Z}_3$.\footnote{In this example we make no statement about the global structure involving the $SU(2)_R$ R-symmetry.} This also agrees with expectations based on the D3-brane probe of an $E_6$ 7-brane.\footnote{For the rank one theory it is possible to construct BPS states on the Coulomb branch as junctions between an $E_6$ stack of 7-branes and a D3-brane. In this scenario the states carry charge under the gauge group of the D3-brane so the symmetry group is $(E_6 \times U(1)_{\text{D3}})/\mathbb{Z}_3$. However since a $U(1)_R$ transformation in this construction is simply a rotation in the space transverse to the 7-branes it can be identified with the $U(1)$ center of mass of the D3-brane; it is possible to identify $U(1)_R \sim U(1)_{\text{D3}}$ with $U(1)_R$ being the symmetry that survives at the conformal point.}

However, this cannot be the full story, since the theory contains, for example, the supercharges which are not charged under any flavor symmetries, but do transform under a discrete R-symmetry twist.
This twist can be naturally cancelled if we include the remaining parts of the superconformal symmetry group \cite{Distler:2020tub,Manschot:2021qqe}.
We will postpone a detailed analysis of this interplay in the above 4d example, and turn our attention to 6d SCFTs for now.

For these, the supercharges are in the fundamental representation of $\mathfrak{su}(2)_R$, but otherwise uncharged under any flavor symmetry.
A natural way to cancel the effects of the $\bbZ_2 = Z(SU(2)_R)$ twist would be to activate a $\bbZ_2 = Z(Spin(1,5))$ twist in the Lorentz group which acts on spinors such as the supercharges.
Therefore, whenever we contemplate turning on an R-symmetry twist, the minimal requirement for the theory to be invariant is if it is accompanied by a $\bbZ_2$ twist of the Lorentz symmetry.

Now, we observe that our tensor branch analysis naturally incorporate both twists, since the topological Green--Schwarz couplings capture the contribution from non-trivial R-symmetry bundles, as well as the tangent bundle which is associated to Lorentz symmetry. Indeed, the Green--Schwarz four-form,
\begin{align}
    I^i \supset y^i \, c_2(R) - (2 + A^{ii}) \tfrac14 p_1(T) \,,
\end{align}
contains the second Chern-class $c_2(R)$ of the R-symmetry bundle and the first Pontryagin class $p_1(T)$ of the tangent bundle.
The coefficient $y^i \equiv h^\vee_{\fkg_i}$ is fixed to be the dual Coxeter number of the $i$th gauge algebra $\fkg_i$ by requiring the cancellation of all mixed gauge-R-symmetry anomalies; if $\fkg_i = \emptyset$ (which requires $A^{ii} = -1$ or $-2$), the coefficient is set to be $y^i = 1$.
For the R-symmetry bundle, the fractionalization is just as for any other $SU(2)/\bbZ_2$ gauge or flavor bundle, $c_2(R) \equiv -\tfrac14 w_R^2 \mod \bbZ$.
To quantify the fractionalization of the tangent bundle, we will work under the assumption that a Wick rotation to Euclidean signature does not affect the results.
Then, $p_1(T) = -\tfrac12 \text{tr}_\text{vec}({\cal R}^2)$, where the trace over the curvature ${\cal R}$ is in the vector, or anti-symmetric representation, of $Spin(6) \cong SU(4)$.
For $SU(4)$, this is the same as the 1-instanton normalized trace, so we conclude that
$\tfrac14 p_1(T) = -\tfrac12 c_2(SU(4))$.
In Euclidean signature, the corresponding $\bbZ_2$ twist (which leaves the vector representation invariant) is generated by $2 \in \bbZ_4 = Z(SU(4))$, for which the fractionalization is
\begin{align}
    \tfrac14 p_1(T) = -\tfrac12 c_2(SU(4)) \equiv \tfrac12 \times 2^2 \times \tfrac38 w_R^2 \mod \bbZ \equiv \tfrac34 w_R^2 \mod \bbZ \, .
\end{align}
Then, a twist by a center-flavor symmetry can occur if the fractionalization of the gauge, flavor, R-symmetry, and tangent bundles cancel out in $I^i$ for every $i$.

We present some examples of R-symmetry / spacetime symmetry mixing for the tensor branch of the $\mathcal{N} = (2,0)$ and E-string SCFTs in Appendix \ref{app:JUSTHEFLUBRO}. These cases are a bit special in that the tensor branch has no gauge group factors. For the sake of illustrating this general phenomenon, we now turn to some examples with center-flavor symmetry mixing, and no additional $U(1)$ factors. In this case, the gauge-global 0-form symmetry is of the general form:
\begin{equation}
G_{\text{gauge-global}} = \frac{\widetilde{G}_{\text{gauge}} \times \widetilde{G}_{\text{flavor}} \times [SU(2)_R \times Spin(1,5)]}{\mathcal{C}},
\end{equation}
where $\mathcal{C}$ is a suitably defined quotienting subgroup.
The global symmetry group that acts on spacetime scalars is then
\begin{align}
    G_\text{global} = \frac{\widetilde{G}_\text{flavor} \times SU(2)_R}{{\cal C}} \, .
\end{align}

As a first example, consider the SCFT with tensor branch description
\begin{align}\label{eq:tensor_branch_-2_example}
    \overset{\mathfrak{su}_N}{2} \, [\mathfrak{su}_{2N}] \, .
\end{align}
In this case, the $p_1(T)$-term drops out of the Green--Schwarz coupling:
\begin{align}
    \Theta \wedge \left( 2 c_2(F_{\text{gauge}}) - c_2 (F_{\text{flavor}}) + N \, c_2(R) \right) \, ,
\end{align}
which fractionalizes, for general center-twisted backgrounds, as
\begin{align}\label{eq:R-sym_example_fractionalization}
    \Theta \wedge \left( - \tfrac{N-1}{N} w_g^2 + \tfrac{2N-1}{4N} w_{f}^2 - \tfrac{N}{4} w_R^2 \right) \mod \bbZ \, .
\end{align}
The well-known flavor symmetry group $SU(2N)/\bbZ_N$ results from a combined twist of the gauge and flavor factor, with trivial R-symmetry twist:
\begin{align}
    w_{f} = -2 w_g = -2 w_N \, , \quad w_R = 0 \, ,
\end{align}
which leads to an overall integer shift in the GS-coupling.

In order to turn on a $\bbZ_2$ twist of the R-symmetry (which, as discussed above, is always accompanied by a Lorentz group twist), we must first make sure that the hypermultiplets are invariant.
Since these transform in the fundamental of $\mathfrak{su}(2)_R$, such a twist acts with a phase $(-1)$, which must be cancelled by a suitable gauge or flavor symmetry twist.
In the present example, we can turn on the $\bbZ_2 \subset \bbZ_{2N} = Z(SU(2N))$ simultaneously to achieve this.
More precisely, we claim that the theory is invariant under the central subgroup with generators
\begin{align}\label{eq:gens_quotient_R-sym_example}
    \left. \begin{array}{ll}
        \bbZ_N: & a = (1,-2,0)  \\
        \bbZ_2: & b = (0, N, 1)
    \end{array} \right\}
    \in \bbZ_N \times \bbZ_{2N} \times \bbZ_2 = Z(SU(N) \times SU(2N) \times SU(2)_R) \, .
\end{align}
At the level of background fields, these twists correspond to the following correlations,
\begin{align}
    w_g = w_N, \quad w_R =w_2, \quad w_{f} = -2 w_{N} + N w_2 \, ,
\end{align}
where $w_N$ and $w_2$ are the background fields associated to the $\bbZ_N$ and the $\bbZ_2$ generator, respectively, in equation \eqref{eq:gens_quotient_R-sym_example}.
This indeed shifts the Green--Schwarz four-form by an integer class,
\begin{align}
    - \tfrac{N-1}{N} w_g^2 + \tfrac{2N-1}{4N} w_{f}^2 - \tfrac{N}{4} w_R^2 = w_N^2 - (2N-1) w_N \cup w_{R} + \tfrac{N(N-1)}{2} w_R^2 \equiv 0 \mod \bbZ \, .
\end{align}

To write down the global symmetry group structure, note that for odd $N$, we have $\bbZ_N \times \bbZ_{2N} \times \bbZ_2 \cong \bbZ_N \times \bbZ_N \times \bbZ_2 \times \bbZ_2$, and $a$ generates the diagonal of the two $\bbZ_N$ factors, while $b$ generates the diagonal $\bbZ_2$.
For even $N = 2n$, on the other hand, we can consider the $\bbZ_2$ generator $na + b = (n, 0, 1) \in \bbZ_{N} \times \bbZ_{2N} \times \bbZ_2$, which maps trivially onto the $\bbZ_{2N}$ factor of the flavor symmetry.
Therefore, the global symmetry group compatible with the large gauge transformations of the tensor is
\begin{align}\label{eq:R-sym_example1_global_form}
\renewcommand{\arraystretch}{2.2}
    G_\text{global} = \left\{ \begin{array}{l  l}
        \displaystyle\frac{SU(2N)/\bbZ_N \times SU(2)_R}{\bbZ_2} \, , & N \ \text{odd,} \\
        \displaystyle\frac{SU(2N)}{\bbZ_N} \times \frac{SU(2)_R}{\bbZ_2} \, , & N \ \text{even.}
    \end{array} \right.
\end{align}

Let us compare this with results known from the Higgs branch chiral ring.
Elements of this ring carry representations of the global symmetry of the SCFT, so a center-flavor symmetry must leave all combinations of flavor and R-symmetry representations  that can be found in the chiral ring invariant.
For the SCFT with tensor branch description as in equation \eqref{eq:tensor_branch_-2_example}, it turns out that the chiral ring generator with non-trivial center charges has representation $(\boldsymbol\wedge^N, {\bf N + 1})$ under $SU(2N) \times SU(2)_R$ \cite{Hanany:2018vph}.
As the $N$-index anti-symmetric representation, $\boldsymbol\wedge^N$, of $SU(2N)$ picks up a phase $(-1)$ under the generator of $\bbZ_{2N} = Z(SU(2N))$, this state is clearly invariant under the $\bbZ_N$ subgroup in equation \eqref{eq:gens_quotient_R-sym_example}.
Therefore, the flavor symmetry group $SU(2N)/\bbZ_N$ is also what the Higgs branch data sees.
Moreover, $(\boldsymbol\wedge^N, {\bf N + 1})$ transforms with phase $(-1)^N$ under $N \in \bbZ_{2N}$ and with phase $(-1)^N$ under $1 \in \bbZ_2 = Z(SU(2)_R)$.
So it is also invariant under the second generator in equation \eqref{eq:gens_quotient_R-sym_example}.
Hence, the global structure in equation \eqref{eq:R-sym_example1_global_form} is also predicted from the Higgs branch chiral ring.

To illustrate the importance of the fractionalization of the tangent bundle, we consider the minimal $(D_k, D_k)$ conformal matter theory, whose tensor branch gauge theory is
\begin{align}
    \overset{\mathfrak{sp}_{k-4}}{1} \, [\mathfrak{so}_{4k}] \, ,
\end{align}
containing a half-hypermultiplet ${\bf h}$ in the bifundamental representation ${\bf R} = \tfrac12 ({\bf 2k-8, {\bf 4k}})$, with ${\bf 4k}$ the vector of $\mathfrak{so}_{4k}$. The hypermultiplet ${\bf h}$ also transforms as the fundamental of $\mathfrak{su}(2)_R$.
The Higgs branch chiral ring is generated by a moment map $\mu$ transforming in the $({\bf adj}, {\bf 3})$ of the $\mathfrak{so}_{4k} \oplus \mathfrak{su}(2)_R$, thus uncharged under the center, and a generator $\mu^+$ transforming in the $({\bf S^+}, {\bf k-1})$ representation, where ${\bf S^+}$ is one of the $\mathfrak{so}_{4k}$ spinor representations \cite{Ferlito:2017xdq,Hanany:2018uhm},\footnote{Aspects of the Higgs branch of minimal $(D_k, D_k)$ conformal matter have recently been explored from the perspective of the conformal bootstrap \cite{Baume:2021chx}.} which we pick to be of positive chirality for definiteness.\footnote{In this case, the choice of chirality of the spinor generator is irrelevant, however, it can be relevant when minimal $(D_k, D_k)$ conformal matter is used as a building block for other 6d SCFTs \cite{Distler:2022yse}.}
Therefore, the action of an element
\begin{align}
    (a_g, (a_+, a_-), a_R) \in \bbZ_2 \times (\bbZ_2^+ \times \bbZ_2^-) \times \bbZ_2 = Z(Sp(k-4)) \times Z(Spin(4k)) \times Z(SU(2)_R) \,,
\end{align}
on the representations of ${\bf h}$ and $\mu^+$ give phases
\begin{align}
    \mu^+ : \ (-1)^{a_+ + k \, a_R} \, , \quad {\bf h}: \ (-1)^{a_g + a_+ + a_- + a_R} \, ,
\end{align}
which must be trivial for any element $(a_g, (a_+, a_-), a_R)$ of the quotienting subgroup $\mathcal{C}$.
For even $k$, this requires $a_+ = 0 \ \text{mod 2}$, and $a_g + a_- + a_R = 0 \ \text{mod 2}$.
This leaves two independent generators,
\begin{align}\label{eq:DkDk_even_twists}
    (a_g, (a_+, a_-), a_R) = (1,(0,1),0) \quad \text{and} \quad (1,(0,0),1) \qquad (k \ \text{even}) \,,
\end{align}
which correspond to the diagonal $\bbZ_2$ of $Z(Sp(k-4)) \times \bbZ_2^-$ and $Z(Sp(k-4)) \times Z(SU(2)_R)$, respectively.
For odd $k$, we instead have $a_+ + a_R = 0 \ \text{mod 2}$, and $a_g + a_- = 0 \ \text{mod 2}$, which has independent solutions corresponding to the generators
\begin{align}\label{eq:DkDk_odd_twists}
    (a_g, (a_+, a_-), a_R) = (1,(0,1),0) \quad \text{and} \quad (0,(1,0),1) \qquad (k \ \text{odd}) \,,
\end{align}
of the diagonal $\bbZ_2$ factors of $Z(Sp(4-k)) \times \bbZ_2^-$ and $\bbZ_2^+ \times Z(SU(2)_R)$, respectively.
Considering the gauge invariant representations that can appear in the SCFT, the Higgs branch therefore predicts the global symmetry group
\begin{equation}\label{eqn:GHBpro}
    G_\text{global}^\text{HB} = \begin{cases}
        Spin(4k) / \mathbb{Z}_2^{-} \times (SU(2)_R/\bbZ_2) \quad &\text{if $k$ even} \,, \\[0.2em]
        [Spin(4k) / \mathbb{Z}_2^{-} \times SU(2)_R]/\mathbb{Z}_2 \quad &\text{if $k$ odd} \, .
    \end{cases}
\end{equation}

This agrees with the analysis from the Green--Schwarz coupling,
\begin{align}
    \Theta \wedge I_4 := \Theta \wedge \left( c_2(F_\mathfrak{sp}) - c_2(F_\mathfrak{so}) + h^\vee_{\mathfrak{sp}_{k-4}} c_2(R) - \tfrac14 p_1(T) \right) \, ,
\end{align}
which, with $h^\vee_{\mathfrak{sp}_{k-4}} = k-3$, fractionalizes as
\begin{align}
    I_4 \equiv - \tfrac{k}{4} w_g^2 + \tfrac{k}{4} (w_+ + w_{-})^2 + \tfrac12 w_+ \cup w_- - \tfrac{k-3}{4} w_R^2 - \tfrac34 w_R^2 \mod \bbZ \, ,
\end{align}
where $w_g$, $(w_+, w_-)$, and $w_R$ are the generic background fields for $Z(Sp(k-4))$, $Z(Spin(4k))$, and $Z(SU(2)_R)$, respectively, and we have already imposed the correlation between the R-symmetry and Lorentz group twist.
Restricting to the twists in equations \eqref{eq:DkDk_even_twists} and \eqref{eq:DkDk_odd_twists} predicted by the Higgs branch, we find
\begin{align}
    \begin{split}
        k \ \text{even:} \quad & w_g = w_- + w_R, \ w_+ = 0 \quad \Rightarrow  I_4 \equiv -\tfrac{k}{2} ( w_R^2 + w_- \cup w_R) \equiv 0 \mod \bbZ \, , \\
        k \ \text{odd:} \quad & w_g = w_-, \ w_+ = w_R \quad \Rightarrow  I_4 \equiv \tfrac{k+1}{2} w_R \cup w_- \equiv 0 \mod \bbZ \, .
    \end{split}
\end{align}

\section{Intermezzo: Orbi-Instanton Theories}\label{sec:e8CF}

Throughout Section \ref{sec:6d}, we have demonstrated that, given the quiver description of the generic point of the tensor branch of a 6d $(1,0)$ SCFT, one can determine the global structure of the flavor symmetry group. Since we also wish to generate 4d theories via Stiefel--Whitney twisted compactifications on a $T^2$, we now turn to a rich class of examples where we can systematically study possible center-flavor symmetry mixing.

The theories we now consider are Higgs branch deformations of the ``orbi-instanton theories'', as obtained from as obtained in M-theory terms from M5-branes probing an ADE singularity wrapped by an $E_8$ nine-brane \cite{DelZotto:2014hpa}. Via a process of fission and fusion, these turn out to be the progenitors for all 6d SCFTs \cite{Heckman:2018pqx} realized in a geometric phase of F-theory. As shown in \cite{Heckman:2015bfa}, a large class of Higgs branch deformations are captured by a nilpotent orbit of $\mathfrak{g}$, $\sigma$, and a homomorphism $\rho:\,\Gamma_{\mathfrak{g}} \rightarrow E_8$. We denote the resulting theories as
\begin{equation}
    \Omega_{\mathfrak{g}, N}(\rho, \sigma) \,.
\end{equation}
It is natural to ask: does the pair $(\rho, \sigma)$ capture the presence or absence of center-flavor symmetry in a straightforward manner? We assume that $N$ is sufficiently large that the Higgsing by $\rho$ and $\sigma$ are uncorrelated on the tensor branch, and in this section we will focus on the case $\mathfrak{g} = \mathfrak{su}_K$. Furthermore, we will assume that $\sigma$ is the maximal nilpotent orbit given by the trivial embedding $\mathfrak{su}_2 \rightarrow \mathfrak{g}$. As we see, the condition on $\rho$ for $\Omega_{\mathfrak{g}, N}(\rho, \sigma)$ to have a non-trivial center flavor symmetry in these cases is straightforward.

We consider the rank $N$ $(\mathfrak{e}_8, \mathfrak{su}_K)$ orbi-instanton 6d SCFT, which has the tensor branch configuration
\begin{equation}
    12\overset{\mathfrak{su}_{2}}{2}\overset{\mathfrak{su}_{3}}{2}\cdots\overset{\mathfrak{su}_{K}}{2}\underbrace{\overset{\mathfrak{su}_{K}}{2}\cdots \overset{\mathfrak{su}_{K}}{2}}_{N-1} \,.
\end{equation}
This theory can be obtained in M-theory as the worldvolume theory of a stack of $N$ M5-branes probing a $\mathbb{C}^2/\mathbb{Z}_K$ orbifold singularity and on top of an M9-plane \cite{DelZotto:2014hpa}.
The non-Abelian part of the flavor symmetry of this theory is generically
\begin{equation}
    \mathfrak{e}_8 \oplus \mathfrak{su}_K \,.
\end{equation}
A Higgsing of the $\mathfrak{e}_8$ flavor is specified by a choice of homomorphism $\rho: \, \Gamma_{\mathfrak{su}_K} \cong \mathbb{Z}_K \rightarrow E_8$. Such homomorphisms, as explained by Kac \cite{MR739850}, are captured by a weighted partition of the Dynkin labels of the affine $E_8$ Dynkin diagram into $K$:
\begin{equation}\label{eqn:Zkembed}
    (a_1, a_2, a_3, a_4, a_5, a_6, a_{4^\prime}, a_{2^\prime}, a_{3^\prime}) \,,
\end{equation}
such that
\begin{equation}\label{eqn:Zkembed2}
    a_1 + 2(a_2 + a_{2^\prime}) + 3(a_3 + a_{3^\prime}) + 4(a_4 + a_{4^\prime}) + 5a_5 + 6a_6 = K \,.
\end{equation}
We find that Higgsing the $\mathfrak{e}_8$ by a homomorphism, represented by a tuple as in equation \eqref{eqn:Zkembed} whose non-zero entries are $\{a_{i^{(')}}\}$, leads to a 6d SCFT with center-flavor symmetry
\begin{equation}
    \mathbb{Z}_\ell = \mathbb{Z}_{\gcd(\{i\})} \, .
\end{equation}
As evident from equation \eqref{eqn:Zkembed2}, this $\bbZ_\ell$ is always a subgroup of $Z(SU(K))$, consistent with the fact that the Higgsed theory has an $\mathfrak{su}_K$ flavor algebra.
For each $E_8$-homomorphism specified by equations \eqref{eqn:Zkembed} and \eqref{eqn:Zkembed2}, there exists an algorithm that determines the tensor branch configuration \cite{Mekareeya:2017jgc}. These tensor branch descriptions, for each of the putative $\mathbb{Z}_\ell$-preserving $E_8$-homomorphisms, are written in Table \ref{tbl:6dscfts};\footnote{In fact, when $t \neq 0$, the tensor branch description on the first line of Table \ref{tbl:6dscfts} corresponds to \emph{two} 6d SCFTs, depending on the choice of $\theta$-angle for the $\mathfrak{sp}_q$ gauge algebra on the $(-1)$-curve. These theories have the same central charges and flavor symmetries, but differ in the spectrum of local operators at large conformal dimension. See \cite{Mekareeya:2017jgc,Distler:2022yse} for more details; we suppress this subtlety in this paper.} in each case one can then use the study of the large gauge transformation anomalies to verify that there is indeed a $\mathbb{Z}_\ell$ center-flavor symmetry. As the tensor branch configurations are rather involved, we explicate the analysis in one example.

\begin{sidewaystable}[]
    \centering
    \footnotesize
    \begin{threeparttable}
    \begin{tabular}{cc}
    \toprule
        $\mathbb{Z}_\ell$ & Tensor branch description of the 6d $(1,0)$ SCFT \\\midrule
        \multirow{3}{*}{$\mathbb{Z}_2$}  & $\overset{\mathfrak{sp}_{q}}{1}
        \underbrace{\overset{\mathfrak{su}_{2q + 8}}{2}
        \cdots\overset{\mathfrak{su}_{2q + 8t}}{2}}_{t}
        \underbrace{\overset{\mathfrak{su}_{2q + 8t + 6}}{2}
        \cdots\overset{\mathfrak{su}_{2q + 8t + 6u}}{2}}_{u}
        \underbrace{\overset{\mathfrak{su}_{2q + 8t + 6u + 4}}{2}
        \cdots\overset{\mathfrak{su}_{2q + 8t + 6u + 4s}}{2}}_{s}
        \underbrace{\overset{\mathfrak{su}_{2q + 8t + 6u + 4s + 2}}{2}
        \cdots\overset{\mathfrak{su}_{2q + 8t + 6u + 4s + 2p}}{2}}_p
        \underbrace{\overset{\mathfrak{su}_{2q + 8t + 6u + 4s + 2p}}{2}\cdots \overset{\mathfrak{su}_{2q + 8t + 6u + 4s + 2p}}{2}}_{N-1}$  \\

        &  $\overset{\mathfrak{su}_{2q+4}}{1}
        \underbrace{\overset{\mathfrak{su}_{2q + 12}}{2}
        \cdots\overset{\mathfrak{su}_{2q + 8t + 4}}{2}}_{t}
        \underbrace{\overset{\mathfrak{su}_{2q + 8t + 10}}{2}
        \cdots\overset{\mathfrak{su}_{2q + 8t + 6u + 4}}{2}}_{u}
        \underbrace{\overset{\mathfrak{su}_{2q + 8t + 6u + 8}}{2}
        \cdots\overset{\mathfrak{su}_{2q + 8t + 6u + 4s + 4}}{2}}_{s}
        \underbrace{\overset{\mathfrak{su}_{2q + 8t + 6u + 4s + 6}}{2}
        \cdots\overset{\mathfrak{su}_{2q + 8t + 6u + 4s + 2p + 4}}{2}}_p
        \underbrace{\overset{\mathfrak{su}_{2q + 8t + 6u + 4s + 2p + 4}}{2}\cdots \overset{\mathfrak{su}_{2q + 8t + 6u + 4s + 2p + 4}}{2}}_{N-1}$  \\\midrule

        \multirow{6}{*}{$\mathbb{Z}_3$}  & $\overset{\mathfrak{su}_{3}}{1}
        \underbrace{\overset{\mathfrak{su}_{12}}{2}
        \cdots\overset{\mathfrak{su}_{9q+3}}{2}}_{q}
        \underbrace{\overset{\mathfrak{su}_{9q+9}}{2}
        \cdots\overset{\mathfrak{su}_{9q+3+6s}}{2}}_{s}
        \underbrace{\overset{\mathfrak{su}_{9q + 6s + 6}}{2}
        \cdots\overset{\mathfrak{su}_{9q + 6s + 3p + 3}}{2}}_{p}
        \underbrace{\overset{\mathfrak{su}_{9q + 6s + 3p + 3}}{2}\cdots \overset{\mathfrak{su}_{9q + 6s + 3p + 3}}{2}}_{N-1}$ \\

         & $1 \underbrace{\overset{\mathfrak{su}_{9}}{2}
         \cdots\overset{\mathfrak{su}_{9q}}{2}}_{q}
         \underbrace{\overset{\mathfrak{su}_{9q+6}}{2}
         \cdots\overset{\mathfrak{su}_{9q+6s}}{2}}_{s}
         \underbrace{\overset{\mathfrak{su}_{9q + 6s + 3}}{2}
         \cdots\overset{\mathfrak{su}_{9q + 6s + 3p}}{2}}_{p}
         \underbrace{\overset{\mathfrak{su}_{9q + 6s + 3p}}{2}\cdots \overset{\mathfrak{su}_{9q + 6s + 3p}}{2}}_{N-1}$ \\

          & $\overset{\mathfrak{su}_{6}^\prime}{1}
         \underbrace{\overset{\mathfrak{su}_{15}}{2}
         \cdots\overset{\mathfrak{su}_{9q+6}}{2}}_{q}
         \underbrace{\overset{\mathfrak{su}_{9q + 12}}{2}
         \cdots\overset{\mathfrak{su}_{9q+6 + 6s}}{2}}_{s}
         \underbrace{\overset{\mathfrak{su}_{9q + 6s + 9}}{2}
         \cdots\overset{\mathfrak{su}_{9q + 6s + 3p + 6}}{2}}_{p}
    \underbrace{\overset{\mathfrak{su}_{9q + 6s + 6s + 3p + 6}}{2}\cdots \overset{\mathfrak{su}_{9q + 6s + 3p + 6}}{2}}_{N-1}$ \\\midrule

        \multirow{4}{*}{$\mathbb{Z}_4$}  & $\overset{\mathfrak{su}_{4}}{1} \underbrace{\overset{\mathfrak{su}_{12}}{2}
        \cdots\overset{\mathfrak{su}_{8q + 4}}{2}}_{q} \underbrace{\overset{\mathfrak{su}_{8q + 8}}{2}
        \cdots\overset{\mathfrak{su}_{8q + 4p + 4}}{2}}_{p}
        \underbrace{\overset{\mathfrak{su}_{8q + 4p + 4}}{2}\cdots \overset{\mathfrak{su}_{8q + 4p + 4}}{2}}_{N-1}$ \\

         & $1 \underbrace{\overset{\mathfrak{su}_{8}}{2}
         \cdots\overset{\mathfrak{su}_{8q}}{2}}_{q} \underbrace{\overset{\mathfrak{su}_{8q + 4}}{2}
         \cdots\overset{\mathfrak{su}_{8q + 4p}}{2}}_{p}
    \underbrace{\overset{\mathfrak{su}_{8q + 4p}}{2}\cdots \overset{\mathfrak{su}_{8q + 4p}}{2}}_{N-1}$ \\\midrule

        $\mathbb{Z}_5$  & $1\underbrace{\overset{\mathfrak{su}_{5}}{2}
        \cdots\overset{\mathfrak{su}_{5p}}{2}}_p\underbrace{\overset{\mathfrak{su}_{5p}}{2}\cdots \overset{\mathfrak{su}_{5p}}{2}}_{N-1} $ \\\midrule

        $\mathbb{Z}_6$ & $1\underbrace{\overset{\mathfrak{su}_{6}}{2}
        \cdots\overset{\mathfrak{su}_{6p}}{2}}_p\underbrace{\overset{\mathfrak{su}_{6p}}{2}\cdots \overset{\mathfrak{su}_{6p}}{2}}_{N-1} $ \\
        \bottomrule
    \end{tabular}
    \end{threeparttable}
    \caption{6d SCFTs that we consider that have discrete center-flavor symmetry. In descending order, the $E_8$-homomorphisms as in equation \eqref{eqn:Zkembed}, are $(0,p,0,s,0, u,2t+1,q,0)$, $(0,p,0,s,0, u,2t,q,0)$, $(0,0,p,0,0, s,0,0,3q+2)$, $(0,0,p,0,0, s,0,0,3q+1)$, $(0,0,p,0,0, s,0,0,3q)$, $(0,0,0,p,0,0,2q+1,0,0)$, $(0,0,0,p,0,0,2q,0,0)$, $(0,0,0,0,p,0,0,0,0)$, and $(0,0,0,0,0,p,0,0,0)$.}
    \label{tbl:6dscfts}
\end{sidewaystable}

The simplest example is the Higgsing induced by a $\bbZ_{6p} \rightarrow E_8$ homomorphism specified by $a_6 = p$, and all other labels being zero.
The resulting tensor branch gauge theory has the quiver description
\begin{equation}\label{eq:Z6-orbi-inst-example}
    [\mathfrak{su}_3 \oplus \mathfrak{su}_2] \, \, 1 \, \, \overset{\mathfrak{su}_{6}}{2} \, \, \overset{\mathfrak{su}_{12}}{2} \cdots \overset{\mathfrak{su}_{6p-6}}{2} \, \, \underset{[\mathfrak{su}_6]}{\overset{\mathfrak{su}_{6p}}{2}} \, \, \underbrace{\overset{\mathfrak{su}_{6p}}{2}\cdots \overset{\mathfrak{su}_{6p}}{2}}_{N-1} \, \, [\mathfrak{su}_{6p}] \, .
\end{equation}
Between each $\mathfrak{su}$ gauge and flavor factor on or next to $2$-nodes, there is a bifundamental hypermultiplet,
\begin{align}
\begin{split}
    & {\bf R}^{(1)} = ({\bf 6}, \overline{\bf 12})_0 \, , \quad {\bf R}^{(2)} = ({\bf 12}, \overline{\bf 18})_0 \, , \quad \cdots , \quad {\bf R}^{(p-1)} = ({\bf 6p-6}, \overline{\bf 6p})_0 \, , \\
    & {\bf R}^{(p)} = ({\bf 6p}, \overline{\bf 6})_{-p} \, ,  \quad {\bf R}^{(p+1)} = ({\bf 6p}, \overline{\bf 6p})_1 \, , \quad \cdots , \quad {\bf R}^{(p+N)} = ({\bf 6p}, \overline{\bf 6p})_1 \, .
\end{split}
\end{align}
In addition, there is a $U(1)_f$ flavor symmetry without ABJ-anomalies \cite{Apruzzi:2020eqi}, which only charges the hypermultiplets between the $\mathfrak{su}_{6p}$ factors (``the plateau'') with the charges indicated in the subscripts. The tensor pairing is the $(N+p) \times (N+p)$ matrix
\begin{align}
    A^{ij} = \begin{pmatrix}
        -1 & 1 & 0 & \cdots \\
        1 & -2 & 1 & \ddots \\
        0 & 1 & -2 & \ddots \\
        \vdots & \ddots & \ddots & \ddots
    \end{pmatrix} \, ,
\end{align}
and the anomaly coefficients of the flavor factors are
\begin{align}
\begin{split}
    & B^{i,\mathfrak{su}_3} = B^{i,\mathfrak{su}_2} = \delta^{i,1} \, , \quad B^{i,\mathfrak{su}_6} = \delta^{i,p+1} \, , \quad B^{i,\mathfrak{su}_{6p}} =  \delta^{i,N+p} \, , \\
    & C^{i;f,f} = \begin{cases}
        0 \, , & i\leq p \, , \\
        6p(p+1) \, , & i = p+1 \, , \\
        12p  \, , & i > p+1 \, .
    \end{cases}
\end{split}
\end{align}
Without taking into consideration the $U(1)_f$, one can easily verify that there is a $\bbZ_3 \times \bbZ_2 \times \bbZ_6 \cong {\bbZ_6}^{(1)} \times {\bbZ_6}^{(2)}$ center-flavor symmetry that leaves all hypermultiplets invariant. These have generators
\begin{align}
\begin{split}
    {\bbZ_6}^{(1)}: \quad & [\overset{1}{\mathfrak{su}_3} \oplus \overset{1}{\mathfrak{su}_2}] \, \, 1 \, \, \overset{\overset{0}{\mathfrak{su}_{6}}}{2} \, \, \overset{\overset{0}{\mathfrak{su}_{12}}}{2} \cdots \overset{\overset{0}{\mathfrak{su}_{6p-6}}}{2} \, \, \underset{[\underset{0}{\mathfrak{su}_6}]}{\overset{\overset{0}{\mathfrak{su}_{6p}}}{2}} \, \, \underbrace{\overset{\overset{0}{\mathfrak{su}_{6p}}}{2}\cdots \overset{\overset{0}{\mathfrak{su}_{6p}}}{2}}_{N-1} \, \, [\overset{0}{\mathfrak{su}_{6p}}] \, , \\
    {\bbZ_6}^{(2)}: \quad & [\overset{0}{\mathfrak{su}_3} \oplus \overset{0}{\mathfrak{su}_2}] \, \, 1 \, \, \overset{\overset{1}{\mathfrak{su}_{6}}}{2} \, \, \overset{\overset{2}{\mathfrak{su}_{12}}}{2} \cdots \overset{\overset{p-1}{\mathfrak{su}_{6p-6}}}{2} \, \, \underset{[\underset{1}{\mathfrak{su}_6}]}{\overset{\overset{p}{\mathfrak{su}_{6p}}}{2}} \, \, \underbrace{\overset{\overset{p}{\mathfrak{su}_{6p}}}{2}\cdots \overset{\overset{p}{\mathfrak{su}_{6p}}}{2}}_{N-1} \, \, [\overset{p}{\mathfrak{su}_{6p}}] \, ,
\end{split}
\end{align}
where we have indicated the embedding $k_g \in Z(G_g)$ by the overset $\overset{k_g}{\fkg_g}$ (or underset for the $[\mathfrak{su}_6]$ flavor factor) on each node of the quiver.
However, the presence of the E-string breaks these individual $\bbZ_6$ factors to the diagonal $\bbZ_6$, with the fractional part of Chern classes given by\footnote{By an abuse of notation, we will write $c_2(\fkg)$ for $c_2(F_\fkg)$.}
\begin{align}
    \begin{split}
        & c_2(\mathfrak{su}_3) \equiv - \tfrac13 w^2 \, , \quad c_2(\mathfrak{su}_2) \equiv -\tfrac14 w^2 \, , \quad c_2(\mathfrak{su}_{6l}) \equiv -\tfrac{l (6l-1)}{12} w^2 \quad (l=1,\ldots,p) \, .
    \end{split}
\end{align}
It is straightforward to verify that these cancel for each tensor:
\begin{align}
    \begin{aligned}
        \Theta_1 : & \qquad -c_2(\mathfrak{su}_3) - c_2(\mathfrak{su}_2) - c_2(\mathfrak{su}_6) \equiv \big( \tfrac13 + \tfrac14 + \tfrac{5}{12} \big) w^2 \equiv 0 \mod \bbZ \, , \\
        \Theta_{1+i} \, (1 \leq i < p) : & \qquad -c_2(\mathfrak{su}_{6(i-1)}) + 2c_2(\mathfrak{su}_{6i}) - c_2(\mathfrak{su}_{6(i+1)}) \\
        & \quad \equiv \tfrac{(i-1)(6i-7) -2i(6i-1) + (i+1)(6i+5)}{12} \, w^2 \equiv \tfrac{12}{12} w^2 \equiv 0 \mod \bbZ \, , \\
        \Theta_{p+1} : & \qquad -c_2(\mathfrak{su}_{6(p-1)}) + 2c_2(\mathfrak{su}_{6p}) - c_2(\mathfrak{su}_{6p}) - c_2(\mathfrak{su}_{6}) \\
        & \quad \equiv \tfrac{(p-1)(6p-7) - 2p(6p-1) + p(6p-1) -5}{12} \, w^2 \equiv \tfrac{12(p-1)}{12} \, w^2 \equiv 0 \mod \bbZ \, , \\
        \Theta_{p+i} \, (1 < i \leq N) : & \qquad -c_2(\mathfrak{su}_{6p}) + 2c_2(\mathfrak{su}_{6p}) - c_2(\mathfrak{su}_{6p}) \equiv 0 \mod \bbZ \, .
    \end{aligned}
\end{align}
Therefore, the non-Abelian structure group is
\begin{align}\label{eq:E8_Z6_example_non-Ab-flavor-group}
    \frac{[SU(3) \times SU(2)] \times \prod_{i< p} SU(6i) \times SU(6p)^N \times [SU(6)] \times [SU(6p)]}{\bbZ_6} \, ,
\end{align}
and the non-Abelian flavor symmetry of the SCFT is
\begin{equation}
    \frac{SU(3) \times SU(2) \times SU(6) \times SU(6p)}{\bbZ_6} \,.
\end{equation}
This $\bbZ_6$ center-flavor symmetry will allow us to perform a Stiefel--Whitney twisted $T^2$ compactification down to 4d, which we will turn to in the next section.

To complete the characterization of the full symmetry structure, we include possible $U(1)_f$ twists, in which case the hypermultiplets are invariant under a further center transformation of order $6p$, with generator
\begin{align}
    {\bbZ_{6p}}: \quad & [\overset{0}{\mathfrak{su}_3} \oplus \overset{0}{\mathfrak{su}_2}] \, \, 1 \, \, \overset{\overset{0}{\mathfrak{su}_{6}}}{2} \, \, \overset{\overset{0}{\mathfrak{su}_{12}}}{2} \cdots \overset{\overset{0}{\mathfrak{su}_{6p-6}}}{2} \, \, \underset{[\underset{-1}{\mathfrak{su}_6}]}{\overset{\overset{0}{\mathfrak{su}^{(0)}_{6p}}}{2}} \, \, \underbrace{\overset{\overset{1}{\mathfrak{su}^{(1)}_{6p}}}{2} \, \, \overset{\overset{2}{\mathfrak{su}^{(2)}_{6p}}}{2} \cdots \overset{\overset{N-1}{\mathfrak{su}^{(N-1)}_{6p}}}{2}}_{N-1} \, \, [\overset{N}{\mathfrak{su}^{(N)}_{6p}}] \quad \text{and} \quad e^{-\frac{2\pi i \hat{q}}{6p}} \in U(1)_f\, ,
\end{align}
where we have enumerated, for convenience, the $\mathfrak{su}_{6p}$ factors.
The resulting non-trivial Chern class fractionalizations are then
\begin{align}
    c_2(\mathfrak{su}_6) \equiv - \tfrac{5}{12}w^2 \, , \quad c_2(\mathfrak{su}_{6p}^{(k)}) \equiv -k^2 \, \tfrac{6p-1}{12p} w^2 \, , \quad c_1(F_f)^2 \equiv \tfrac{1}{36p^2} w^2 + \tfrac{1}{3p} w \cup \chi \mod \bbZ \, .
\end{align}
For the tensors of the $\mathfrak{su}_{6p}^{(k)}$ factors with $k \geq 1$, the cancellation of the fractionalizations is analogous to that appearing in equation \eqref{eq:u1-center-anomaly-cancellation_A-type_quiver} for the simple A-type quiver example. For $k=0$, the cancellation is due to
\begin{align}
\begin{split}
    &-c_2(\mathfrak{su}_{6p-6}) + 2 c_2(\mathfrak{su}_{6p}^{(0)}) - c_2(\mathfrak{su}_6) - c_2(\mathfrak{su}_{6p}^{(1)}) + 3p(p+1) c_1(F_f)^2 \\
    &\quad\equiv \,  \big( 0 + 0 + \tfrac{5}{12} + \tfrac{6p-1}{12p} + \tfrac{p+1}{12p} \big) \, w^2 \equiv 0 \mod \bbZ \, .
\end{split}
\end{align}

Lastly, we also incorporate the R-symmetry.
Again, the main constraint is to have the bifundamental hypermultiplets being invariant, which all transform in the fundamental representation of $\mathfrak{su}(2)_R$.
To cancel the phase $(-1)$ which these states acquire upon a $\bbZ_2 = Z(SU(2)_R)$ twist, we turn on a corresponding $\bbZ_2 \subset Z(SU(6k))$ in every \emph{second} $\mathfrak{su}$-factor.
These two a priori different twists are related by adding the $\bbZ_2$ subgroup of the $\bbZ_6$ center-flavor symmetry responsible for the non-Abelian flavor group structure in equation \eqref{eq:E8_Z6_example_non-Ab-flavor-group}, so do not give rise to two new and independent center-flavor symmetries when we include the R-symmetry, as expected.
For concreteness, we take the generator that compensates the $Z(SU(2)_R)$ twist to be
\begin{align}
\begin{split}
    p \ \text{even}: \quad & [\overset{0}{\mathfrak{su}_3} \oplus \overset{0}{\mathfrak{su}_2}] \, \, 1 \, \, \overset{\overset{0}{\mathfrak{su}_{6}}}{2} \, \, \overset{\overset{6}{\mathfrak{su}_{12}}}{2} \, \, 
    \overset{\overset{0}{\mathfrak{su}_{18}}}{2} \, \, 
    \overset{\overset{12}{\mathfrak{su}_{24}}}{2} \, \, \cdots \overset{\overset{0}{\mathfrak{su}_{6p-6}}}{2} \, \, \underset{[\underset{0}{\mathfrak{su}_6}]}{\overset{\overset{3p}{\mathfrak{su}_{6p}}}{2}} \, \, \underbrace{\overset{\overset{0}{\mathfrak{su}_{6p}}}{2} \, \, \overset{\overset{3p}{\mathfrak{su}_{6p}}}{2} \cdots }_{N-1} \, \, [\overset{*}{\mathfrak{su}_{6p}}] \, , \\
    p \ \text{odd}: \quad & [\overset{0}{\mathfrak{su}_3} \oplus \overset{0}{\mathfrak{su}_2}] \, \, 1 \, \, \overset{\overset{0}{\mathfrak{su}_{6}}}{2} \, \, \overset{\overset{6}{\mathfrak{su}_{12}}}{2} \, \, 
    \overset{\overset{0}{\mathfrak{su}_{18}}}{2} \, \, 
    \overset{\overset{12}{\mathfrak{su}_{24}}}{2} \, \, \cdots \overset{\overset{3p-3}{\mathfrak{su}_{6p-6}}}{2} \, \, \underset{[\underset{3}{\mathfrak{su}_6}]}{\overset{\overset{0}{\mathfrak{su}_{6p}}}{2}} \, \, \underbrace{\overset{\overset{3p}{\mathfrak{su}_{6p}}}{2} \, \, \overset{\overset{0}{\mathfrak{su}_{6p}}}{2} \cdots }_{N-1} \, \, [\overset{*}{\mathfrak{su}_{6p}}] \, ,
\end{split}
\end{align}
where the $*$ is either $3p$ if $p+N$ is even, or $0$ if $p+N$ is odd.
For this quiver, the tangent bundle enters only in the first tensor multiplet $t_1$ associated to the E-string, whose corresponding Green--Schwarz four-form contains $c_2(R)$ and $p_1(T)$:
\begin{align}
    \Theta_1 : \quad -c_2(\mathfrak{su}_3) -c_2(\mathfrak{su}_2) - c_2(\mathfrak{su}_6) + c_2(R) - \tfrac14 p_1(T) \equiv -\tfrac14 w^2 - \tfrac34 w^2 \equiv 0 \mod \bbZ \, .
\end{align}
For the other tensors, the topological coupling to the R-symmetry bundle is through the term $h^\vee c_2(R)$, where $h^\vee (\mathfrak{su}_{6k}) = 6k$.
Since these tensors all have $A^{ii}=-2$, the coupling to $p_1(T)$ is trivial.
Let us first examine those on a generic position on the ramp (i.e., a 2-node with $\mathfrak{su}_{6i<6p}$).
Here, we have
\begin{align}
    \begin{split}
        \Theta_{1+i} \, (i \ \text{even}): \quad & -c_2(\mathfrak{su}_{6(i-1)}) + 2c_2(\mathfrak{su}_{6i}) - c_2(\mathfrak{su}_{6(i+1)}) + 6i \, c_2(R) \\
        \equiv & \left( 2 \times \tfrac{9i (6i-1)}{12} +  \tfrac32 i \right) w^2 \equiv 0 \mod \bbZ \, ,
    \end{split}\\
    \begin{split}
        \Theta_{1+i} \, (i \ \text{odd}): \quad & -c_2(\mathfrak{su}_{6(i-1)}) + 2c_2(\mathfrak{su}_{6i}) - c_2(\mathfrak{su}_{6(i+1)}) + 6i \, c_2(R) \\
        \equiv & \left( - \tfrac{9(i-1)(6i-7) + 9(i+1)(6i+5)}{12} +  \tfrac32 i \right) w^2 \equiv 0 \mod \bbZ \, ,
    \end{split}
\end{align}
where the fractional part of $c_2(\mathfrak{su}_{6(i-1)})$ is automatically zero for $i=1$.
For the node that connects the ramp to the plateau (i.e., the first node with $\mathfrak{su}_{6p}$ gauge algebra), we have
\begin{align}
\begin{split}
    \Theta_{p+1} \, (p \ \text{even}) : \quad & -c_2(\mathfrak{su}_{6p-6}) + 2 c_2(\mathfrak{su}_{6p}) - c_2(\mathfrak{su}_{6p}) - c_2(\mathfrak{su}_6) + 6p \, c_2(R) \\
    \equiv & \left( 2\times \tfrac{9p(6p-1)}{12} + \tfrac{3}{2}p \right) w^2 \equiv 0 \mod \bbZ \, , 
\end{split}\\
\begin{split}
    \Theta_{p+1} \, (p \ \text{odd}) : \quad & -c_2(\mathfrak{su}_{6p-6}) + 2 c_2(\mathfrak{su}_{6p}) - c_2(\mathfrak{su}_{6p}) - c_2(\mathfrak{su}_6) + 6p \, c_2(R) \\
    \equiv & \left( - \tfrac{9(p-1)(6p-7) + 9p(6p-1) + 9 (6-1)}{12} + \tfrac{3}{2}p \right) w^2 \equiv 0 \mod \bbZ \, .
\end{split}
\end{align}
For the other nodes on the ramp, there is either a $\bbZ_2$ twist only in the corresponding gauge factor, or only in the two adjacent gauge / flavor factors:
\begin{align}
    \begin{split}
        \Theta_{p+i} \, (p+i \ \text{odd}) : \quad & -c_2(\mathfrak{su}_{6p}) + 2 c_2(\mathfrak{su}_{6p}) - c_2(\mathfrak{su}_{6p}) + 6p \, c_2(R) \equiv \\
        & \left( 2 \times \tfrac{9p(6p-1)}{12} + \tfrac32 p \right) w^2 \equiv 0 \mod \bbZ \, ,
    \end{split}\\
    \begin{split}
        \Theta_{p+i} \, (p+i \ \text{even}) : \quad & -c_2(\mathfrak{su}_{6p}) + 2 c_2(\mathfrak{su}_{6p}) - c_2(\mathfrak{su}_{6p}) + 6p \, c_2(R) \equiv \\
        & \left( - 2 \times \tfrac{9p(6p-1)}{12} + \tfrac32 p \right) w^2 \equiv 0 \mod \bbZ \, .
    \end{split}
\end{align}

Again, we omit the straightforward, but somewhat tedious crosscheck that we can activate simultaneously the $\bbZ_6$ twist in the non-Abelian flavor factors, the $\bbZ_{6p}$ twist involving the $U(1)_f$ flavor, and the $\bbZ_2$ R-symmetry twist.
From this analysis, we conclude that the 6d SCFT with tensor branch description as in equation \eqref{eq:Z6-orbi-inst-example} has global symmetry group
\begin{align}
    \frac{SU(3) \times SU(2) \times SU(6) \times SU(6p) \times U(1)_f \times SU(2)_R}{\bbZ_6 \times \bbZ_{6p} \times \bbZ_2} \,.
\end{align}
The analysis of the structure for the global symmetry for the other tensor branch descriptions in Table \ref{tbl:6dscfts} follows directly from the application of the methods described in this example.

\section{4d \texorpdfstring{\boldmath{$\mathcal{N}=2$}}{N=2} SW-folds}\label{sec:sfolds}

Having shown how to extract the global symmetry group of 6d SCFTs, we now turn to a specific application in the context of constructing 4d $\mathcal{N} = 2$ SCFTs. To reach such a theory from a 6d $\mathcal{N} = (1,0)$ SCFT, one can consider compactification on a $T^2$. Activating background gauge bundle configurations with vanishing flux provides a general template for realizing 4d $\mathcal{N} = 2$ SCFTs. In fact, one can also consider compactifications which are sensitive to the global topology of the 6d global symmetries, namely by switching on an 't Hooft magnetic flux \cite{tHooft:1979rtg} in the $T^2$ directions \cite{Ohmori:2018ona}.\footnote{One can in principle consider various generalizations, as obtained from compactifying on a more general genus $g$ Riemann surface with marked points, with non-trivial contributions from the R-symmetry bundle also switched on. In this broader setting, one would expect to get 4d $\mathcal{N} = 1$ SCFTs, along the lines of \cite{Morrison:2016nrt, Razamat:2016dpl} (see also, for example, \cite{Gaiotto:2015usa,Franco:2015jna,Coman:2015bqq,Heckman:2016xdl,Bah:2017gph,Bourton:2017pee,Kim:2017toz,Apruzzi:2018oge,Razamat:2018zus,Kim:2018lfo,Razamat:2018gro,Kim:2018bpg,Razamat:2018gbu,Chen:2019njf,Pasquetti:2019hxf,Sela:2019nqa,Razamat:2019mdt,Razamat:2019ukg,Razamat:2020bix,Sabag:2020elc,Bourton:2020rfo,Nazzal:2021tiu,Hwang:2021xyw,Bourton:2021das,Razamat:2022gpm} and references therein).}
In \cite{Ohmori:2018ona} this was referred to as a Stiefel--Whitney (SW) twisted compactification. These are a specific class of configurations involving background flat bundle configurations with non-trivial holonomies which commute in $G_{\text{global}} = \widetilde{G}_{\text{global}} / \mathcal{C}$, but which would not have commuted as holonomies of bundles with structure group $\widetilde{G}$. Treated as a bundle with structure group $\widetilde{G}$, we would have a non-zero flux valued in a subgroup of $\mathcal{C}$, i.e., the holonomies commute up to a specific element of this flux. We shall loosely speaking refer to such holonomies as being ``charged under an element of $\mathcal{C}$'' since this has a clear meaning when treating these backgrounds as $\widetilde{G}$ bundles. Owing to their similarities with S-fold constructions, we often refer to these theories as ``SW-folds'' in what follows.

We consider SW-folds obtained from the 6d theories considered in Section \ref{sec:e8CF}, namely the theories of the form $\Omega_{\mathfrak{su}_K, N}(\rho, 1)$, where $\rho$ is an $E_8$-homomorphism that leads to a $\mathbb{Z}_\ell$ center-flavor symmetry. The tensor branch descriptions of such 6d $(1,0)$ SCFTs were given in Table \ref{tbl:6dscfts}. Compactification on a torus with a $\mathbb{Z}_\ell$ Stiefel--Whitney twist then leads to the 4d SCFTs we consider herein; furthermore, many of the properties of the 4d theories can be obtained from a knowledge of the 6d $(1,0)$ parent theory. We emphasize that when we say we turn on a $\mathbb{Z}_\ell$ Stiefel--Whitney twist, we are turning on a non-commuting holonomy charged under the element $p$ of $\mathbb{Z}_\ell$ such that $\gcd(p, \ell) = 1$. All of these Stiefel--Whitney twisted theories are listed in Table \ref{tbl:genSfolds}.

As a general comment, while we could in principle extract the global symmetry \textit{group} of the resulting 4d theory, there can be additional structures which emerge from extended objects which can now wrap on the $T^2$ directions. For this reason, we primarily focus on just the global symmetry \textit{algebra} of the resulting 4d theories, leaving a more complete analysis of their global structure group to future work.

The rest of this section is organized as follows. We first explain how to extract the central charges and flavor symmetries for the resulting SW-fold theories. This is followed by an extensive list of examples, as given in Table \ref{tbl:genSfolds}. As an independent cross-check, we also directly study the Coulomb branch operator spectrum for these theories. In some cases, there are alternative ways to generate some of these theories.\footnote{A recent and detailed review of both the features of $\mathcal{N}=2$ SCFTs, and of the various different constructions, is \cite{Akhond:2021xio}.} We discuss some examples of this in the context of class $\mathcal{S}$ constructions, as well as 4d $\mathcal{N} = 2$ S-folds \cite{Apruzzi:2020pmv,  Giacomelli:2020jel, Heckman:2020svr,Giacomelli:2020gee}, and we comment on the overlap as well as differences from these other methods of generating 4d $\mathcal{N} = 2$ SCFTs.

\begin{table}[t!]
    \centering
    \renewcommand{\arraystretch}{1.4}
    \begin{threeparttable}
    \begin{tabular}{cccc}
    \toprule
         SW-fold SCFT & Orbi-instanton & $E_8$-Homomorphism & SW Twist \\\midrule
         $\mathcal{S}_2^{(N)}(p,s,u,2t+1,q)$ &  $(\mathfrak{e}_8, \mathfrak{su}_{2q + 8t + 6u + 4s + 2p + 4})$  & $(0,p,0,s,0,u,2t+1,q,0)$ & \multirow{2}{*}{$\mathbb{Z}_2$} \\
         $\mathcal{T}_2^{(N)}(p,s,u,2t,q)$ &  $(\mathfrak{e}_8, \mathfrak{su}_{2q + 8t + 6u + 4s + 2p})$  & $(0,p,0,s,0,u,2t,q,0)$ &  \\\midrule
         $\mathcal{R}_3^{(N)}(p,s,3q+2)$ &  $(\mathfrak{e}_8, \mathfrak{su}_{9q+6s+3p+6})$  & $(0,0,p,0,0,s,0,0,3q+2)$ & \multirow{3}{*}{$\mathbb{Z}_3$} \\
         $\mathcal{S}_3^{(N)}(p,s,3q+1)$ &  $(\mathfrak{e}_8, \mathfrak{su}_{9q+6s+3p+3})$  & $(0,0,p,0,0,s,0,0,3q+1)$ &  \\
         $\mathcal{T}_3^{(N)}(p,s,3q)$ &  $(\mathfrak{e}_8, \mathfrak{su}_{9q+6s+3p})$  & $(0,0,p,0,0,s,0,0,3q)$ &  \\\midrule
         $\mathcal{S}_4^{(N)}(p, 2q+1)$ &  $(\mathfrak{e}_8, \mathfrak{su}_{8q+4p+4})$  & $(0,0,0,p,0,0,2q+1,0,0)$ & \multirow{2}{*}{$\mathbb{Z}_4$} \\
         $\mathcal{T}_4^{(N)}(p, 2q)$ &  $(\mathfrak{e}_8, \mathfrak{su}_{8q+4p})$  & $(0,0,0,p,0,0,2q,0,0)$ &  \\\midrule
         $\mathcal{T}_5^{(N)}(p)$ &  $(\mathfrak{e}_8, \mathfrak{su}_{5p})$  & $(0,0,0,0,p,0,0,0,0)$ & $\mathbb{Z}_5$ \\\midrule
         $\mathcal{T}_6^{(N)}(p)$ &  $(\mathfrak{e}_8, \mathfrak{su}_{6p})$  & $(0,0,0,0,0,p,0,0,0)$ & $\mathbb{Z}_6$  \\\bottomrule
    \end{tabular}
    \end{threeparttable}
    \caption{The 4d $\mathcal{N}=2$ SW-folds that we consider in this paper. Each SCFT is obtained by starting with the 6d rank $N$ orbi-instanton SCFT of type $(\mathfrak{e}_8, \mathfrak{su}_K)$, where $K$ is as in the second column. Higgsing the $\mathfrak{e}_8$ flavor symmetry by the homomorphism $\mathbb{Z}_K \rightarrow E_8$, given via $(a_1, a_2, a_3, a_4, a_5, a_6, a_{4^\prime}, a_{2^\prime}, a_{3^\prime})$ in the third column, yields each of the 6d SCFTs in Table \ref{tbl:6dscfts}, which have a $\mathbb{Z}_\ell$ center-flavor symmetry. Compactifying the resulting 6d SCFT on a $T^2$ with a $\mathbb{Z}_\ell$ Stiefel--Whitney twist, where $\ell$ is as in the fourth column, produces the 4d $\mathcal{N}=2$ SW-fold SCFT which we denote by the naming that appears in the first column.}
    \label{tbl:genSfolds}
\end{table}

\subsection{Central Charges and Flavor Symmetries}\label{sec:ccs}

Having specified a construction for an infinite family of 4d $\mathcal{N} = 2$, we now turn to some of their properties. As each of the SW-folds we study arises from the compactification of a 6d SCFT that is very Higgsable, we can apply the methods from \cite{Ohmori:2018ona} to determine the central charges and the flavor central charges.

To determine the central charges of the SW-folds we carry out the following procedure. First we compute the anomaly polynomial of the origin 6d SCFT, $I_8$. Next, we compute the 1-loop contribution on the full tensor branch from just the vector multiplets, tensor multiplets and hypermultiplets, and refer to this as $I_{8}^{\text{fields}}$:\footnote{We note that especially in the case of generalized quivers with conformal matter one sometimes refers to this as a ``1-loop'' contribution as well. Here, we are referring to the full tensor branch, where the conformal matter has also been decomposed into standard 6d $\mathcal{N} = (1,0)$ supermultiplets.}
\begin{equation}
I_{8}^{\text{fields}} \equiv I_{8}^{\text{1-loop,vector}} + I_{8}^{\text{1-loop,tensor}} + I_{8}^{\text{1-loop,hyper}}.
\end{equation}
Both $I_8$ and $I_{8}^{\text{fields}}$ is a formal eight-form polynomial in the characteristic classes of the symmetries of the 6d SCFT. As required, the anomaly polynomial does not contain any terms proportional to the characteristic classes of the gauge symmetries on the tensor branch, as the 6d SCFT is non-anomalous, but the quantity $I_8^\text{fields}$ does contain such gauge-anomalous terms. We write
\begin{equation}\label{eqn:fields}
    I_8 - I_8^\text{fields} = A p_1(T)^2 + B c_2(R) p_1(T) + \sum_a C_a p_1(T)\operatorname{Tr}F_a^2 + \cdots \,,
\end{equation}
where $p_1(T)$ is the first Pontryagin class of the spacetime tangent bundle, $c_2(R)$ is the $SU(2)$ R-symmetry bundle, and $\operatorname{Tr}F_a^2$ is the curvature of the flavor symmetry bundles. The sum is over the simple non-Abelian flavor symmetries of the SCFT. In terms of these quantities, $A$, $B$, and $C_a$,\footnote{As we will not include holonomies of 6d Abelian flavor factors, their anomaly coefficients $C^{i;f,f'}$ will not appear in the following, and $C_a$ will exclusively denote the coefficients in the anomaly polynomial \eqref{eqn:fields}.} we can write the central charges of the SW-fold SCFTs as
\begin{equation}\label{eqn:ack}
  \begin{aligned}
    a - a_\text{generic} &= 32\left(\frac{3}{2\ell} - \frac{3}{4}\right)A - \frac{12}{\ell}B \,, \\
    c - c_\text{generic} &= 32\left(\frac{3}{\ell} - 1\right)A - \frac{12}{\ell}B \,, \\
    \kappa_a - \kappa^a_\text{generic} &= \frac{192}{\ell} C_a I_a \,,
  \end{aligned}
\end{equation}
where $\ell$ is the order of the Stiefel--Whitney twist, and $I_a$ is the Dynkin index of the embedding of the 4d flavor symmetry as a subalgebra of the 6d flavor symmetry. Here $a_\text{generic}$, $c_\text{generic}$, and $\kappa_\text{generic}^a$ are the central charges and flavor central charges of the 4d theory at the generic point of the Coulomb branch. We can rewrite the central charges in terms of the numbers of vector and (full) hypermultiplets at the generic point of the Coulomb branch as
\begin{equation}
    a_\text{generic} = \frac{5}{24}n_V + \frac{1}{24}n_H \,, \quad c_\text{generic} = \frac{1}{6}n_V + \frac{1}{12}n_H \,.
\end{equation}
For all SW-folds the quantities $A$, $B$, $C_a$ and the generic central charges can be determined from the 6d origin and the knowledge of the $\mathbb{Z}_\ell$ center-flavor symmetry. Thus we can always determine the central charges of the SW-fold SCFT.

In this section, we are interested in specific 6d SCFTs, those that appear in Table \ref{tbl:6dscfts}, which all have tensor branch configurations of the form
\begin{equation}\label{eqn:quivquiv}
    \overset{\mathfrak{g}}{1}\overset{\mathfrak{su}_{\ell k_1}}{2}\overset{\mathfrak{su}_{\ell k_2}}{2}\cdots\overset{\mathfrak{su}_{\ell k_r}}{2} \,,
\end{equation}
where the possible choices for $\mathfrak{g}$ and the $k_i$ are specified via their 6d origin in Table \ref{tbl:6dscfts}.  For each $\ell$, we summarize the possible $\mathfrak{g}$, together with their below-mentioned numerical data, in Table \ref{tbl:onedata}. First, we will discuss the contributions to $I_8 - I_8^\text{fields}$ for such 6d SCFTs. The result differs depending on whether $\mathfrak{g}$ is trivial or not. Let us first consider the simplest case where $\mathfrak{g} \neq \varnothing$. Then
\begin{equation}\label{eqn:GS1}
    I_8 - I_8^\text{fields} = I_8^\text{GS} = -\frac{1}{2}A_{ij}I^i I^j \,,
\end{equation}
where the tensor indices $i$, $j$ run over the nodes of the tensor branch configuration in equation \eqref{eqn:quivquiv} from left to right.\footnote{We emphasize that the tensor pairing matrix $A^{ij}$, whose diagonal entries are the negative of the values attached to the nodes in equation \eqref{eqn:quivquiv}, is negative-definite.}
Recall from equation \eqref{eq:green-schwarz-coupling} that in the four-form $I^i$,
\begin{equation}
    I^i = \frac{1}{4}\bigg(-A^{ij}\operatorname{Tr} F_j^2 - B^{ia}\operatorname{Tr} F_a^2 - (2+A^{ii})p_1(T) \bigg)+ y^i c_2(R) \,.
\end{equation}
the index $i$ in the $p_1(T)$ term is not summed over.
In this case, we have $y^i = h_{\mathfrak{g}_i}^\vee$, the dual Coxeter number of the gauge algebra associated to the $i$th tensor. We see that the only contributions to $p_1(T)^2$ arise when $i = j = 1$, and thus
\begin{equation}
    A = \frac{r+1}{32} \,.
\end{equation}
The $c_2(R)p_1(T)$ term is rather more involved to determine, but it can be found to be:
\begin{equation}
  \begin{aligned}
    B &= \frac{1}{4} A_{ij} \left( 2 + A^{ii}\right)y^j  \\
    &= -\frac{1}{4}(r+1)h_{\mathfrak{g}}^\vee - \frac{1}{4}\ell \sum_{j=1}^r (r+1-j)k_j \,,
  \end{aligned}
\end{equation}
where we again need to be careful with the sum over $i$, and we emphasize that the tensor pairing matrix $A^{ij}$ and its inverse $A_{ij}$ are both symmetric. Finally, we consider the terms proportional to $p_1(T)\operatorname{Tr}F_a^2$. We find
\begin{equation}
  \begin{aligned}
    C_a &=  -\frac{1}{16} A_{ij} \left( (2 + A^{ii})B^{ja} \right) \\
    &= \frac{r+1 - k(a)}{16} \,,
  \end{aligned}
\end{equation}
where $k(a)$ is the position of the node in the tensor branch quiver diagram that ``intersects'' the flavor factor $\fkg_a$, i.e., $B^{ja} = 0$ for $j \neq k(a)$.\footnote{This would not apply to baryonic $\mathfrak{su}_2$ flavor symmetries, which we are not considering in this work. Such flavor factors are only relevant for very specific SW-folds, which have already been worked out in \cite{Giacomelli:2020jel}.} In all cases under consideration we have $B^{k(a)a} = 1$. Thus, we have determined $A$, $B$, and $C_a$ for tensor branch configurations of the form in equation \eqref{eqn:quivquiv} when $\mathfrak{g} \neq \varnothing$.

Let us now consider the slightly more complicated configuration where $\mathfrak{g} = \varnothing$, in which case the left-most node in equation \eqref{eqn:quivquiv} becomes an E-string.
For this configuration we have
\begin{equation}
    I_8 - I_8^\text{fields} = I_8^\text{E-string} - I_8^\text{tensor} + I_8^\text{GS} \,,
\end{equation}
where
\begin{equation}
    I_8^\text{GS} = -\frac{1}{2}\widetilde{A}_{ij} I^i I^j \,, \quad I^i = \frac{1}{4}\bigg(-\widetilde{A}^{ij}\operatorname{Tr} F_j^2 - \widetilde{B}^{ia}\operatorname{Tr} F_a^2 - (2+\widetilde{A}^{ii})p_1(T) \bigg)+ y^i c_2(R) \,.
\end{equation}
Here, the matrix of coefficients $\widetilde{A}$ and $\widetilde{B}$ can be interpreted as the contributions from a generalized quiver, where we allow conformal matter between nodes of the quiver. In this case, the indices now run over $i,j = 1, \cdots, r$. The coefficients $y^i$ remain $h_{\mathfrak{su}_{\ell k_i}}^\vee$, except for $y^1$ which is now $1 + h_{\mathfrak{su}_{\ell k_1}}^\vee$.
We can see immediately that
\begin{equation}
    A = \frac{1}{32} + \frac{r}{32} = \frac{r+1}{32} \,.
\end{equation}
Furthermore, the $c_2(R)p_1(T)$ coefficient is
\begin{equation}
  \begin{aligned}
    B &= -\frac{1}{4} + \frac{1}{4} \widetilde{A}_{ij} \left( (2 + \widetilde{A}^{ii})y^j \right) \\
    &= -\frac{1}{4} - \frac{1}{4}(r)(\ell k_1 + 1) - \frac{1}{4}\ell \sum_{j=2}^r (r+1-j) k_j \\
    &= -\frac{1}{4}(r+1) - \frac{1}{4}\ell \sum_{j=1}^r (r+1-j)k_j \,.
  \end{aligned}
\end{equation}
Finally, we need to discuss the flavor symmetry terms. The coefficient $C_a$ of $p_1(T)\operatorname{Tr}F_a^2$ is
\begin{equation}
    C_a = \frac{r+1 - k(a)}{16} \,,
\end{equation}
where $k(a)$ is the index of the quiver node that intersects the flavor symmetry indexed by $a$. Further, we have again used that, in all cases of relevance of this work, $B^{k(a)a} = 1$.

\begin{table}[ht]
    \centering
    \renewcommand{\arraystretch}{1.2}
    \begin{threeparttable}
    \begin{tabular}{cccccc}
    \toprule
         $\ell$ & $\mathfrak{g}$ & $n_V^0$ & $n_H^0$ & $d_0$ & $k_0$ \\\midrule
         \multirow{2}{*}{$2$} & $\mathfrak{sp}_{q\geq 0}$ & $\frac{q(q-1)}{2}$ & $q(q+4)$ & $q+1$ & $q$ \\
         & $\mathfrak{su}_{2q+4\geq 4}$ & $q^2 + 4q + 3$ & $\frac{3}{2}(q+2)(q+5)$ & $2q + 4$ & $q+2$ \\\midrule
         \multirow{3}{*}{$3$} & $\mathfrak{su}_6^\prime$ & $3$ & $12$ & $6$ & $2$ \\
         & $\mathfrak{su}_3$ & $0$  & $4$ & $3$ & $1$ \\
         & $\varnothing$ & $0$ & $0$ & $1$ & $0$ \\\midrule
         \multirow{2}{*}{$4$} & $\mathfrak{su}_4$ & $0$ & $3$ & $4$ & $1$\\
         & $\varnothing$ & $0$ & $0$ & $1$ & $0$ \\\midrule
         $5$ & $\varnothing$ & $0$ & $0$ & $1$ & $0$ \\\midrule
         $6$ & $\varnothing$ & $0$ & $0$ & $1$ & $0$ \\\bottomrule
    \end{tabular}
    \end{threeparttable}
    \caption{The possible decorations on the (left-most) tensor with self-pairing $1$ in equation \eqref{eqn:quivquiv}.}
    \label{tbl:onedata}
\end{table}

While it was necessary that we do the calculation slightly differently for the cases where $\mathfrak{g} \neq \varnothing$ and $\mathfrak{g} = \varnothing$, we see that the resulting coefficients appearing in $I_8 - I_8^\text{fields}$ relevant for the central charges of the compactification can be written succinctly as
\begin{equation}\label{eqn:B}
    \begin{aligned}
      A &= \frac{r+1}{32} \,, \\
      B &= -\frac{1}{4}\bigg( (r+1)d_0 + \ell \sum_{j=1}^r (r + 1 - j)k_j \bigg) \,, \\
      C_a &= \frac{r+1 - k(a)}{16} \,,
    \end{aligned}
\end{equation}
where $d_0$ is as written in Table \ref{tbl:onedata}; it is $1$ if $\mathfrak{g} = \varnothing$ and $h_{\mathfrak{g}}^\vee$ otherwise.

Next, let us determine the numbers of vector and hypermultiplets at the generic point of the 4d Coulomb branch. We recall here how the Stiefel--Whitney twist acts on the weakly coupled 6d spectrum on the tensor branch. When doing a $\mathbb{Z}_\ell$ Stiefel--Whitney twist, we need to know the following 6d $\rightarrow$ 4d transformations:
\begin{equation}\label{eqn:6d4d}
    \begin{aligned}
      \mathfrak{su}_{\ell k} \text{ vector multiplet } &\rightarrow \mathfrak{su}_{k} \text{ vector multiplet } \,, \\
      \mathfrak{su}_{\ell k_1} \oplus \mathfrak{su}_{\ell k_2} \text{ bifund. hypermultiplet } &\rightarrow \mathfrak{su}_{k_1} \oplus \mathfrak{su}_{k_2} \text{ bifund. hypermultiplet } \,, \\
      \text{ tensor multiplet } &\rightarrow \text{ vector multiplet } \,.
    \end{aligned}
\end{equation}
The subtleties arise from the possible ``decorations'' $\fkg$ on the node $\overset{\fkg}{1}$ in equation \eqref{eqn:quivquiv}, which we will often refer to as the $1$-node of the quiver.\footnote{In geometric terms that describe F-theory constructions of 6d SCFTs, such a node is usually called a $(-1)$-curve.}
How the SW-fold acts on such a tensor with the possibilities for the gauge algebra $\fkg$ has been studied in \cite{Ohmori:2018ona}. Putting all this together we see that the number of vector multiplets and hypermultiplets at the generic point of the Coulomb branch is
\begin{equation}\label{eqn:nv}
    \begin{aligned}
        n_V &= 1 + n_V^0 + \sum_{i=1}^r  k_i^2 \,, \\
        n_H &= n_H^0 + \sum_{i=1}^r k_i(2k_i - k_{i-1})  \,.
    \end{aligned}
\end{equation}
Here, $1 + n_V^0$ is the number of vector multiplets that are associated to $\fkg$ and survive the SW-fold. Similarly, $k_0$ is the dimension of the fundamental representation of this gauge algebra after SW-folding, and $k_1 + n_H^0$ is the total number of surviving hypermultiplets from the $\overset{\fkg}{1}$-node. These quantities follow directly from the action of the Stiefel--Whitney twist and they are summarized in Table \ref{tbl:onedata}. Finally, we determine the contribution to the flavor central charges at the generic point of the 4d Coulomb branch. We have
\begin{equation}
    \kappa_a^\text{generic} = 2k_{k(a)} \,,
\end{equation}
where, again, $k(a)$ is the index of the tensor that intersects the $a$th flavor factor. This follows from the existence of the bifundamental (full) hypermultiplet after the Stiefel--Whitney twist described in equation \eqref{eqn:6d4d}.\footnote{When considering the flavor algebras attached to the $\overset{\fkg}{1}$-node, the matter many not simply be a bifundamental hypermultiplet, but some other bi-representation. In these cases, the contribution from a generic hypermultiplet on the 4d Coulomb branch must be worked out individually.}
In the case of flavor symmetries that intersect the $\overset{\fkg}{1}$-node, the value of $k_0$ is written in Table \ref{tbl:onedata}; it comes from the surviving gauge algebra on that node after the Stiefel--Whitney twist.

\subsection{Examples}

Putting everything together, we can see that the 4d $\mathcal{N}=2$ SW-fold SCFT obtained via the Stiefel--Whitney twist of 6d $(1,0)$ tensor branch configuration as in equation \eqref{eqn:quivquiv} has central charges $a$, $c$, and $\kappa_a$ given as in equation \eqref{eqn:ack}. These quantities can thus be worked out for each of the theories listed in Table \ref{tbl:genSfolds}, and we now do so. The central charges $a$ and $c$ become rather lengthy expressions, especially as one decreases the order of the Stiefel--Whitney twist, $\ell$, which thus gives rise to to more parameters describing the discrete homomorphism $\bbZ_\ell \rightarrow E_8$. Therefore, we have attached a {\tt Mathematica} notebook containing these expressions to the {\tt arXiv} submission of this paper for the ease of the reader.

\subsubsection[\texorpdfstring{$\bbZ_6$}{Z6} SW-folds: \texorpdfstring{$\mathcal{T}_6^{(N)}(p)$}{T6N(p)}]{\boldmath{$\mathbb{Z}_6$} SW-folds: \boldmath{$\mathcal{T}_6^{(N)}(p)$}}

We begin by studying the $\mathbb{Z}_6$ SW-folds: $\mathcal{T}_6^{(N)}(p)$. The 6d SCFT origin, with the flavor symmetry included, is
\begin{equation}\label{eqn:Z6TB}
[\mathfrak{su}_3]\underset{[\mathfrak{su}_2]}{1}\overbrace{\overset{\mathfrak{su}_{6}}{2}
        \cdots\underset{[\mathfrak{su}_6]}{\overset{\mathfrak{su}_{6p}}{2}}}^p\overbrace{\overset{\mathfrak{su}_{6p}}{2}\cdots \overset{\mathfrak{su}_{6p}}{2}}^{N-1} [\mathfrak{su}_{6p}] \,.
\end{equation}
As we determined in Section \ref{sec:e8CF}, the non-Abelian flavor symmetry of the 6d SCFT is generically
\begin{equation}
    G_\text{flavor} = [SU(3) \times SU(2) \times SU(6) \times SU(6p)]/\mathbb{Z}_6 \,,
\end{equation}
however in the special case where $N = 1$, the last two factors combine and one has \begin{equation}
    G_\text{flavor} = [SU(3) \times SU(2) \times SU(6(p+1))] / \mathbb{Z}_6 \,.
\end{equation}
In terms of the quiver written in equation \eqref{eqn:quivquiv}, here we have $\mathfrak{g} = \varnothing$ and
\begin{equation}\label{eqn:k6s}
    k_i = (1, 2, \cdots, p, \underbrace{\,p, \cdots, p\,}_{N-1}) \,.
\end{equation}
In this case, we shall write each of the quantities, $A$, $B$, $C_a$, $n_V$, $n_H$, and $\kappa_a^\text{generic}$ necessary to determine the central charges.
After the Stiefel--Whitney twist the only surviving flavor symmetry is either $\mathfrak{su}_p$ arising from the $\mathfrak{su}_{6p}$ in the case of generic $N$, or $\mathfrak{su}_{p+1}$ coming from the $\mathfrak{su}_{6(p+1)}$ factor when $N = 1$; as there is only one simple non-Abelian flavor algebra we shall drop the index $a$. Note, when $p = 1$ and $N > 1$ there is no surviving flavor symmetry. For the quantities determined from the 6d anomaly polynomial we find
\begin{equation}
    A = \frac{p + N}{32} \,, \quad B = -\frac{1}{4}(p^3 + 3Np^2 +  3N^2p + N) \,, \quad C = \frac{1}{16} \,.
\end{equation}
At the generic point of the 4d Coulomb branch we have
\begin{equation}
    n_V = \frac{1}{6}(6 + p - 3p^2 + 6Np^2 + 2p^3) \,, \quad n_H = \frac{1}{3}p(2 + 3Np + p^2) \,, \quad \kappa^\text{generic} = 2p \,.
\end{equation}
Plugging these values into equation \eqref{eqn:ack}, we find that the central charges of the resulting 4d $\mathcal{N}=2$ SCFTs are
\begin{align}\label{eqn:l6ccs}
      a &= \frac{1}{48}\bigg(28p^3 + 84Np^2 - 5p^2 + 72N^2p - 21p + 10 \bigg) \,, \\
      c &= \frac{1}{12}\bigg(7p^3 + 21 N p^2 - p^2 + 18 N^2 p - 5p + 2 \bigg) \,, \\
      \kappa &= 12p+2 \,.
\end{align}
We emphasize that, regardless of whether the residual flavor symmetry algebra is $\mathfrak{su}_p$ or $\mathfrak{su}_{p+1}$, the flavor central charge is identical. As it is required often throughout this section, we will explain the Dynkin indices for the special subalgebras that we consider. We have
\begin{equation}
    \mathfrak{su}_{\ell k} \rightarrow \mathfrak{su}_\ell \oplus \mathfrak{su}_k \,,
\end{equation}
such that
\begin{equation}
    \bm{\ell k} \rightarrow (\bm{\ell}, \bm{k}) \,.
\end{equation}
The index of the embedding can be worked out from this decomposition of the fundamental representation,\footnote{See \cite{Esole:2020tby} for an explanation of the embedding indices applicable to the special subalgebras.} and we find that the $\mathfrak{su}_\ell$ factor has index $k$, and the $\mathfrak{su}_k$ factor has index $\ell$.

The theory $\mathcal{T}^{(N)}_6(p=1)$ has been previously studied in \cite{Giacomelli:2020gee}. In that case, there is no remaining flavor symmetry and we can see from equation \eqref{eqn:l6ccs} that the central charges are
\begin{equation}
    a = c = \frac{1}{4}(6N + 1)(N + 1) \,.
\end{equation}
The result for this special case matches that found in \cite{Giacomelli:2020gee}.\footnote{To aid in comparison, we note that our $\mathcal{T}^{(N)}_6(p=1)$ theory is equivalent to the $\mathcal{T}^{(r+1)}_{\varnothing, 6}$ theory of \cite{Giacomelli:2020gee}.} In this particular case the central charges are equal as the theory enjoys supersymmetry enhancement, either to $\mathcal{N}=4$ supersymmetry when $N = 1$, or else to $\mathcal{N}=3$ when $N > 1$. When $p = 1$ and $N=1$ the theory is $\mathcal{N}=4$ super-Yang--Mills with gauge group $G_2$; in this case there is an $\mathfrak{su}_2$ flavor symmetry and we can see that the flavor central charge is $\kappa = 14 = \operatorname{dim}G_2$, as expected. In the generic case where $p > 1$ there is no such supersymmetry enhancement.

\subsubsection[\texorpdfstring{$\mathbb{Z}_5$}{Z5} SW-folds: \texorpdfstring{$\mathcal{T}_5^{(N)}(p)$}{T5N(p)}]{\boldmath{$\mathbb{Z}_5$} SW-folds: \texorpdfstring{$\mathcal{T}_5^{(N)}(p)$}}

We now study the $\mathbb{Z}_5$ SW-folds: $\mathcal{T}_5^{(N)}(p)$.
The tensor branch configuration describing the 6d SCFT origins of these 4d theories are
\begin{equation}
[\mathfrak{su}_5]1\overbrace{\overset{\mathfrak{su}_{5}}{2}
        \cdots\underset{[\mathfrak{su}_5]}{\overset{\mathfrak{su}_{5p}}{2}}}^p\overbrace{\overset{\mathfrak{su}_{5p}}{2}\cdots \overset{\mathfrak{su}_{5p}}{2}}^{N-1} [\mathfrak{su}_{5p}] \,.
\end{equation}
We have here written the flavor algebras that exist for generic values of $p$ and $N$, however there is a flavor symmetry enhancement when considering a single M5-brane, $N=1$. The full global structure of the non-Abelian part of the flavor symmetry group was determined in Section \ref{sec:e8CF} and we find
\begin{equation}
    G_\text{flavor} = \begin{cases}
      (SU(5) \times SU(5(p+1)))/\mathbb{Z}_5 \qquad &\text{ when } \quad N = 1  \,, \\
      (SU(5) \times SU(5) \times SU(5p))/\mathbb{Z}_5 \qquad &\text{ when } \quad N > 1\,.
    \end{cases}
\end{equation}
To determine the central charges we must determine the $k_i$ when the tensor branch configuration is written in the form in equation \eqref{eqn:quivquiv}; observe that these $k_i$ are the same as those appearing in equation \eqref{eqn:k6s} in the $\mathcal{T}_6^{(N)}(p)$ case. Using the formula in equation \eqref{eqn:ack} leads to the following central charges:
\begin{align}
        a &= \frac{1}{240}\left(140p^3 + 420 N p^2 - 25p^2 + 360 N^2 p - 69p + 36N + 50 \right) \,, \\
        c &= \frac{1}{60}\left(35p^3 + 105N p^2 - 5p^2 + 90N^2 p- 13p + 12 N + 10 \right) \,.
\end{align}
Finally, we determine the non-Abelian flavor algebra that survives after the Stiefel--Whitney twisted compactification.
Denoting the central charges by subscripts, one finds
\begin{equation}
    \mathfrak{g}_\text{flavor}^\text{4d} = \begin{cases}
      \big(\mathfrak{su}_{p+1}\big)_{12p+2} \qquad &\text{ when } \quad N = 1 \,, \\
      \big(\mathfrak{su}_p\big)_{12p+2} \qquad &\text{ when } \quad N > 1 \,.
    \end{cases}
\end{equation}
Similarly to the $\ell = 6$ case, the theories that we have $\mathcal{T}_5^{(N)}(p=1)$ have been previously studied in \cite{Giacomelli:2020gee}, where they were referred to as the $\mathcal{T}^{(r+1)}_{\varnothing, 5}$ theories. As we can see, the $\mathbb{Z}_5$ SW-folds that are written here constitute a broad generalization of the hitherto known theories.

\subsubsection[\texorpdfstring{$\mathbb{Z}_4$}{Z4} SW-folds: \texorpdfstring{$\mathcal{T}_4^{(N)}(p, 2q)$}{T4N(p,2q)} and \texorpdfstring{$\mathcal{S}_4^{(N)}(p, 2q+1)$}{S4N(p,2q+1)}]{\boldmath{$\mathbb{Z}_4$} SW-folds: \boldmath{$\mathcal{T}_4^{(N)}(p, 2q)$} and \boldmath{$\mathcal{S}_4^{(N)}(p, 2q+1)$}}

There are two classes of $\ell = 4$ SW-folds: $\mathcal{T}_4^{(N)}(p, 2q)$ and $\mathcal{S}_4^{(N)}(p, 2q+1)$. Recall that a Higgs-branch deformation by the homomorphism $\mathbb{Z}_K \rightarrow E_8$ preserves a $\mathbb{Z}_4$ center-flavor symmetry of the 6d SCFT only if the only non-zero entries in equation \eqref{eqn:Zkembed} are $a_4$ and $a_{4^\prime}$. The distinction between the $\mathcal{T}$ and $\mathcal{S}$ theories depends on whether $a_{4^\prime}$ is even or odd, respectively. First, we consider the $\mathcal{T}_4^{(N)}(p, 2q)$ theories, which arise from 6d $(1,0)$ SCFTs with tensor branch configuration
\begin{equation}\label{eqn:T4g}
1\overbrace{\overset{\mathfrak{su}_{8}}{2}
         \cdots\underset{[\mathfrak{su}_4]}{\overset{\mathfrak{su}_{8q}}{2}}}^{q} \overbrace{\overset{\mathfrak{su}_{8q + 4}}{2}
         \cdots\underset{[\mathfrak{su}_4]}{\overset{\mathfrak{su}_{8q + 4p}}{2}}}^{p}
    \overbrace{\overset{\mathfrak{su}_{8q + 4p}}{2}\cdots \overset{\mathfrak{su}_{8q + 4p}}{2}}^{N-1} [\mathfrak{su}_{8q+4p}] \,.
\end{equation}
In this generalized quiver, we have written the flavor algebras for generic values of the parameters $p$, $q$, and $N$. From the analysis in Section \ref{sec:6d}, we can see that the non-Abelian part of the global symmetry group is
\begin{equation}
    G_\text{flavor} = \begin{cases}
    (SU(8q+8))/\mathbb{Z}_4 \qquad &\text{ when } \quad p = 0, q \geq 1, N = 1 \,, \\
    (SU(8) \times SU(8q))/\mathbb{Z}_4 \qquad &\text{ when } \quad p = 0, q \geq 1, N > 1 \,, \\
    (Spin(10) \times SU(4p+4))/\mathbb{Z}_4 \qquad &\text{ when } \quad p \geq 1, q = 0, N = 1 \,, \\
    (Spin(10) \times SU(4) \times SU(4p))/\mathbb{Z}_4 \qquad &\text{ when } \quad p \geq 1, q = 0, N > 1 \,, \\
    (SU(4) \times SU(8q + 4p + 4))/\mathbb{Z}_4 \qquad &\text{ when } \quad p \geq 1, q \geq 1, N = 1 \,, \\
    (SU(4) \times SU(4) \times SU(8q + 4p))/\mathbb{Z}_4 \qquad &\text{ when } \quad p \geq 1, q \geq 1, N > 1 \,.
    \end{cases}
\end{equation}
We can see that if we write the models in equation \eqref{eqn:T4g} in the generic form for the tensor branch configurations that we study, as in equation \eqref{eqn:quivquiv}, then the $k_i$ are given by
\begin{equation}
    k_i = (\underbrace{\,2, 4, \cdots, 2q\,}_q, \underbrace{\,2q + 1, 2q + 2, \cdots, 2q + p\,}_p,  \underbrace{\,2q + p, \cdots, 2q + p\,}_{N-1}) \,.
\end{equation}
We observe that there is a steep ramp, of length $q$, where $k_i$ increases by $2$ each step, followed by a shallower length $p$ ramp where the $k_i$ increases by $1$, and finally a plateau of length $N - 1$. Although it starts to become somewhat tedious, it is straightforward to work out the central charges using equation \eqref{eqn:ack}. We find
\begin{align}
        a &= \frac{1}{48}\bigg( 64q^3 + 192pq^2 + 192Nq^2 - 20q^2 + 168p^2 q + 336Npq - 20pq  \\&\qquad\quad + 144N^2q - 18q + 28p^3 + 84Np^2 - 5p^2 + 72N^2p - 3p + 18N + 10\bigg) \,, \cr
        c &= \frac{1}{12}\bigg( 16q^3 + 48pq^2 + 48Nq^2 - 4q^2 + 42p^2q + 84Npq - 4pq \\&\qquad\quad + 36N^2q - 2q + 7p^3 + 21Np^2 - p^2 + 18N^2p + p + 6N + 2 \bigg) \nonumber \,.
\end{align}
To determine the flavor symmetry that survives the Stiefel--Whitney twisting procedure, it is necessary to understand how, in the cases with $q=0$, the $\mathfrak{so}_{10}$ flavor algebra intersecting the E-string is acted on by the $\mathbb{Z}_4$ center-flavor symmetry. Writing
\begin{equation}\label{eqn:so10d}
    \mathfrak{so}_{10} \rightarrow \mathfrak{su}_4 \oplus \mathfrak{su}_2 \oplus \mathfrak{su}_2 \,,
\end{equation}
we can see that the $\mathbb{Z}_4$ is embedded via the generator $(1,1,0)$ inside of the combined $\mathbb{Z}_4 \times \mathbb{Z}_2 \times \mathbb{Z}_2$ center \cite{Giacomelli:2020gee}. As such, the only surviving subalgebra from the $\mathfrak{so}_{10}$ factor is an $\mathfrak{su}_2$, with embedding index $1$. The Stiefel--Whitney twisting of the remaining flavor symmetry factors can be determined as for the $\ell = 5, 6$ cases. The flavor central charges (denoted in subscript) can also be computed using the formula in equation \eqref{eqn:ack}; the result is
\begin{equation}
    \mathfrak{g}_\text{flavor}^\text{4d} = \begin{cases}
        (\mathfrak{su}_{2q + 2})_{4q + 12} &\quad p = 0, q \geq 1, N = 1 \\
        (\mathfrak{su}_{2})_{4q + 12N} \oplus (\mathfrak{su}_{2q})_{4q+12} &\quad p = 0, q \geq 1, N > 1 \\
        (\mathfrak{su}_2)_{3(p+1)} \oplus (\mathfrak{su}_{p+1})_{2p+12} &\quad p \geq 1, q = 0, N = 1 \\
        (\mathfrak{su}_2)_{3(p+N)} \oplus (\mathfrak{su}_{p})_{2p+12} &\quad p \geq 1, q = 0, N > 1 \\
        (\mathfrak{su}_{2q + p + 1})_{4q+2p+12} &\quad p \geq 1, q \geq 1, N = 1 \\
        (\mathfrak{su}_{2q + p})_{4q+2p+12} &\quad p \geq 1, q \geq 1, N > 1 \,.
    \end{cases}
\end{equation}
Again, similarly to the $\ell = 5$ and $\ell = 6$ SW-folds, the theories $\mathcal{T}_4^{(N)}(p=1, 2q=0)$ have been studied afore in \cite{Giacomelli:2020gee}, where they are called the $\mathcal{T}_{A_2, 4}^{(r+1)}$ theories.

There is another class of $\ell = 4$ SW-folds, which are obtained by starting with the 6d A-type orbi-instanton SCFT Higgsed by a $\mathbb{Z}_4$ center-flavor symmetry preserving $E_8$-homomorphism where the embedding into $a_4^\prime$, as in equation \eqref{eqn:Zkembed}, is odd. The SW-fold SCFTs obtained from the $\mathbb{Z}_4$ Stiefel--Whitney twist of these 6d SCFTs are referred to as $\mathcal{S}_\ell^{(N)}(p, 2q+1)$. The tensor branch configurations of these 6d SCFTs have the form
\begin{equation}
\overset{\mathfrak{su}_{4}}{1} \overbrace{\overset{\mathfrak{su}_{12}}{2}
        \cdots\underset{[\mathfrak{su}_4]}{\overset{\mathfrak{su}_{8q + 4}}{2}}}^{q} \overbrace{\overset{\mathfrak{su}_{8q + 8}}{2}
        \cdots\underset{[\mathfrak{su}_4]}{\overset{\mathfrak{su}_{8q + 4p + 4}}{2}}}^{p}
        \overbrace{\overset{\mathfrak{su}_{8q + 4p + 4}}{2}\cdots \overset{\mathfrak{su}_{8q + 4p + 4}}{2}}^{N-1} [\mathfrak{su}_{8q + 4p + 4}] \,,
\end{equation}
where, as usual, the flavor symmetry can enhance when the parameters $p$, $q$, and $N$ obtain their limiting values. The flavor groups, including how the $\mathbb{Z}_4$ quotient acts, can be determined from the algorithm described in Section \ref{sec:6d}. Computing the central charges is a straightforward application of equation \eqref{eqn:ack}:
\begin{align}
        a &= \frac{1}{48}\bigg( 64q^3 + 192pq^2 + 192Nq^2 + 76q^2 + 168p^2 q + 336Npq \cr&\qquad\quad+ 172pq  + 144N^2q + 192Nq   + 10q + 28p^3 + 84Np^2 \\&\qquad\quad+ 79p^2 + 72N^2p + 168Np + 35p + 72N^2 + 66N + 4\bigg) \,, \cr
        c &= \frac{1}{12}\bigg( 16q^3 + 48pq^2 + 48Nq^2 + 20q^2 + 42p^2q + 84Npq \cr&\qquad\quad + 44pq  + 36N^2q +48Nq  + 6q  + 7p^3 + 21Np^2 \\&\qquad\quad+ 20p^2 + 18N^2p + 42Np + 11p + 18N^2 + 18N + 2 \bigg) \nonumber \,.
\end{align}
Similarly, the flavor algebras that survive the Stiefel--Whitney twist can be determined, and their flavor central charges are again given by equation \eqref{eqn:ack}. We find
\begin{equation}
    \mathfrak{g}_\text{flavor}^\text{4d} = \begin{cases}
        (\mathfrak{su}_{2q+3})_{4q + 14} &\quad p = 0, q \geq 0, N = 1 \\
        (\mathfrak{su}_2)_{4q + 12N + 2} \oplus (\mathfrak{su}_{2q+1})_{4q + 14} &\quad p = 0, q \geq 0, N > 1 \\
        (\mathfrak{su}_{2q + p + 2})_{4q + 2p + 14} &\quad p \geq 1, q \geq 0, N = 1 \\
        (\mathfrak{su}_{2q + p + 1})_{4q + 2p + 14} &\quad p \geq 1, q \geq 0, N > 1 \,.
    \end{cases}
\end{equation}
The theories $\mathcal{S}_4^{(N)}(p=0, 2q+1 =1)$ were studied in \cite{Giacomelli:2020gee}, where they were called the $\mathcal{S}^{(r)}_{A_2, 4}$ theories. The central charges and flavor symmetries that we compute here agree with what was found in that particular limiting case. We have similarly labelled these generalized S-fold SCFTs by $\mathcal{S}$ and $\mathcal{T}$ to match with the notation for the special cases that have been previously studied.

In this paper, we have mainly been concerned with the identification of the 4d non-Abelian flavor symmetry that descends from the 6d non-Abelian flavor symmetry. In fact, the 6d SCFTs under consideration also contain Abelian symmetries that arise from the ABJ-anomaly-free combinations of the $\mathfrak{u}(1)$s rotating the bifundamental hypermultiplets. Under certain circumstances, these $\mathfrak{u}(1)$s can enhance, and then we expect a further non-Abelian factor in the 4d flavor symmetry. This occurs when the 4d Coulomb branch description of the SW-fold contains a plateau of neighboring $\mathfrak{su}_2$ gauge algebras: then the $\mathfrak{u}(1)$ enhances to an $\mathfrak{su}_2$ under which the gauge bifundamentals are charged. The SW-folds with this extra, enhanced, baryonic $\mathfrak{su}_2$ flavor symmetry are
\begin{equation}\label{eqn:baryenc}
    \begin{gathered}
        \mathcal{T}_4^{(N)}(0,2) \,, \quad \mathcal{S}_4^{(N)}(1,1) \,, \quad \mathcal{T}_3^{(N)}(0,1,0) \,, \\ \mathcal{S}_3^{(N)}(1,0,1) \,, \quad \mathcal{R}_3^{(N)}(0,0,2) \,, \quad \mathcal{S}_2^{(N)}(0,0,0,1,0) \\
        \mathcal{T}_2^{(N)}(0,0,0,0,2) \,, \quad
        \mathcal{T}_2^{(N)}(0,1,0,0,0) \,, \quad
        \mathcal{T}_2^{(N)}(2,0,0,0,0) \,, \quad
        \mathcal{T}_2^{(N)}(1,0,0,0,1) \,.
    \end{gathered}
\end{equation}
In each case, we can see that they correspond to theories obtained from an orbi-instanton theory involving M5-branes probing $\mathbb{C}^2/\mathbb{Z}_{2\ell}$; after the $\mathbb{Z}_\ell$ Stiefel--Whitney twist, the orbifold is reduced to $\mathbb{C}^2/\mathbb{Z}_2$, and the additional $\mathfrak{su}_2$ global symmetry comes from the exceptional isometry of this particular orbifold. For low values of $N$ we expect that this baryonic $\mathfrak{su}_2$ can combine with other non-Abelian factors in the flavor symmetry, and cause further enhancement. In rare cases there can also be dehancement. We discuss some instances where this enhancement occurs in Section \ref{sec:classs}.

\subsubsection[\texorpdfstring{$\mathbb{Z}_3$}{Z3} SW-folds: \texorpdfstring{$\mathcal{T}_3^{(N)}(p,s,3q)$}{T3N(p,s,3q)}, \texorpdfstring{$\mathcal{S}_3^{(N)}(p,s,3q+1)$}{S3N(p,s,3q+1)}, and \texorpdfstring{$\mathcal{R}_3^{(N)}(p,s,3q+2)$}{R3N(p,s,3q+2)}]{\boldmath{$\mathbb{Z}_3$} SW-folds: \boldmath{$\mathcal{T}_3^{(N)}(p,s,3q)$}, \boldmath{$\mathcal{S}_3^{(N)}(p,s,3q+1)$}, and \boldmath{$\mathcal{R}_3^{(N)}(p,s,3q+2)$}}

There are three distinct ways that one can construct a Higgs-branch flow from the rank $N$ $(\mathfrak{e}_8, \mathfrak{su}_K)$ orbi-instanton theory such that the resulting SCFT has a $\mathbb{Z}_3$ center-flavor symmetry. From Section \ref{sec:e8CF}, we see that $K$ must be a multiple of three and the homomorphism $\mathbb{Z}_K \rightarrow E_8$ must be specified by the vector
\begin{equation}
    (a_1, a_2, a_3, a_4, a_5, a_6, a_{4^\prime}, a_{2^\prime}, a_{3^\prime}) = (0, 0, a_3, 0, 0, a_6, 0, 0, a_{3^\prime}) \,.
\end{equation}
In all cases, the resulting 6d SCFTs have tensor branch configurations of the form in equation \eqref{eqn:quivquiv}, however the algebra $\mathfrak{g}$ associated to the left-most $1$-node in equation \eqref{eqn:quivquiv}, depends on the parity modulo three of $a_{3^\prime}$. Respectively, we find that the $1$-node has no gauge algebra; $\mathfrak{su}_3$ with twelve fundamental and one antisymmetric hypermultplets; and an $\mathfrak{su}_6$ algebra with fifteen fundamental hypermultiplets and one further hypermultiplet in the triple-antisymmetric representation. As in Table \ref{tbl:genSfolds}, we label these theories by $\mathcal{T}$, $\mathcal{S}$, and $\mathcal{R}$, respectively. The tensor branch configurations for each of these configurations are shown in Table \ref{tbl:6dscfts}, and we do not repeat them here.

We begin our journey into the $\mathbb{Z}_3$ SW-folds with the $\mathcal{T}_3^{(N)}(p,s,3q)$ theory, whose 6d origins have $\fkg = \emptyset$ for the $1$-node.
The central charges can be determined straightforwardly from the tensor branch configuration by application of the formulae in equation \eqref{eqn:ack}. One finds
\begingroup
\allowdisplaybreaks
\begin{align}
        a &= \frac{1}{48} \bigg(72 N^2 p+216 N^2 q+144 N^2 s+84 N p^2+504 N p q+336 N p s \cr&\qquad\quad+324 N q^2  +576 N q s+192
   N s^2+36 N+252 p^2 q+168 p^2 s \\&\qquad\quad +28 p^3  -5 p^2+324 p q^2  +576 p q s-30 p q+192 p s^2-20 p s \cr&\qquad\quad +15 p+324
   q^2 s+108 q^3  -45 q^2+288 q s^2-60 q s-9 q+64 s^3-20 s^2+10\bigg) \,, \cr
        c &= \frac{1}{12} \bigg(18 N^2 p+54 N^2 q+36 N^2 s+21 N p^2+126 N p q+84 N p s \cr&\qquad\quad +81 N q^2  +144 N q s+48 N
   s^2+12 N+63 p^2 q+42 p^2 s+7 p^3 \\&\qquad\quad -p^2+81 p q^2  +144 p q s-6 p q+48 p s^2-4 p s+7 p+81 q^2 s \cr&\qquad\quad +27 q^3-9
   q^2+72 q s^2  -12 q s+3 q+16 s^3-4 s^2+4 s+2\bigg) \nonumber \,.
\end{align}
\endgroup
To determine the flavor algebra for the theory after Stiefel--Whitney twist, we need to understand how the $\mathbb{Z}_3$ quotient acts on the $\mathfrak{e}_6$ flavor symmetry. The decomposition is
\begin{equation}\label{eqn:e6ss}
  \begin{aligned}
    \mathfrak{e}_6 &\rightarrow \mathfrak{g}_2 \oplus \mathfrak{su}_3 \\
    \bm{27} &\rightarrow (\bm{7,3}) \oplus (\bm{1,\overline{6}}) \,,
  \end{aligned}
\end{equation}
where the $\mathbb{Z}_3$ acts on the $\mathfrak{su}_3$ factor and only the $\mathfrak{g}_2$ survives. From the decomposition of the fundamental representation in equation \eqref{eqn:e6ss}, we see that the Dynkin index of the $\mathfrak{g}_2$ subalgebra is one. Similarly, when there is an $\mathfrak{su}_3 \oplus \mathfrak{su}_2$ flavor algebra attached to the undecorated $1$-node, we note that the $\mathbb{Z}_3$ acts only on the $\mathfrak{su}_3$ factor and leaves the $\mathfrak{su}_2$ factor untouched. The remaining flavor factors are quotiented by the Stiefel--Whitney twist exactly as in the $\ell > 3$ cases that we have discussed. In the end, one discovers that the flavor symmetries, and the flavor central charges of these 4d SCFTs are:
\begin{equation}\label{eqn:T4flav}
    \begin{aligned}
        q = s = 0, p \geq 1, N \geq 1 \,&: \quad (\mathfrak{g}_2)_{4(N + p)} \oplus (\mathfrak{su}_1)_{12N+2p} \oplus (\mathfrak{su}_p)_{12+2p} \\
        q = 0, s \geq 1, p \geq 0, N \geq 1 \,&: \quad (\mathfrak{su}_2)_{4(N + p + s)} \oplus (\mathfrak{su}_1)_{12(N+p)+4s} \\ &\qquad\qquad \oplus (\mathfrak{su}_1)_{12N+4s+2p} \oplus (\mathfrak{su}_{2s+p})_{12+4s+2p} \\
        q \geq 1, s \geq 0,  p \geq 0, N \geq 1 \,&:\quad
        (\mathfrak{su}_1)_{12(N+p+s)+6q} \oplus (\mathfrak{su}_1)_{12(N+p)+6q+4s} \\ &\qquad\qquad
        \oplus (\mathfrak{su}_1)_{12N+6q+4s+2p}  \oplus (\mathfrak{su}_{3q+2s+p})_{12+6q+4s+2p}
         \,.
    \end{aligned}
\end{equation}
We have introduced a shorthand notation here as the number of combinations of $p$, $q$, $s$, and $N$ where there are flavor symmetry enhancements becomes large. In this way, if we write the flavor symmetry as $(\mathfrak{su}_{k_1})_{\kappa_1} \oplus (\mathfrak{su}_{k_2})_{\kappa_2}$ then for $\kappa_1 \neq \kappa_2$ the flavor symmetry is as written, but if $\kappa_1 = \kappa_2$ then there is an enhancement to $(\mathfrak{su}_{k_1 + k_2})_{\kappa_1}$. Of course, if there is an $\mathfrak{su}_1$ factor where the flavor central charge is such that it does not combine with another flavor symmetry factor, then that symmetry is, of course, trivial.

Next, we turn to the $\mathcal{S}_3^{(N)}(p,s,3q+1)$ SCFTs, originating from a 6d theory with $\fkg = \mathfrak{su}_3$.
From the tensor branch description of the 6d origin and the formulae in equation \eqref{eqn:ack}, one can determine the central charges. As these expressions are rather lengthy, we remind the reader that they also appear in the {\tt Mathematica} notebook attached to the {\tt arXiv} submission for this paper. The central charges for these theories are
\begin{align}
        a &= \frac{1}{48} \bigg( 72 N^2 p+216 N^2 q+144 N^2 s+72 N^2+84 N p^2+504 N p q  +336 N p s+168 N p\nonumber\\&\qquad\quad +324 N
   q^2+576 N q s+216 N q+192 N s^2  +192 N s+72 N+252 p^2 q+168 p^2 s \nonumber \\&\qquad\quad +28 p^3+79 p^2+324 p q^2  +576 p q
   s+186 p q+192 p s^2+172 p s+41 p \\ &\qquad\quad +324 q^2 s+108 q^3 +63 q^2+288 q s^2+156 q s-3 q+64 s^3+76 s^2+16
   s+6 \bigg) \,, \nonumber\\
        c &= \frac{1}{12} \bigg( 18 N^2 p+54 N^2 q+36 N^2 s+18 N^2+21 N p^2+126 N p q  +84 N p s \cr&\qquad\quad+42 N p  +81 N q^2+144
   N q s+54 N q+48 N s^2+48 N s+21 N \\&\qquad\quad +63 p^2 q  +42 p^2 s+7 p^3  +20 p^2+81 p q^2+144 p q s+48 p q+48 p
   s^2+44 p s \cr&\qquad\quad +14 p+81 q^2 s+27 q^3  +18 q^2+72 q s^2+42 q s+6 q+16 s^3+20 s^2+9 s+3 \bigg) \nonumber \,.
\end{align}
As expected, the generic four-dimensional flavor algebra experiences enhancement at the lower limits of the parameters describing the $E_8$-homomorphism, $p$, $q$, and $s$, and also when one has only a single M5-brane, $N = 1$. The resulting flavor symmetries, together with the flavor central charges, are
\begin{equation}
    \begin{aligned}
        q,s,p \geq 0, N \geq 1 \,&:\quad
        (\mathfrak{su}_1)_{12(N+p+s)+6q+2} \oplus (\mathfrak{su}_1)_{12(N+p)+6q+4s+2} \\ &\qquad\qquad
        \oplus (\mathfrak{su}_1)_{12N+6q+4s+2p+2}  \oplus (\mathfrak{su}_{3q+2s+p+1})_{12+6q+4s+2p+2}
         \,.
    \end{aligned}
\end{equation}
Here, we use an F-theoretic convention for keeping track of trivial symmetry factors such as ``$\mathfrak{su}_1$'' since the parameters can sometimes conspire such that two of the $\mathfrak{su}$ flavor factors have the same flavor central charges. In such situations, the flavor symmetry enhances as described around equation \eqref{eqn:T4flav}.

Finally, we turn to the third class of $\ell = 3$ SW-folds, which we refer to as the $\mathcal{R}_3^{(N)}(p,s,3q+2)$ SW-folds. The central charges are again determined from the tensor branch configuration of the 6d SCFT of which these SW-folds are the $\mathbb{Z}_3$ Stiefel--Whitney twisted torus compactification. They are
\begingroup
\allowdisplaybreaks
\begin{align}\label{eqn:3Rccs}
        a &= \frac{1}{48} \bigg( 72 N^2 p+216 N^2 q+144 N^2 s+144 N^2+84 N p^2+504 N p q \cr&\qquad\quad +336 N p s  +336 N p+324 N
   q^2+576 N q s+432 N q+192 N s^2 \cr&\qquad\quad +240 N s+180 N  +252 p^2 q+168 p^2 s+28 p^3+163 p^2+324 p q^2 \\&\qquad\quad +576 p q
   s+402 p q+192 p s^2  +220 p s+139 p+324 q^2 s+108 q^3\cr&\qquad\quad+171 q^2+288 q s^2+372 q s+75 q+64 s^3+100 s^2+176 s+16 \bigg) \,, \cr
        c &= \frac{1}{12} \bigg( 18 N^2 p+54 N^2 q+36 N^2 s+36 N^2+21 N p^2+126 N p q  +84 N p s+84 N p \cr&\qquad\quad +81 N q^2+144
   N q s+108 N q+48 N s^2  +60 N s+48 N+63 p^2 q+42 p^2 s \\&\qquad\quad +7 p^3+41 p^2+81 p q^2 +144 p q s+102 p q+48 p
   s^2+56 p s+39 p+81 q^2 s \cr&\qquad\quad +27 q^3  +45 q^2+72 q s^2+96 q s+27 q+16 s^3+26 s^2+50 s+8 \bigg) \nonumber \,.
\end{align}
\endgroup
The flavor symmetries and flavor central charges can also be worked out using equation \eqref{eqn:ack}, and we find the following result:
\begin{equation}
    \begin{aligned}
        q,s,p \geq 0, N \geq 1 \,&:\quad
        (\mathfrak{su}_1)_{12(N+p+s)+6q+4} \oplus (\mathfrak{su}_1)_{12(N+p)+6q+4s+4} \\ &\qquad\qquad
        \oplus (\mathfrak{su}_1)_{12N+6q+4s+2p+4}  \oplus (\mathfrak{su}_{3q+2s+p+2})_{12+6q+4s+2p+4}
         \,.
    \end{aligned}
\end{equation}
While the $\mathcal{T}_3^{(N)}(p=1,s=0,3q=0)$ and $\mathcal{S}_3^{(N)}(p=0,s=0,3q+1=1)$ S-fold SCFTs have been studied before in \cite{Giacomelli:2020gee}, where they are referred to as the $\mathcal{T}_{D_4,3}^{(r+1)}$ and $\mathcal{S}_{D_4,3}^{(r)}$ theories, respectively, the theories $\mathcal{R}_3^{(N)}(p,s,3q+2)$ have not been studied in the context of S-folds before. The theory $\mathcal{R}_3^{(1)}(p=0,s=0,3q+2)$ has appeared previously in \cite{Ohmori:2018ona}, where the authors point out that, at the generic point of the 4d Coulomb branch, there is a half-hypermultiplet transforming in the $\bm{4}$ of the rightmost $\mathfrak{su}_2$ gauge algebra; this arises from the action of the Stiefel--Whitney twist on the triple-antisymmetric representation of the $\mathfrak{su}_6$ associated to the $1$-node. We refer the reader to Appendix \ref{app:lit}, where we list all of the limiting cases of the Stiefel--Whitney twists and S-folds that have previously appeared in the literature.

\subsubsection[\texorpdfstring{$\mathbb{Z}_2$}{Z2} SW-folds: \texorpdfstring{$\mathcal{T}_2^{(N)}(p,s,u,2t,q)$}{T2N(p,s,u,2t,q)} and \texorpdfstring{$\mathcal{S}_2^{(N)}(p,s,u,2t+1,q)$}{S2N(p,s,u,2t+1,q)}]{\boldmath{$\mathbb{Z}_2$} SW-folds: \boldmath{$\mathcal{T}_2^{(N)}(p,s,u,2t,q)$} and \boldmath{$\mathcal{S}_2^{(N)}(p,s,u,2t+1,q)$}}

The last class of SW-folds which we wish to consider are those involving a $\mathbb{Z}_2$ Stiefel--Whitney twist. These theories depend on five $E_8$-homomorphism parameters, associated to the five different nodes of the $E_8$ Dynkin diagram with even Dynkin label, and one positive integer counting the number of M5-branes. This proliferation of parameters leads to very complicated and unwieldy expressions for the central charges, which are typically degree three polynomials in these parameters. For posterity, we present these expressions here, however, we refer the reader to the attached {\tt Mathematica} notebook for a more practical format.

We begin with the $\mathcal{T}_2^{(N)}(p,s,u,2t,q)$ theories, which arise from 6d SCFTs obtained from $E_8$-homomorphisms of the rank $N$ $(\mathfrak{e}_8, \mathfrak{su}_K)$ orbi-instanton where the parameter $a_{4^\prime}$, as in equation \eqref{eqn:Zkembed} is even. The generalized quivers describing the 6d SCFTs are depicted in Table \ref{tbl:6dscfts}, and from there one can determine the central charges using equation \eqref{eqn:ack}. We find
\begingroup
\allowdisplaybreaks
\begin{align}
        a &= \frac{1}{48} \bigg( 72 N^2 p+72 N^2 q+144 N^2 s+288 N^2 t+216 N^2 u+84 N p^2+168 N p q \cr&\qquad\quad +336 N p s+672 N
   p t+504 N p u+12 N q^2+192 N q s+240 N q t+216 N q u \cr&\qquad\quad +192 N s^2+768 N s t+576 N s u+480 N t^2+864 N
   t u+324 N u^2+72 N \cr&\qquad\quad +84 p^2 q+168 p^2 s+336 p^2 t+252 p^2 u+28 p^3-5 p^2+12 p q^2+192 p q s \cr&\qquad\quad +240 p q
   t  +216 p q u-10 p q+192 p s^2+768 p s t+576 p s u-20 p s+480 p t^2 \\&\qquad\quad +864 p t u  -40 p t+324 p u^2-30 p
   u+51 p+12 q^2 s+12 q^2 t+12 q^2 u-5 q^2 \cr&\qquad\quad +96 q s^2  +240 q s t+216 q s u-20 q s+120 q t^2+240 q t u-40
   q t+108 q u^2 \cr&\qquad\quad -30 q u  +3 q  +384 s^2 t+288 s^2 u+64 s^3-20 s^2+480 s t^2+864 s t u-80 s t \cr&\qquad\quad +324 s
   u^2 -60 s u  +36 s+480 t^2 u+160 t^3-80 t^2+432 t u^2-120 t u \cr&\qquad\quad +24 t+108 u^3-45 u^2+27 u +10
   \bigg) \,, \cr
        c &= \frac{1}{12} \bigg( 18 N^2 p+18 N^2 q+36 N^2 s+72 N^2 t+54 N^2 u+21 N p^2+42 N p q  +84 N p s \cr&\qquad\quad +168 N p
   t  +126 N p u+3 N q^2+48 N q s+60 N q t+54 N q u+48 N s^2 \cr&\qquad\quad +192 N s t  +144 N s u+120 N t^2+216 N t u+81
   N u^2+24 N+21 p^2 q+42 p^2 s \cr&\qquad\quad +84 p^2 t  +63 p^2 u+7 p^3-p^2+3 p q^2+48 p q s+60 p q t+54 p q u-2 p
   q+48 p s^2 \cr&\qquad\quad +192 p s t  +144 p s u-4 p s+120 p t^2+216 p t u-8 p t+81 p u^2-6 p u+19 p  \\&\qquad\quad +3 q^2 s +3 q^2
   t  +3 q^2 u-q^2+24 q s^2+60 q s t+54 q s u-4 q s+30 q t^2+60 q t u \cr&\qquad\quad -8 q t+27 q u^2  -6 q u  + 3 q+96 s^2
   t+72 s^2 u+16 s^3-4 s^2+120 s t^2+216 s t u \cr&\qquad\quad -16 s t+81 s u^2-12 s u  +16 s+120 t^2 u+40 t^3-16
   t^2+108 t u^2 \cr&\qquad\quad -24 t u+16 t+27 u^3-9 u^2+15 u+2 \bigg) \nonumber \,.
\end{align}
\endgroup
To determine the flavor symmetries after the $\mathbb{Z}_2$ Stiefel--Whitney twist, it is necessary to know how the $\mathbb{Z}_2$ acts on the flavor factor that is attached to the $1$-node in the description of the 6d origin. First, we consider the special case where $q = 0$, then the flavor symmetry attached to the undecorated $1$-node (i.e., with $\fkg = \emptyset$) is either $\mathfrak{e}_7$, $\mathfrak{so}_{10}$, $\mathfrak{su}_3 \oplus \mathfrak{su}_2$, or $\varnothing$, depending on which combinations of parameters $t$, $u$, $s$, $p$ attain their lower limits, if any.
When we have $\mathfrak{e}_7$, we consider the special subalgebra
\begin{equation}
    \mathfrak{e}_7 \rightarrow \mathfrak{f}_4 \oplus \mathfrak{su}_2 \,,
\end{equation}
where the Dynkin index of the $\mathfrak{f}_4$ factor is one, and the $\mathbb{Z}_2$ acts as the center of the $\mathfrak{su}_2$. When the flavor algebra is $\mathfrak{so}_{10}$ we have the decomposition
\begin{equation}
    \begin{aligned}
      \mathfrak{so}_{10} &\rightarrow \mathfrak{su}_4 \oplus \mathfrak{su}_2 \oplus \mathfrak{su}_2 \\
      \bm{10} &\rightarrow (\bm{6,1,1}) \oplus (\bm{1,2,2})  \,.
    \end{aligned}
\end{equation}
The $\mathbb{Z}_2$ quotient is generated by the element $(2,0,0)$ of the combined center $\mathbb{Z}_4 \times \mathbb{Z}_2 \times \mathbb{Z}_2$ Note that this is the $\mathbb{Z}_2$ subgroup of the $\mathbb{Z}_4$ discussed around equation \eqref{eqn:so10d}. Thus we can see that the surviving flavor algebra in 4d is
\begin{equation}
    \mathfrak{su}_2^{(2)} \oplus \mathfrak{su}_2^{(1)} \oplus \mathfrak{su}_2^{(1)} \,,
\end{equation}
where we have written the Dynkin indices as superscripts, as determined from the decomposition of the vector representation. We can see from the Dynkin indices and the decomposition of the representations that the surviving flavor algebra from the $\mathfrak{so}_{10}$ after turning on this particular $\mathbb{Z}_2$ is in fact the enhanced $\mathfrak{so}_7^{(1)}$. Finally, when the 6d flavor symmetry attached to the $1$-node is $\mathfrak{su}_3 \oplus \mathfrak{su}_2$, we are considering the same decomposition as we did in the $\ell = 3$ case; the $\mathbb{Z}_2$ quotient acts on the $\mathfrak{su}_2$, and leaves the $\mathfrak{su}_3$ as a flavor symmetry of the Stiefel--Whitney twisted theory. When $q \geq 1$, the flavor symmetry attached to the $1$-node is $\mathfrak{so}_{4n}$, where $n$ is fixed in terms of the parameters $p$, $q$, $s$, $t$, $u$. The relevant decomposition appears in \cite{Ohmori:2018ona}, and we have
\begin{equation}
    \begin{aligned}
      \mathfrak{so}_{4n} &\rightarrow \mathfrak{su}_2^{(n)} \oplus \mathfrak{sp}_n^{(1)} \\
      \bm{4n} &\rightarrow (\bm{2, 2n}) \,.
    \end{aligned}
\end{equation}
The action of the $\mathbb{Z}_2$ is on the $\mathfrak{su}_2$ factor, and the flavor symmetry left after the Stiefel--Whitney twist is $\mathfrak{sp}_n$. As we can see, the embedding index of the surviving factor is one.

To proceed further, it is helpful to split up our analysis into the cases $q > 0$ and $q = 0$.

\paragraph{\boldmath{$q>0$}}

We begin by studying the flavor symmetry when $q > 0$. Due to the presence of the symplectic gauge algebra the flavor symmetries surviving after the Stiefel--Whitney twist have rather complex dependences on the $E_8$-homomorphism parameters. The most generic case occurs when $t > 0$, and we find
\begin{equation}
    \begin{aligned}
        &(\mathfrak{su}_1)_{12(N+p+s+u)+2q+8t} \oplus (\mathfrak{su}_1)_{12(N+p+s)+2q+8t+6u}
        \oplus (\mathfrak{su}_1)_{12(N+p)+2q+8t+6u+4s} \\ &\qquad \oplus
        (\mathfrak{su}_1)_{12N+2q+8t+6u+4s+2p}  \oplus (\mathfrak{su}_{q+4t+3u+2s+p})_{12+2q+8t+6u+4s+2p}
         \,.
    \end{aligned}
\end{equation}
When $t = 0$, but $u > 0$, the flavor symmetry is
\begin{equation}
    \begin{aligned}
        &(\mathfrak{sp}_1)_{6(N+p+s+u) + q} \oplus (\mathfrak{su}_1)_{12(N+p+s)+2q+6u}
        \oplus (\mathfrak{su}_1)_{12(N+p)+2q+6u+4s} \\ &\qquad \oplus
        (\mathfrak{su}_1)_{12N+2q+6u+4s+2p}  \oplus (\mathfrak{su}_{q+3u+2s+p})_{12+2q+6u+4s+2p}
         \,.
    \end{aligned}
\end{equation}
Next, we must consider the case where $t = u = 0$, but $s > 0$. The flavor symmetry becomes
\begin{equation}
    \begin{aligned}
        &(\mathfrak{sp}_2)_{6(N+p+s) + q}
        \oplus (\mathfrak{su}_1)_{12(N+p)+2q+4s}  \oplus
        (\mathfrak{su}_1)_{12N+2q+4s+2p}  \oplus (\mathfrak{su}_{q+2s+p})_{12+2q+4s+2p}
         \,.
    \end{aligned}
\end{equation}
When $t = u = s = 0$ and $p > 0$, one finds that the flavor algebra is
\begin{equation}
    \begin{aligned}
        &(\mathfrak{sp}_3)_{6(N+p) + q}
        \oplus (\mathfrak{su}_1)_{12N+2q+2p}  \oplus (\mathfrak{su}_{q+p})_{12+2q+2p}
         \,.
    \end{aligned}
\end{equation}
Finally, when all of the $E_8$-homomorphism parameters, except $q$, vanish, the flavor is
\begin{equation}\label{eqn:swiper}
    \begin{cases}
        (\mathfrak{sp}_4)_{6N + q} \oplus (\mathfrak{su}_{q})_{12+2q} \,, \quad &\text{when} \quad N > 1 \,, \\
        (\mathfrak{sp}_{q+4})_{6+q} \,, \quad &\text{when} \quad N = 1 \,.
    \end{cases}
\end{equation}

\paragraph{\boldmath{$q =0$}}
A similar analysis can be carried out when $q = 0$, and again one finds a variety of special cases. When $t > 0$ the flavor symmetry is
\begin{equation}
    \begin{aligned}
        &(\mathfrak{su}_1)_{12(N+p+s+u)+8t} \oplus (\mathfrak{su}_1)_{12(N+p+s)+8t+6u}
        \oplus (\mathfrak{su}_1)_{12(N+p)+8t+6u+4s} \\ &\qquad \oplus
        (\mathfrak{su}_1)_{12N+8t+6u+4s+2p}  \oplus (\mathfrak{su}_{4t+3u+2s+p})_{12+8t+6u+4s+2p}
         \,.
    \end{aligned}
\end{equation}
When $t = 0$, but $u > 0$, the flavor symmetry is
\begin{equation}
    \begin{aligned}
        &(\mathfrak{su}_3)_{6(N + u + s + p)} \oplus (\mathfrak{su}_1)_{12(N+p+s)+6u}
        \oplus (\mathfrak{su}_1)_{12(N+p)+6u+4s} \\ &\qquad \oplus
        (\mathfrak{su}_1)_{12N+6u+4s+2p}  \oplus (\mathfrak{su}_{3u+2s+p})_{12+6u+4s+2p}
         \,.
    \end{aligned}
\end{equation}
Next, we consider the case where $t = u = 0$, but $s > 0$. The flavor symmetry is
\begin{equation}
    \begin{aligned}
        &(\mathfrak{so}_7)_{6(N+p+s)}
        \oplus (\mathfrak{su}_1)_{12(N+p)+4s}  \oplus
        (\mathfrak{su}_1)_{12N+4s+2p}  \oplus (\mathfrak{su}_{2s+p})_{12+4s+2p}
         \,.
    \end{aligned}
\end{equation}
Finally, when $t = u = s = 0$ and $p > 0$, one finds that the flavor algebra is
\begin{equation}\label{eqn:dora}
    \begin{aligned}
        &(\mathfrak{f}_4)_{6(N+p)}
        \oplus (\mathfrak{su}_1)_{12N+2p}  \oplus (\mathfrak{su}_{p})_{12+2p}
         \,.
    \end{aligned}
\end{equation}
Note that this last expression is not valid when $p = N = 1$ due to an exceptional enhancement of the associated 6d SCFT, which we discuss below. This analysis exhausts the non-Abelian flavor symmetries of the $\mathcal{T}_2^{(N)}(p,s,u,2t,q)$ SW-fold SCFTs. In rare occasions, the flavor symmetry can be enhanced further, either because an Abelian $\mathfrak{u}(1)$ flavor symmetry of the 6d SCFT can enhance to an $\mathfrak{su}_2$ as described around equation \eqref{eqn:baryenc}, or else because the 6d SCFT has a baryonic $\mathfrak{su}_2$ flavor symmetry, in addition to the flavor symmetries that we have considered here. We discuss this latter case at the end of this subsection. Finally, it appears that in a small number of exceptional circumstances, there can also be flavor symmetry \emph{dehancement}; we explore these examples further in Section \ref{sec:classs}.

To conclude this subsection, we turn to the last class of SW-folds that we wish to consider. These are the $\mathbb{Z}_2$ SW-folds: $\mathcal{S}_2^{(N)}(p,s,u,2t+1,q)$. As usual, the central charges can be worked out from the formulae in equation \eqref{eqn:ack}, and one finds:
\begingroup
\allowdisplaybreaks
\begin{align}
        a &= \frac{1}{48} \bigg( 72 N^2 p+72 N^2 q+144 N^2 s+288 N^2 t+216 N^2 u+144 N^2+84 N p^2 \cr&\qquad\quad +168 N p q+336 N p
   s+672 N p t+504 N p u+336 N p+12 N q^2+192 N q s \cr&\qquad\quad +240 N q t+216 N q u+120 N q+192 N s^2+768 N s
   t+576 N s u+384 N s \cr&\qquad\quad +480 N t^2+864 N t u+480 N t+324 N u^2+144 N u+192 N+84 p^2 q+168 p^2 s \cr&\qquad\quad +336 p^2
   t+252 p^2 u+28 p^3+163 p^2+12 p q^2+192 p q s+240 p q t+216 p q u \\&\qquad\quad +110 p q+192 p s^2+768 p s t+576
   p s u+364 p s+480 p t^2+864 p t u+440 p t \cr&\qquad\quad +324 p u^2+114 p u+151 p + q^2 +12 q^2 s+12 q^2 t+12 q^2 u+96 q
   s^2+240 q s t \cr&\qquad\quad +216 q s u+100 q s+120 q t^2+240 q t u+80 q t+108 q u^2+98 q u+13 q+384 s^2 t \cr&\qquad\quad +288 s^2
   u+64 s^3+172 s^2+480 s t^2+864 s t u+400 s t+324 s u^2+84 s u+116 s \cr&\qquad\quad +480 t^2 u+160 t^3+160 t^2+432
   t u^2+392 t u+64 t+108 u^3 \cr&\qquad\quad +39 u^2+239 u+22 \bigg) \,, \cr
    c &= \frac{1}{12} \bigg( 18 N^2 p+18 N^2 q+36 N^2 s+72 N^2 t+54 N^2 u+36 N^2+21 N p^2+42 N p q \cr&\qquad\quad +84 N p s+168
   N p t+126 N p u+84 N p+3 N q^2+48 N q s+60 N q t+54 N q u \cr&\qquad\quad +30 N q+48 N s^2+192 N s t+144 N s u+96 N
   s+120 N t^2  +216 N t u \cr&\qquad\quad +120 N t+81 N u^2+36 N u+54 N+21 p^2 q+42 p^2 s+84 p^2 t+63 p^2 u+7 p^3 \cr&\qquad\quad +41
   p^2+3 p q^2+48 p q s+60 p q t+54 p q u+28 p q+48 p s^2+192 p s t+144 p s u \\&\qquad\quad +92 p s+120 p t^2+216 p
   t u+112 p t+81 p u^2+30 p u+45 p + \frac{1}{2}q^2 +3 q^2 s+3 q^2 t \cr&\qquad\quad +3 q^2 u+24 q s^2+60 q s t+54 q s u+26 q s+30 q
   t^2+60 q t u+22 q t+27 q u^2 \cr&\qquad\quad +28 q u+ \frac{13}{2}q+96 s^2 t+72 s^2 u+16 s^3+44 s^2+120 s t^2+216 s t u+104 s
   t \cr&\qquad\quad +81 s u^2+24 s u+38 s+120 t^2 u+40 t^3+44 t^2+108 t u^2+112 t u+30 t \cr&\qquad\quad +27 u^3+15 u^2+73 u+11 \bigg) \nonumber \,.
\end{align}
\endgroup
With generic values for the number of M5-branes and the $E_8$-homomorphism parameters, we find that the flavor symmetries of the resulting 4d $\mathcal{N}=2$ SCFTs are as follows:
\begin{equation}
    \begin{aligned}
        q > 0; t,u,s,p \geq 0; N \geq 1 \,&:\quad
        (\mathfrak{su}_1)_{12(N+p+s+u)+2q+8t+4} \oplus (\mathfrak{su}_1)_{12(N+p+s)+2q+8t+6u+4} \\ &\qquad
        \oplus (\mathfrak{su}_1)_{12(N+p)+2q+8t+6u+4s+4} \\ &\qquad \oplus
        (\mathfrak{su}_1)_{12N+2q+8t+6u+4s+2p+4}  \\&\qquad \oplus (\mathfrak{su}_{q+4t+3u+2s+p+2})_{12+2q+8t+6u+4s+2p+4}
         \,.
    \end{aligned}
\end{equation}
Again, we use a compact notation where $\mathfrak{su}_a \oplus \mathfrak{su}_b$ enhances to $\mathfrak{su}_{a+b}$ if the flavor central charges are identical.
When $q = 0$ there is an additional $\mathfrak{su}_2$ flavor symmetry in the 6d SCFT as the anti-symmetric hypermultiplet attached to the $\mathfrak{su}_4$ gauge algebra on the $1$-node is pseudo-real. The $\mathbb{Z}_2$ center-flavor symmetry does not embed inside of this $\mathfrak{su}_2$, and thus this flavor factor survives the Stiefel--Whitney twist intact. We find that the non-Abelian flavor symmetry of the SW-fold in the $q=0$ case is
\begin{equation}
    \begin{aligned}
        t,u,s,p \geq 0; N \geq 1 \,&:\quad
        (\mathfrak{su}_2)_{6(N+p+s+u+t)+3} \oplus
        (\mathfrak{su}_1)_{12(N+p+s+u)+8t+4} \\ &\qquad \oplus (\mathfrak{su}_1)_{12(N+p+s)+8t+6u+4}
        \oplus (\mathfrak{su}_1)_{12(N+p)+8t+6u+4s+4} \\ &\qquad \oplus
        (\mathfrak{su}_1)_{12N+8t+6u+4s+2p+4}  \\&\qquad \oplus (\mathfrak{su}_{4t+3u+2s+p+2})_{12+8t+6u+4s+2p+4}
         \,.
    \end{aligned}
\end{equation}
As discussed around equation \eqref{eqn:baryenc}, there can be further enhancement when $p = s = u = t = 0$, as in those cases it is expected that the baryonic $\mathfrak{u}(1)$ global symmetry enhances to an $\mathfrak{su}_2$, and this factor can further combine with the other non-Abelian factors for low values of $N$. We discuss some of these baryonic enhancements further in Section \ref{sec:classs}.

There are also two classes of $\mathbb{Z}_2$ SW-folds where the 6d SCFT origin itself has a baryonic $\mathfrak{su}_2$ flavor symmetry. These 6d theories correspond to the tensor branch configurations
\begin{equation}
    \overset{\mathfrak{sp}_1}{1}\overset{\mathfrak{su}_2}{2}\cdots\overset{\mathfrak{su}_2}{2} \,, \quad \text{ and } \quad 1\overset{\mathfrak{su}_2}{2}\cdots\overset{\mathfrak{su}_2}{2} \,.
\end{equation}
The $\mathbb{Z}_2$ center-flavor symmetry embeds trivially inside of the center of the baryonic $\mathfrak{su}_2$, and thus this additional non-Abelian factor is unbroken by the Stiefel--Whitney twist. In addition to the above flavor symmetries, these SW-folds then have the following additional flavor algebras:
\begin{equation}
    \begin{aligned}
        \mathcal{T}_2^{(N>1)}(0,0,0,0,1) &\,: \quad (\mathfrak{su}_2)_{6N^2 + N} \,, \\
        \mathcal{T}_2^{(N>1)}(1,0,0,0,0) &\,: \quad (\mathfrak{su}_2)_{6N^2 + 7N + 1} \,.
    \end{aligned}
\end{equation}
In the latter case, there is a further exceptional enhancement when $N = 1$.\footnote{For the former case, the flavor symmetry when $N=1$ is captured by equation \eqref{eqn:swiper}.} Let us discuss this special case of $\mathcal{T}_2^{(1)}(1,0,0,0,0)$, which arises from the 6d SCFT with tensor branch configuration
\begin{equation}
    1\overset{\mathfrak{su}_2}{2} \,.
\end{equation}
This 6d SCFT has an $\mathfrak{e}_7 \oplus \mathfrak{so}_7$ flavor symmetry, instead of the naively expected $\mathfrak{e}_7 \oplus \mathfrak{so}_8$ flavor algebra. Taking into account the embedding of the $\mathbb{Z}_2$ center-flavor symmetry inside of the $\mathfrak{so}_7$, we find that the flavor carried by the resulting SW-fold theory is
\begin{equation}
    (\mathfrak{f}_4)_{12} \oplus (\mathfrak{su}_{2})_{7} \oplus (\mathfrak{su}_2)_{7} \,.
\end{equation}
The SW-folds $\mathcal{T}_2^{(N)}(1,0,0,0,0)$ and $\mathcal{T}_2^{(N)}(0,0,0,0,1)$ have been discussed previously in \cite{Giacomelli:2020gee} where they were referred to as $\mathcal{T}_{E_6,2}^{(r+1)}$ and $\mathcal{S}_{E_6,2}^{(r)}$, respectively.

\subsection{Coulomb Branch Operator Spectrum}\label{sec:cb}

In \cite{Ohmori:2018ona}, the authors developed a heuristic method for determining the scaling dimensions of the Coulomb branch operators of the 4d SCFT obtained from the Stiefel--Whitney twisted compactification of a very Higgsable 6d SCFT. In this section, we test the consistency of the method described therein when applied to the SW-folds. In \cite{Ohmori:2018ona}, the authors verify their method by testing that the Coulomb branch spectrum they obtain agrees with the known spectrum from the dual twisted-class $\mathcal{S}$ theory; in our cases no such class $\mathcal{S}$ theory is known, and we rely on the weaker test that \begin{equation}
    4(2a - c) = \sum_i 2\Delta(u_i) - 1 \,,
\end{equation}
where the LHS is determined using the anomaly polynomial as in Section \ref{sec:ccs}. From the anomaly polynomial we expect that this quantity is:
\begin{equation}\label{eqn:acANOM}
    4(2a - c) = n_V - 64A - \frac{48}{\ell}B \,.
\end{equation}
We find that the method in \cite{Ohmori:2018ona} is consistent with this formula in most cases, but that it must be extended in a few special cases beyond the ones considered in \cite{Ohmori:2018ona}. These special cases occur for the E-type SW-folds that we discuss in Section \ref{sec:Esfolds}. We hope that this analysis will be useful in the determination of a closed-form top-down method for understanding the Coulomb branch spectrum from the 6d origin.

We are interested in 6d SCFTs with tensor branch configurations of the form given in equation \eqref{eqn:quivquiv}. We recall that in equation \eqref{eqn:quivquiv} we index the quiver nodes from $0$ to $r$ going from left-to-right. The algorithm presented in Appendix B of \cite{Ohmori:2018ona} for the Coulomb branch operator dimensions of the 4d $\mathcal{N}=2$ theory obtained from the $\mathbb{Z}_\ell$ Stiefel--Whitney twist of 6d SCFTs of the form in equation \eqref{eqn:quivquiv} is as follows. From each node of the form $\overset{\mathfrak{su}_{\ell k_i}}{2}$ the spectrum of operator dimensions is
\begin{equation}
    \Delta_i = \{ 6(r+1-i) \} \cup \{ 6(r+1-i) + d \,|\, d = 2, \cdots, k_i \} \,,
\end{equation}
where $i$ is the index of that quiver node. We can directly work out the contribution to $4(2a - c)$ from each of these $2$-nodes:
\begin{equation}
    4(2a - c)_i = \sum_{u \in \Delta_i} (2u - 1) = k_i^2 + 12k_i(r+1-i) - 2 \,.
\end{equation}
Summing over all contributions to $4(2a - c)$, we find
\begin{equation}\label{eqn:acCB}
    4(2a - c) = 4(2a - c)_0 + \left(1 - n_V^0 - \frac{12(r+1)d_0}{\ell}\right) + n_V  - 64A - \frac{48B}{\ell} \,,
\end{equation}
where we have used the expressions for $n_V$, $A$, and $B$ in equations \eqref{eqn:nv} and \eqref{eqn:B}. It remains for us to determine the contribution from the 1-nodes in equation \eqref{eqn:quivquiv}; there are four distinct cases which we must consider.
For each of the following $\mathfrak{g}$, the Coulomb branch operator dimensions coming from the 1-node with gauge algebra $\mathfrak{g}$ are proposed to be
\begin{equation}
    \begin{aligned}
      \mathfrak{g} = \varnothing &: \quad \left\{ \frac{6(r+1)}{\ell} \right\} \,, \\
      \mathfrak{g} = \mathfrak{su}_{\ell k_0} &: \quad \{ 6(r+1) \} \cup \{ 6(r+1) + d \,|\, d = 2, \cdots, k_0 \}  \,, \\
      \mathfrak{g} = \mathfrak{sp}_{2m+1} &: \quad \{ 6(r+1) \} \cup \{ 6(r+1) + 2d \,|\, d = 1, \cdots, m \}
      \,, \\
      \mathfrak{g} = \mathfrak{sp}_{2m} &: \quad \{ 6(r+1) \}
      \cup \{ 6(r+1) + 2d \,|\, d = 1, \cdots, m - 1 \}
      \cup \left\{ 3(r+1) + m \right\}
      \,,
    \end{aligned}
\end{equation}
where we remind the reader that the latter two options can only occur when $\ell = 2$. In each of the four cases, we can see that
\begin{equation}
    4(2a - c)_0 =  \frac{12(r+1)d_0}{\ell} + n_V^0 -1 \,,
\end{equation}
and thus we see that equation \eqref{eqn:acCB} determining $4(2a - c)$ from the putative Coulomb branch spectrum proposed in \cite{Ohmori:2018ona} matches the value of $4(2a - c)$ determined in equation \eqref{eqn:acANOM} from the 6d anomaly polynomial and tensor branch configuration.

\subsection{Alternative Constructions}

In some cases, the theories we can generate via SW-folds have alternative constructions. In this section we compare with class $\mathcal{S}$ constructions, as well as methods based 4d $\mathcal{N} = 2$ S-folds.

\subsubsection[Exceptional Twisted Class \texorpdfstring{$\mathcal{S}$}{S} and Flavor Symmetry]{Exceptional Twisted Class \boldmath{$\mathcal{S}$} and Flavor Symmetry}\label{sec:classs}

In contrast to constructing 4d $\mathcal{N}=2$ SCFTs via compactification of 6d $(1,0)$ SCFTs on a $T^2$, one can also explore the class $\mathcal{S}$  construction \cite{Gaiotto:2009we,Gaiotto:2009hg}. This class of theories is obtained via the twisted-compactification of the 6d $(2,0)$ SCFT of type $\mathfrak{g}$ on a punctured Riemann surface.

It has been established in \cite{Ohmori:2015pua,Ohmori:2015pia} that the 4d $\mathcal{N}=2$ SCFT that arises from compactifying minimal $(\mathfrak{e}_6, \mathfrak{e}_6)$ conformal matter on a $T^2$ with $\mathbb{Z}_3$ Stiefel--Whitney twist is dual to a class $\mathcal{S}$ theory. The latter is obtained from compactification of the 6d $(2,0)$ SCFT of type $\mathfrak{so}_8$ on a sphere with two maximal $\mathbb{Z}_3$-twisted punctures and one simple puncture. Similarly, minimal $(\mathfrak{e}_7, \mathfrak{e}_7)$ conformal matter on a $T^2$ with $\mathbb{Z}_2$ Stiefel--Whitney twist is dual to the 6d $(2,0)$ SCFT of type $\mathfrak{e}_6$ on a sphere with two maximal $\mathbb{Z}_2$-twisted punctures and one simple puncture. In rare limiting cases, some of the 6d SCFTs written in Table \ref{tbl:6dscfts} can also be obtained by starting from minimal $(\mathfrak{e}_{n}, \mathfrak{e}_{n})$ conformal matter, with $n = 6, 7$, and performing nilpotent Higgs branch deformations of the $\mathfrak{g} \oplus \mathfrak{g}$ flavor symmetry. Compactifying these theories with Stiefel--Whitney twist then gives rise to 4d $\mathcal{N}=2$ SCFTs that can also be obtained by partial closure of the maximal punctures in the aforementioned class $\mathcal{S}$ construction.\footnote{See \cite{Baume:2021qho} for an in depth analysis of the relationship between the nilpotent Higgs branch deformations and the partial closure of the punctures in the untwisted case.}

Class $\mathcal{S}$ theories of type $\mathfrak{so}_8$ with $\mathbb{Z}_3$-twisted punctures have been studied in \cite{Chacaltana:2016shw}. Similarly, class $\mathcal{S}$ of type $\mathfrak{e}_6$ with $\mathbb{Z}_2$-twisted punctures has been explored in \cite{Chacaltana:2015bna}.
We list the SW-fold SCFTs from Table \ref{tbl:genSfolds} that can be realized in class $\mathcal{S}$, as described, in Table \ref{tbl:cs}. In all cases, bar one, the flavor symmetry determined from the dual class $\mathcal{S}$ construction matches the flavor symmetry that was determined from the Stiefel--Whitney twisted description in Section \ref{sec:ccs}. There is one special case, the SW-fold theory $\mathcal{T}_2^{(1)}(0,1,0,0,0)$, for which the analysis in Section \ref{sec:ccs} predicts that the non-Abelian flavor symmetry should be
\begin{equation}
    (\mathfrak{so}_7)_{12} \oplus (\mathfrak{so}_7)_{16} \,,
\end{equation}
but the dual class $\mathcal{S}$ theory has non-Abelian flavor algebra
\begin{equation}
    (\mathfrak{so}_7)_{12} \oplus (\mathfrak{g}_2)_{16} \,.
\end{equation}
This kind of dehancement occurs in the context of 6d SCFTs when one has an $\mathfrak{su}_2$ gauge algebra associated to a tensor with self-pairing 2 \cite{Morrison:2016djb}. In this case, the Coulomb branch description of the 4d SW-fold SCFT has a single $\mathfrak{su}_2$ gauge algebra, and a parallel argument to that in 6d may explain why the flavor symmetry is smaller than expected. We would similarly suspect that the SW-fold SCFTs $\mathcal{T}_4^{(1)}(0,1)$ and $\mathcal{T}_3^{(1)}(0,1,0)$, whose 4d Coulomb branch descriptions also involve a single $\mathfrak{su}_2$ gauge algebra coming from a 2-node decorated algebra in 6d, to evince similar dehancement. It would be interesting to understand the physical mechanism behind this rare but curious effect.

\begin{table}[ht]
    \centering
    \begin{threeparttable}
    \begin{tabular}{ccccc}
    \toprule
        SW-fold & SW Twist
        & Class $\mathcal{S}$ Type & Punctures & Flavor
        \\\midrule
        $\mathcal{T}_2^{(1)}(2,0,0,0,0)$ & \multirow{9}{*}{$\mathbb{Z}_2$} & \multirow{9}{*}{$\mathfrak{e}_6$}
        & $[0, A_2]_I$ & $(\mathfrak{f}_4)_{18} \oplus (\mathfrak{su}_3)_{16}$ 
        \\
        $\mathcal{T}_2^{(2)}(1,0,0,0,0)$ & &
        & $[0, A_2 + \widetilde{A}_1]_I$ & $(\mathfrak{f}_4)_{18} \oplus (\mathfrak{su}_2)_{39}$  
        \\
        $\mathcal{T}_2^{(1)}(1,0,0,0,1)$ & &
        & $[A_2, A_1]_I$ &  $(\mathfrak{sp}_3)_{13} \oplus (\mathfrak{su}_3)_{16}$ 
        \\
        $\mathcal{T}_2^{(2)}(0,0,0,0,1)$ & &
        & $[A_2 + \widetilde{A}_1, A_1]_I$ & $(\mathfrak{sp}_4)_{13} \oplus (\mathfrak{su}_2)_{26}$ 
        \\
        $\mathcal{T}_2^{(1)}(0,1,0,0,0)$ &  &
        & $[A_2, \widetilde{A}_1]_I$ & $(\mathfrak{so}_7)_{12} \oplus (\mathfrak{g}_2)_{16}$ 
        \\
        $\mathcal{T}_2^{(1)}(1,0,0,0,0)$ & &
        & $[A_2 + \widetilde{A}_1, \widetilde{A}_1]_M$ & $(\mathfrak{f}_4)_{12} \oplus 2(\mathfrak{su}_2)_7$ 
        \\
        $\mathcal{S}_2^{(1)}(0,0,0,0,0)$ & &
        & $[A_2, A_1 + \widetilde{A}_1]_M$ & $(\mathfrak{su}_6)_{16} \oplus (\mathfrak{su}_2)_9$ 
        \\
        $\mathcal{T}_2^{(1)}(0,0,0,0,1)$ & &
        & $[A_2 + \widetilde{A}_1, A_1 + \widetilde{A}_1]_M$ & $(\mathfrak{sp}_5)_7$ 
        \\\midrule
        $\mathcal{T}_3^{(1)}(1,0,0)$  & \multirow{2}{*}{$\mathbb{Z}_3$} & \multirow{2}{*}{$\mathfrak{so}_8$} & $[0, A_1]_I$ & $(\mathfrak{g}_2)_{8} \oplus (\mathfrak{su}_2)_{14}$ 
        \\
        $\mathcal{S}_3^{(1)}(0,0,1)$  & & & $[A_1, A_1]_I$ & $(\mathfrak{su}_4)_{14}$ 
        \\\bottomrule
    \end{tabular}
    \end{threeparttable}
    \caption{Twisted punctures are usually denoted with an \uline{underline}, however, since all of the punctures that we write in this table are twisted, we have chosen to drop this notational feature. The subscripts $I$ (interacting) and $M$ (mixed) denote whether the class $\mathcal{S}$ theory is an interacting SCFT, or whether it is coupled to free hypermultiplets, respectively. In the latter case, the SW-fold SCFT matches the interacting part of the class $\mathcal{S}$ theory. In the flavor column we write the non-Abelian flavor algebra as determined from the class $\mathcal{S}$ perspective.}
    \label{tbl:cs}
\end{table}

\subsubsection{Relation to 4d $\mathcal{N} = 2$ S-fold Theories}\label{sec:seqsw}

In the previous sections we studied the properties of 4d $\mathcal{N} = 2$ SW-folds, and we also observed that in some cases, the resulting theories can be realized via 4d $\mathcal{N} = 2$ S-fold theories. In this section we discuss some suggestive hints that such a top-down correspondence may be at work, but leave a more complete treatment for future work.

To frame the discussion to follow, recall that an S-fold in Type IIB / F-theory backgrounds
is a non-perturbative generalization of an orientifold plane in which a quotient on the target space is combined with a group action from the
$SL(2,\mathbb{Z})$ duality group of Type IIB string theory. Now, for such a quotient to exist we must work at specific values of the axio-dilaton compatible with this group action, e.g. $\tau = i$ and $\tau = \exp(2 \pi i / 6)$. In the presence of a probe D3-brane, this can be used to realize $\mathcal{N} = 3$ SCFTs, as noted in \cite{Garcia-Etxebarria:2015wns} (see also \cite{Aharony:2016kai}).
One can also introduce 7-branes provided they are compatible with a specific value of $\tau$, and this leads to 4d $\mathcal{N} = 2$ S-folds. D3-brane probes of such systems then realize 4d $\mathcal{N} = 2$ SCFTs \cite{Apruzzi:2020pmv,Giacomelli:2020jel,Heckman:2020svr,Giacomelli:2020gee,Bourget:2020mez}. As a general comment, the global symmetry in these systems also depends on the presence (or absence) of a torsional flux, and this effect can be detected via open string junctions which extend from the D3-brane to the 7-brane flavor stack \cite{Heckman:2020svr}.

As we now explain, there are reasons to suspect that the 4d SW-fold theories considered in this paper, and 4d S-fold theories are potentially related by a chain of dualities. To see why, it is helpful to first consider some of the different top-down realizations of the rank $N$ $E_8$ Minahan--Nemeschansky theory \cite{Minahan:1996fg, Minahan:1996cj}. One way is to first start with the rank $N$ E-string theory 6d SCFT. Compactification on a $T^2$ then yields the 4d $\mathcal{N} = 2$ SCFT. Observe that in M-theory, this is engineered from the $T^2$ compactification of $N$ M5-branes probing an $E_8$ nine-brane in M-theory. On the other hand, we can also directly relate this to Type IIB/F-theory backgrounds with $N$ D3-branes probing an $E_8$ seven-brane. Intuitively, there is a generalized notion of T-duality at play which allows us trade the $E_8$ nine-brane of M-theory for the $E_8$ seven-brane of F-theory.\footnote{Indeed, this figures prominently in the standard Fourier--Mukai transformation of heterotic vector bundles on an elliptically fibered Calabi--Yau threefold and their characterization in the associated spectral cover construction for gauge theory on the base K\"ahler surface (see, e.g., \cite{Donagi:2000fw}).}

There is a natural extension of this generalized T-duality which makes any proposed correspondence quite suggestive. On the M-theory side, our 6d SCFT orbi-instanton theories were realized by small instantons probing an ADE singularity wrapped by an $E_8$ nine-brane. Likewise, we note that D3-branes probing an ADE singularity wrapped by an $E_8$ seven-brane will give rise to 4d $\mathcal{N} = 2$ SCFTs. In both cases, the worldvolume theory of the probe brane is specified as an instanton solution in the directions filled by the ambient brane. As such, we can generate a wide class of examples by specifying the boundary data of a flat connection at the boundary $S^3 / \Gamma_\text{ADE}$, which are in turn captured by discrete group homomorphisms $\text{Hom}(\Gamma_\text{ADE} \rightarrow E_8)$ \cite{DelZotto:2014hpa,Heckman:2015bfa}. So, from this perspective, we see that the $T^2$ compactification of the 6d orbi-instanton theories provides us with a direct way to match the two sets of theories.

So far, our discussion has made no reference to switching on an SW-fold on the orbi-instanton side of this correspondence. Now, on the SW-fold side we consider a pair of holonomies which commute in $\widetilde{G}$ up to a flux valued in the quotienting subgroup. These profiles make direct reference to the $T^2$ direction on which we have compactified the orbi-instanton theory. To make sense of such deformations in the D3-brane probe theories, we would need to have a notion of generalized T-duality which extends to such configurations as well. The fact that there are known examples where SW-folds and S-folds produce the same conformal fixed point is of course suggestive \cite{Giacomelli:2020jel}, but without
a suitable generalization of T-duality, it is unclear whether it should be expected to persist for all SW-folds and S-folds, or just some subset. Exploring this issue further would be of great interest and would likely lead to a better understanding of both sorts of constructions.

\section{SW-folds of Type DE}\label{sec:Esfolds}

In Section \ref{sec:e8CF}, we have enumerated the 6d SCFTs that have non-trivial center-flavor symmetry and arise from a Higgsing, by homomorphisms $\mathbb{Z}_K \rightarrow E_8$, of the rank $N$ orbi-instanton theory of type $(\mathfrak{e}_8, \mathfrak{su}_K)$. In Section \ref{sec:sfolds}, we considered the compactification of the 6d SCFTs found in Section \ref{sec:e8CF} on a $T^2$ together with a Stiefel--Whitney twist in the center-flavor symmetry. We refer to the resulting 4d $\mathcal{N}=2$ SCFTs as the A-type SW-folds, due to the $\mathfrak{su}_K$ factor in the orbi-instanton origin. In this section we consider the rank $N$ orbi-instanton theories of type $(\mathfrak{e}_8, \mathfrak{g})$, where $\mathfrak{g}$ is any ADE Lie algebra. We consider homomorphisms $\Gamma \rightarrow E_8$, where $\Gamma$ is the finite ADE group of the same type as $\mathfrak{g}$, and such that the Higgsed 6d SCFT has a non-trivial center-flavor symmetry. For generic values of $N$, this center-flavor symmetry can be, at most
\begin{equation}
    \begin{aligned}
      \mathbb{Z}_4 \qquad &\text{ for } \qquad \mathfrak{g} = \mathfrak{so}_{4k+2} \,, \\
      \mathbb{Z}_2 \times \mathbb{Z}_2 \qquad &\text{ for } \qquad  \mathfrak{g} = \mathfrak{so}_{4k} \,, \\
      \mathbb{Z}_3 \qquad &\text{ for } \qquad  \mathfrak{g} = \mathfrak{e}_6 \,, \\
      \mathbb{Z}_2 \qquad &\text{ for } \qquad \mathfrak{g} = \mathfrak{e}_7 \,.
    \end{aligned}
\end{equation}
We can now consider the $T^2$ compactifications of these 6d SCFTs with a non-trivial Stiefel--Whitney class inside of the center-flavor symmetry turned on. This opens up a vast new vista of D-type and E-type SW-folds. We will not consider all such families of SW-folds here, but we highlight a few choice examples; the remaining cases can be determined straightforwardly from the methods utilized throughout this paper.

\subsection[\texorpdfstring{$E_6$}{E6}-type SW-folds]{\boldmath{$E_6$}-type SW-folds}\label{sec:e6sfolds}

We begin with the $(\mathfrak{e}_8, \mathfrak{e}_6)$ orbi-instanton, of rank $N$. The tensor branch configuration has the form
\begin{equation}
    1 \, 2 \overset{\mathfrak{su}_2}{2} \overset{\mathfrak{g}_2}{3} 1 \overset{\mathfrak{f}_4}{5} 1 \overset{\mathfrak{su}_3}{3} 1 \overset{\mathfrak{e}_6}{6}
      1 \overset{\mathfrak{su}_3}{3} 1
      \overbrace{\overset{\mathfrak{e}_6}{6} 1 \overset{\mathfrak{su}_3}{3} 1 \cdots \overset{\mathfrak{e}_6}{6} 1 \overset{\mathfrak{su}_3}{3} 1}^{N-1} \,,
\end{equation}
which has an $\mathfrak{e}_8 \oplus \mathfrak{e}_6$ flavor symmetry, and no center-flavor symmetry. We will consider the 6d SCFTs obtained by the finite group homomorphism:
\begin{equation}
    \Gamma_{\mathfrak{e}_6} \rightarrow E_8 \,,
\end{equation}
with $\Gamma_{E_6}$ the binary tetrahderal finite subgroup of $SU(2)$. The Higgs branch flows induced by such homomorphisms have been studied in \cite{Frey:2018vpw}. There are fifty-two such SCFTs, however we are only interested in those that have a non-trivial center-flavor symmetry. There are only seven such Higgsings which give rise to a center-flavor symmetry, which is always a $\mathbb{Z}_3$.\footnote{Much as in Section \ref{sec:e8CF}, we emphasize that if $N = 1$ then many more of the $E_8$-homomorphisms lead to theories with center-flavor symmetry, and it is not restricted to be $\mathbb{Z}_3$.} These correspond to the seven tensor branch geometries
\begingroup
\allowdisplaybreaks
\begin{align}
        \overset{\mathfrak{su}_3}{3} 1 \overset{\mathfrak{su}_3}{3} 1 \overset{\mathfrak{e}_6}{6}
      1 \overset{\mathfrak{su}_3}{3} 1
      \underbrace{\overset{\mathfrak{e}_6}{6} 1 \overset{\mathfrak{su}_3}{3} 1 \cdots \overset{\mathfrak{e}_6}{6} 1 \overset{\mathfrak{su}_3}{3} 1}_{N-1} &\,, \label{eqn:e61} \\
      \overset{\mathfrak{su}_3}{2} \overset{\mathfrak{su}_3}{2} 1 \overset{\mathfrak{e}_6}{6}
      1 \overset{\mathfrak{su}_3}{3} 1
      \underbrace{\overset{\mathfrak{e}_6}{6} 1 \overset{\mathfrak{su}_3}{3} 1 \cdots \overset{\mathfrak{e}_6}{6} 1 \overset{\mathfrak{su}_3}{3} 1}_{N-1} &\,, \\
      1\overset{\mathfrak{su}_3}{3} 1 \underset{1}{\overset{\mathfrak{e}_6}{6}}
      1 \overset{\mathfrak{su}_3}{3} 1
      \underbrace{\overset{\mathfrak{e}_6}{6} 1 \overset{\mathfrak{su}_3}{3} 1 \cdots \overset{\mathfrak{e}_6}{6} 1 \overset{\mathfrak{su}_3}{3} 1}_{N-1} &\,, \\
      \overset{\mathfrak{su}_3}{2} 1 \underset{1}{\overset{\mathfrak{e}_6}{6}}
      1 \overset{\mathfrak{su}_3}{3} 1
      \underbrace{\overset{\mathfrak{e}_6}{6} 1 \overset{\mathfrak{su}_3}{3} 1 \cdots \overset{\mathfrak{e}_6}{6} 1 \overset{\mathfrak{su}_3}{3} 1}_{N-1} &\,, \\
      1 \overset{1}{\underset{1}{\overset{\mathfrak{e}_6}{6}}}
      1 \overset{\mathfrak{su}_3}{3} 1
      \underbrace{\overset{\mathfrak{e}_6}{6} 1 \overset{\mathfrak{su}_3}{3} 1 \cdots \overset{\mathfrak{e}_6}{6} 1 \overset{\mathfrak{su}_3}{3} 1}_{N-1} &\,, \\
      \overset{\mathfrak{e}_6}{3}
      1 \overset{\mathfrak{su}_3}{3} 1
      \underbrace{\overset{\mathfrak{e}_6}{6} 1 \overset{\mathfrak{su}_3}{3} 1 \cdots \overset{\mathfrak{e}_6}{6} 1 \overset{\mathfrak{su}_3}{3} 1}_{N-1} &\,, \\
      \overset{\mathfrak{su}_6}{2}
      \overset{\mathfrak{su}_3}{2} 1
      \underbrace{\overset{\mathfrak{e}_6}{6} 1 \overset{\mathfrak{su}_3}{3} 1 \cdots \overset{\mathfrak{e}_6}{6} 1 \overset{\mathfrak{su}_3}{3} 1}_{N-1} &\,. \label{eqn:e67}
\end{align}
\endgroup
To work out the central charges of the 4d $\mathcal{N}=2$ SW-folds obtained from the Stiefel--Whitney twisted compactification of these 6d SCFTs we will again use the formulae of \cite{Ohmori:2018ona}, which we have summarized in equations \eqref{eqn:fields} and \eqref{eqn:ack}.

We consider the compactification of the tensor branch configuration in equation \eqref{eqn:e61} in detail. The 6d SCFT has an $(SU(3)^2 \times E_6)/\mathbb{Z}_3$ flavor symmetry group, and after the Stiefel--Whitney twist there remains only a $G_2$ subgroup of the $E_6$. For the 6d SCFT from which this SW-fold originates, the relevant terms in the anomaly polynomial are
\begin{equation}
    I_8 \supset \frac{1}{24}\left(\left(\frac{7N}{8} + \frac{35}{12} \right)p_1(T)^2 - \left(72N^2 + 209N + 102\right)c_2(R)p_1(T)\right) + \frac{3}{16}p_1(T)\operatorname{Tr}F^2 \,,
\end{equation}
where we have only written the mixed-gravitational-flavor anomaly for the $\mathfrak{e}_6$ flavor algebra. Next, we find that the contribution from the weakly-coupled multiplets is
\begin{equation}
    I_8^\text{fields} \supset \left(-\frac{17N}{192} - \frac{1}{288} \right)p_1(T)^2 + \left(-\frac{41N}{24} - \frac{1}{4} \right)c_2(R)p_1(T) \,.
\end{equation}
Putting this altogether we find that
\begin{equation}
    a - a_\text{generic} = 12N^2 + 27N + 15 \,, \quad c - c_\text{generic} = 12N^2 + 28N + 16 \,, \quad \kappa - \kappa_\text{generic} = 12 \,,
\end{equation}
where we have used that the Dynkin index of the $\mathfrak{g}_2$ inside of the $\mathfrak{e}_6$ is one, as explained around equation \eqref{eqn:e6ss}.
It remains for us to determine what the contributions to the central charges are from the 4d theory at the generic point of the Coulomb branch. The $\mathbb{Z}_3$ Stiefel--Whitney twist breaks the $\mathfrak{su}_3$ gauge algebras completely, and it breaks each $\mathfrak{e}_6$ down to a $\mathfrak{g}_2$. As such, at the generic point of the Coulomb branch, we have $(4(N-1)+8)$ vector multiplets from the 6d tensors, and $N\operatorname{dim}(\mathfrak{g}_2)$ vector multiplets from the surviving $\mathfrak{g}_2$ gauge symmetries. We end up with
\begin{equation}
    a_\text{generic} = \frac{5}{24}(18N + 4) \,, \qquad c_\text{generic} = \frac{1}{6}(18N + 4) \,.
\end{equation}
Furthermore, since there are no hypermultiplets charged under the residual $\mathfrak{g}_2$ flavor symmetry we find that
\begin{equation}
    \kappa_\text{generic} = 0 \,.
\end{equation}
The central charges of the SW-fold are thus:
\begin{equation}\label{eqn:acE6}
    a = 12N^2 + \frac{123N}{4} + \frac{95}{6} \,, \qquad c = 12N^2 + 31N + \frac{50}{3} \,,
\end{equation}
and the flavor symmetry and flavor central charge is
\begin{equation}
    (\mathfrak{g}_2)_{12} \,.
\end{equation}
The central charges from each of the 6d SCFTs with tensor branch descriptions given in equations \eqref{eqn:e61} to \eqref{eqn:e67} can be determined, and we do not belabor the computation here. The central charges, flavor symmetries, and flavor central charges for the seven families of $E_6$-type $\mathbb{Z}_3$ SW-folds are given in Table \ref{tbl:e6Sfolds}.

\begin{table}[ht]
    \centering
    \small
    \begin{threeparttable}
    \begin{tabular}{cccc}
    \toprule
         6d Origin & $a$ & $c$ & Flavor \\\midrule
         $\overset{\mathfrak{su}_3}{3} 1 \overset{\mathfrak{su}_3}{3} 1 \overset{\mathfrak{e}_6}{6}
      1 \overset{\mathfrak{su}_3}{3} 1
      \underbrace{\overset{\mathfrak{e}_6}{6} 1 \overset{\mathfrak{su}_3}{3} 1 \cdots \overset{\mathfrak{e}_6}{6} 1 \overset{\mathfrak{su}_3}{3} 1}_{N-1}$ & $12N^2 + \frac{123N}{4} + \frac{95}{6}$ & $12N^2 + 31N + \frac{50}{3}$ & $(\mathfrak{g}_2)_{12}$ \\

      $\overset{\mathfrak{su}_3}{2} \overset{\mathfrak{su}_3}{2} 1 \overset{\mathfrak{e}_6}{6}
      1 \overset{\mathfrak{su}_3}{3} 1
      \underbrace{\overset{\mathfrak{e}_6}{6} 1 \overset{\mathfrak{su}_3}{3} 1 \cdots \overset{\mathfrak{e}_6}{6} 1 \overset{\mathfrak{su}_3}{3} 1}_{N-1}$ & $12N^2 + \frac{111N}{4} + \frac{97}{8}$ & $12N^2 + 28N + 13$ & $(\mathfrak{g}_2)_{12}$ \\

      $1\overset{\mathfrak{su}_3}{3} 1 \underset{\displaystyle 1}{\overset{\mathfrak{e}_6}{6}}
      1 \overset{\mathfrak{su}_3}{3} 1
      \underbrace{\overset{\mathfrak{e}_6}{6} 1 \overset{\mathfrak{su}_3}{3} 1 \cdots \overset{\mathfrak{e}_6}{6} 1 \overset{\mathfrak{su}_3}{3} 1}_{N-1}$ & $12N^2 + \frac{91N}{4} + \frac{47}{6}$ & $12N^2 + 23N + \frac{26}{3}$ & $(\mathfrak{g}_2)_{12} \oplus (\mathfrak{g}_2)_{4N+8}$ \\

      $\overset{\mathfrak{su}_3}{2} 1 \underset{\displaystyle 1}{\overset{\mathfrak{e}_6}{6}}
      1 \overset{\mathfrak{su}_3}{3} 1
      \underbrace{\overset{\mathfrak{e}_6}{6} 1 \overset{\mathfrak{su}_3}{3} 1 \cdots \overset{\mathfrak{e}_6}{6} 1 \overset{\mathfrak{su}_3}{3} 1}_{N-1}$ & $12N^2 + \frac{87N}{4} + \frac{145}{24}$ & $12N^2 + 22N + \frac{41}{6}$ & $(\mathfrak{g}_2)_{12} \oplus (\mathfrak{su}_2)_{12N+20}$  \\

      $1 \overset{\displaystyle 1}{\underset{\displaystyle 1}{\overset{\mathfrak{e}_6}{6}}}
      1 \overset{\mathfrak{su}_3}{3} 1
      \underbrace{\overset{\mathfrak{e}_6}{6} 1 \overset{\mathfrak{su}_3}{3} 1 \cdots \overset{\mathfrak{e}_6}{6} 1 \overset{\mathfrak{su}_3}{3} 1}_{N-1}$ & $12N^2 + \frac{75N}{4} + \frac{23}{8}$ & $12N^2 + 19N + \frac{7}{2}$ & $(\mathfrak{g}_2)_{12}$  \\

      $\overset{\mathfrak{e}_6}{3}
      1 \overset{\mathfrak{su}_3}{3} 1
      \underbrace{\overset{\mathfrak{e}_6}{6} 1 \overset{\mathfrak{su}_3}{3} 1 \cdots \overset{\mathfrak{e}_6}{6} 1 \overset{\mathfrak{su}_3}{3} 1}_{N-1}$ & $12N^2 + \frac{63N}{4} + \frac{7}{12}$ & $12N^2 + 16N + \frac{7}{6}$ & $(\mathfrak{g}_2)_{12}$  \\

      $\overset{\mathfrak{su}_6}{2}
      \overset{\mathfrak{su}_3}{2} 1
      \underbrace{\overset{\mathfrak{e}_6}{6} 1 \overset{\mathfrak{su}_3}{3} 1 \cdots \overset{\mathfrak{e}_6}{6} 1 \overset{\mathfrak{su}_3}{3} 1}_{N-1}$ & $12N^2 + \frac{27N}{4} - \frac{31}{12}$ & $12N^2 + 7N - \frac{5}{3}$ & $(\mathfrak{g}_2)_{12} \oplus (\mathfrak{su}_3)_{12N + 16}$ \\\bottomrule
    \end{tabular}
    \end{threeparttable}
    \caption{In this table, we write the central charges, non-Abelian flavor algebras, and flavor central charges of the $E_6$-type SW-folds.}
    \label{tbl:e6Sfolds}
\end{table}

\subsection{Coulomb Branch Scaling Dimensions}\label{sec:cbe6}

In Section \ref{sec:cb}, we determined the conformal dimensions of the spectrum of Coulomb branch operators of the 4d $\mathcal{N}=2$ SCFTs arising from the Stiefel--Whitney twisted torus compactifications of the (Higgsed) rank $N$ $(\mathfrak{e}_8, \mathfrak{su}_K)$ orbi-instanton theories. In such cases, the Coulomb branch spectrum was determined by following the heuristic proposal in Appendix B of \cite{Ohmori:2018ona}; therein the scaling dimensions were determined in terms of each curve/algebra combination, $\overset{\mathfrak{g}}{m}$, in the tensor branch description, together with the knowledge of the residual gauge algebra after the $\mathbb{Z}_\ell$ Stiefel--Whitney twist. The contributions were proposed on a case-by-case basis for certain combinations of $(\mathfrak{g}, \ell)$, however, theories involving $(\mathfrak{e}_6, 3)$ and $(\mathfrak{e}_7, 2)$ were not explored in \cite{Ohmori:2018ona}.

When studying the E-type SW-folds, as we are doing here, it is necessary to extend the proposal of \cite{Ohmori:2018ona} to include the $(\mathfrak{e}_6, 3)$ and $(\mathfrak{e}_7, 2)$ cases. We make the following, again heuristic, proposal for the Coulomb branch scaling dimensions of the operators that arise in the $(\mathfrak{e}_6, 3)$ case\footnote{The quantity $J$ is a number associated to each curve in the tensor branch configuration which, roughly, counts where that curves lies in the order of blow-downs required to reach the origin of the tensor branch. This was referred to as $n$ in Appendix B of \cite{Ohmori:2018ona}, and we refer the reader there for the definition.}
\begin{equation}\label{eqn:6curveCOR}
    6J \quad 6J \times 1 + 2 \quad 6J \times 2 + 6 \,.
\end{equation}
Here, the multiplicative factors of $1$ and $2$ that we have introduced are the comarks of the residual $\mathfrak{g}_2$ gauge algebra; furthermore, the additive factors of $2$ and $6$ are the degrees of the Casimir invariants of $\mathfrak{g}_2$.\footnote{We note that, because we are only checking the matching of $4(2a - c)$, which is given by equation \eqref{eqn:2ac}, then $6J$, $6J \times 2 + 2$, and $6J \times 1 + 6$ also work equally well.} Similarly, when the 6d tensor branch contains a curve/algebra combination of the form $\overset{\mathfrak{e}_7}{8}$, then, after a $\mathbb{Z}_2$ SW-twist one obtains a residual $\mathfrak{f}_4$ gauge algebra on the Coulomb branch. We propose that the contribution from this curve/algebra combination to the 4d Coulomb branch consists of five operators with scaling dimensions:
\begin{equation}\label{eqn:f4comark}
    6J \quad 6J \times 1 + 2 \quad 6J \times 2 + 6 \quad 6J \times 3 + 8 \quad 6J \times 2 + 12
\end{equation}
Here the multiplicative factors $1$, $2$, $3$, and $2$ are the comarks, and $2$, $6$, $8$, and $12$ are the degrees of the Casimir invariants, of the surviving $\mathfrak{f}_4$ gauge algebra.\footnote{Again, we emphasize that the level of analysis here is insensitive to which comark is paired with which Casimir degree.} 

We first consider the 6d $(1,0)$ SCFT with tensor branch configuration as given in equation \eqref{eqn:e61}. We have determined that the central charges of the $\mathbb{Z}_3$ SW-twisted torus compactification satisfy
\begin{equation}\label{eqn:2acanom}
    4(2a - c) = 48N^2 + 122N + 60 \,.
\end{equation}
This quantity can also be recovered from the scaling dimensions of the Coulomb branch operators:
\begin{equation}\label{eqn:2ac}
    4(2a - c) = \sum_{i = 1}^r (2D(u_i) - 1) \,,
\end{equation}
where $r$ is the rank of the Coulomb branch and $u_i$ are the Coulomb branch operators. Combining the analysis in Appendix B of \cite{Ohmori:2018ona} with our proposal in equation \eqref{eqn:6curveCOR}, we conjecture that the Coulomb branch operators dimensions are
\begin{equation}\label{eqn:e6CB}
    \begin{aligned}
        &\left.6\right. \\
        &\left.12\right. \\
        &\left.8\right. \\
        &\left.\begin{aligned}
        &6J \quad 6J \times 1 + 2 \quad 6J \times 2 + 6 \\
        &6J + 6 \\
        &12J + 12 \\
        &6J + 8 \\
        \end{aligned}\,\, \right\} \quad J = 1, \cdots, N \\
        &\left.6(N+1)\right. \,.
    \end{aligned}
\end{equation}
In this way, we find that $4(2a-c)$ as worked out from the anomaly, as written in Table \ref{tbl:e6Sfolds}, matches with $4(2a - c)$ as worked out from the Coulomb branch spectrum using equation \eqref{eqn:2ac}. In fact, this matching occurs for all of the SW-twisted theories appearing in Table \ref{tbl:e6Sfolds}. Unfortunately, we do not know of any dual class $\mathcal{S}$ description of a 4d $\mathcal{N}=2$ SCFT obtained from a $\mathbb{Z}_3$ Stiefel--Whitney twist of the 6d theory containing such an $\mathfrak{e}_6$ algebra, and thus we do not have any independent verification of the proposal given in equation \eqref{eqn:6curveCOR}.

To further explore the association between the tensor branch configuration and the dimensions of the Coulomb branch operators of the Stiefel--Whitney twisted theory, we now study one example of an $E_7$-type SW-fold. The tensor branch configuration
\begin{equation}
    1 \overset{\mathfrak{su}_2}{2} \overset{\mathfrak{so}_7}{3} \overset{\mathfrak{su}_2}{2} 1 \underset{\displaystyle 1}{\overset{\mathfrak{e}_7}{8}} 1 \overset{\mathfrak{su}_2}{2} \overset{\mathfrak{so}_7}{3} \overset{\mathfrak{su}_2}{2} 1
      \underbrace{\overset{\mathfrak{e}_7}{8} 1 \overset{\mathfrak{su}_2}{2} \overset{\mathfrak{so}_7}{3} \overset{\mathfrak{su}_2}{2} 1  \cdots \overset{\mathfrak{e}_7}{8} 1 \overset{\mathfrak{su}_2}{2} \overset{\mathfrak{so}_7}{3} \overset{\mathfrak{su}_2}{2} 1}_{N-1} \,,
\end{equation}
has a $\mathbb{Z}_2$ center-flavor symmetry, and arises via Higgsing the $\mathfrak{e}_8$ flavor symmetry of the rank $N$ $(\mathfrak{e}_8, \mathfrak{e}_7)$ orbi-instanton by a homomorphism $\Gamma_{\mathfrak{e}_7} \rightarrow E_8$. The flavor group of this 6d SCFT is $(E_7 \times E_7 \times SU(2))/\mathbb{Z}_2$. Using the anomaly polynomial of the 6d SCFT associated to this tensor branch configuration, and the Coulomb branch theory in 4d after $\mathbb{Z}_2$ Stiefel--Whitney twist, one finds from equation \eqref{eqn:ack} that
\begin{equation}\label{eqn:e7sfoldanom}
    4(2a - c) = 144N^2 + 316N + 120 \,.
\end{equation}
We propose that the scaling dimensions of the Coulomb branch operators are
\begin{equation}
    \begin{aligned}
        &\left.9N + 9\right. \\
        &\left.12N + 12\right. \\
        &\left.6N + 6 \quad 6N + 8 \quad 3N + 5\right. \\
        &\left.6N + 12\right. \\
        &\left.3N + 9\right. \\
        &\left.\begin{aligned}
        &6J \quad 6J \times 1 + 2 \quad 6J \times 2 + 6 \quad 6J \times 3 + 8 \quad 6J \times 2 + 12 \\
        &12J + 6 \\
        &18J + 6 \\
        &12J \quad 12J + 2 \quad 6J + 2 \\
        &18J \\
        &12J \\
        \end{aligned}\,\, \right\} \quad J = 1, \cdots, N \\
        &\left.3(N+1)\right. \,,
    \end{aligned}
\end{equation}
where we have written the contributions from different curves on different lines. Here, we have used our proposal in equation \eqref{eqn:f4comark} for the curves with residual $\mathfrak{f}_4$ gauge algebras. We can see that
\begin{equation}
    \sum_u 2D(u) - 1 = 144N^2 + 316N + 120 \,,
\end{equation}
where the sum is taken over all of the Coulomb branch operators. As we can see, this matches the anomaly polynomial result in equation \eqref{eqn:e7sfoldanom}.

A uniform expression for the Coulomb branch scaling dimensions associated to a pair $(\mathfrak{g}, \ell)$, combining the proposals in Appendix B of \cite{Ohmori:2018ona} and equations \eqref{eqn:6curveCOR} and \eqref{eqn:f4comark}, has been observed in \cite{YTpriv}.\footnote{We thank Y.~Tachikawa for sharing this observation, and for encouraging us to include it here.} As discussed in \cite{Ohmori:2018ona}, when considering a $\mathbb{Z}_\ell$ Stiefel--Whitney twist that breaks the gauge algebra, $\mathfrak{g}$, to a residual gauge algebra, $\mathfrak{h}$, then the coefficients that appear in the Coulomb branch scaling dimensions may be expected to be some $\widetilde{c}_i$ satisfying
\begin{equation}
    1 + \sum_{i=1}^{\operatorname{rank}(\mathfrak{h})} \widetilde{c}_i = \frac{h_\mathfrak{g}^\vee}{\ell} \,.
\end{equation}
For a pair $(\mathfrak{g}, \ell)$, such $\widetilde{c}_i$ have been studied from the perspective of the supersymmetric index of 4d $\mathcal{N}=1$ pure Yang--Mills in \cite{Witten:2000nv}, where the mathematical results on almost commuting holonomies for compact Lie groups\cite{Borel:1999bx} were utilized, which we now review briefly.

Consider $(\mathfrak{g}, \ell)$, where $\mathbb{Z}_\ell$ is a subgroup of the center of the simply-connected Lie group $\widetilde{G}$ associated to $\mathfrak{g}$. The subgroup $\mathbb{Z}_\ell$ can be identified with a particular graph automorphism of the extended Dynkin diagram of $\mathfrak{g}$, $\Gamma$. One can construct a second extended Dynkin diagram, $\Gamma^\prime$, via the action of $\mathbb{Z}_\ell$ on $\Gamma$; each collection of nodes of $\Gamma$ lying within the same $\mathbb{Z}_\ell$ orbit maps to the same node of $\Gamma^\prime$. Furthermore, each node of $\Gamma^\prime$ has a ``generalized comark'', obtained by summing the comarks of all the nodes of $\Gamma$ which map to that particular node of $\Gamma^\prime$. These $\Gamma^\prime$ together with the generalized comarks are shown explicitly in the appendix of \cite{Borel:1999bx}. We refer to these generalized comarks as $c_i$, and then $\widetilde{c}_i = c_i/\ell$. The uniform expression for the dimensions of the Coulomb branch operators arising from the residual gauge algebra is then
\begin{equation}\label{eqn:YTconj}
    6J \times \frac{c_1}{\ell} + d_1 \quad \cdots \quad 6J \times \frac{c_r}{\ell} + d_r \,,
\end{equation}
where $d_i$ are the Casimirs of the residual gauge algebra.\footnote{Recall that there, in addition, exists a Coulomb branch operator arising from the torus reduction of the 6d tensor multiplet associated to each curve.} We note that, while the $c_i/\ell$ sometimes are identical with the comarks of the residual gauge algebra, this is not always the case. Equation \eqref{eqn:YTconj} appears to produce the correct answer in all known cases of Stiefel--Whitney twisted torus compactifications of very Higgsable 6d $(1,0)$ SCFTs; we consider it an interesting open question to understand such a formula from a top-down perspective.

\section{Conclusion}\label{sec:conc}

The global symmetries of a quantum field theory constitute some of its most basic data. In this paper we have presented a general prescription for reading off the continuous zero-form symmetry group for 6d SCFTs based on the topological structure of the effective field theory on the tensor branch. Using this, we have determined the continuous part of the zero-form symmetry group on the tensor branch, including the center-flavor symmetry, the contribution from Abelian symmetry factors, as well as possible mixing with R-symmetry factors. Using this data, we have also determined the continuous zero-form symmetry group for a large class of orbi-instanton theories as obtained from small instantons probing an $E_8$ nine-brane. Making use of this global structure, we have also shown that such theories provide a fruitful starting point for generating a large class of 4d $\mathcal{N} = 2$ SCFTs via Stiefel--Whitney twisted compactifications on a $T^2$. In the remainder of this section we discuss some avenues of future investigation.

In this work we have primarily focused on the structure of the global zero-form symmetries, but one can in principle also study higher symmetries that act on extended objects. For example, the one-form symmetries of some 6d SCFTs were recently studied in \cite{Bhardwaj:2020phs, Hubner:2022kxr}, and the corresponding 0-form, 1-form and 2-group symmetries of the 5d theories obtained from a reduction on an $S^1$ were recently calculated using the geometry of the associated non-compact elliptically fibered Calabi--Yau threefold \cite{Cvetic:2022imb}. Some aspects of these issues have also been explored in \cite{Cvetic:2021sxm,Apruzzi:2021mlh,Apruzzi:2022dlm}. It would be interesting to use our bottom up approach based on the effective field theory on the tensor branch to provide an independent cross-check on these results.

One of the operating assumptions in much of our work is that the effective field theory on the tensor branch provides an accurate characterization of the resulting flavor symmetries of an SCFT. In some cases, the SCFT may have enhanced flavor symmetry, and in others, there can even be a dehancement. For example, $\mathfrak{su}_2$ gauge theory on a $-2$ curve with eight half hypermultiplets in the fundamental representation has an $\mathfrak{so}_8$ flavor symmetry algebra on the tensor branch, but only a $\mathfrak{so}_7$ flavor symmetry at the fixed point (see, e.g., \cite{Heckman:2015bfa,Ohmori:2015pia,Morrison:2016djb,Hanany:2018vph,Baume:2021qho}). Similarly, when one of the half-hypermultiplets is eaten up by a neighboring undecorated self-pairing $2$ tensor, the naive $\mathfrak{so}_7$ flavor symmetry is dehanced to a $\mathfrak{g}_2$. Geometrically, this curiosity is related to the complicated nature of the $I_0^*$ singular fiber, which engineers $\mathfrak{so}_8$, $\mathfrak{so}_7$, and $\mathfrak{g}_2$ algebras; some of the geometric properties and subtleties in these cases have been studied in \cite{Bertolini:2015bwa,Esole:2017qeh,Esole:2018mqb}. In this paper, we have seen evidence that 6d SCFTs of the form
\begin{equation}\label{eqn:app1}
    \cdots \overset{\mathfrak{su}_{2\ell}}{2} \,,
\end{equation}
compactified on a $T^2$ with a $\mathbb{Z}_\ell$ Stiefel--Whitney twist also feature these type of dehancements. These observations have mainly come from dual class $\mathcal{S}$ descriptions, as in Section \ref{sec:classs}, where there are alternative methods to calculate the exact superconformal flavor symmetry; in configurations of the form in equation \eqref{eqn:app1} without a known class $\mathcal{S}$ dual, the flavor symmetry is at present not convincingly known. It would be worthwhile to understand both the field theoretic and the geometric origin of these rare and exceptional dehancements directly from a 6d or F-theory perspective.

We have explicitly shown that the global form of the R-symmetry can potentially mix with the flavor symmetries of a 6d SCFT.
Now, in the context of compactification to lower-dimensional spaces, a partial topological twist is often used to correctly
capture the resulting supersymmetries which are retained. It would be quite interesting to track this data in the resulting compactifications of theories. Indeed, in the broader context of generating 4d SCFTs from compactification 6d SCFTs, it is natural to consider Stiefel--Whitney twists on a genus $g$ curve with marked points. Here, we can in principle consider more than just a single pair of holonomies which commute up to a center-valued flux in $\widetilde{G}$. Since we now have a large class of 6d SCFTs which can generate such theories, it is natural to consider this more general situation.


\section*{Acknowledgements}

We thank Y.~Tachikawa for comments on an earlier draft of this manuscript and for sharing an observation on the coefficients entering into the Coulomb branch scaling dimensions. We further thank M.~H\"ubner, K.~Intriligator, M.J.~Kang, G.~Moore, E.~Sharpe, and E.~Torres for helpful discussions. Part of this work was performed at the conference ``Geometrization
of (S)QFTs in $D \leq 6$'' held at the Aspen Center for Physics, which is supported by National
Science Foundation grant PHY-1607611. The work of JJH is supported by the DOE (HEP) Award DE-SC0013528. CL acknowledges support from DESY (Hamburg, Germany), a member of the Helmholtz Association HGF. HYZ acknowledges support from the Simons Foundation Collaboration grant \#724069 on ``Special Holonomy in Geometry, Analysis and Physics''. The work of GZ is supported by the FWO project G.0926.17N.

\appendix

\section{Symmetries of E-String and \texorpdfstring{\boldmath{$\mathcal{N} = (2,0)$}}{N = (2,0)} Theories} \label{app:JUSTHEFLUBRO}

In this Appendix, we examine the global symmetry structure of the E-string and $(2,0)$ theories, including the R-symmetry. As 
earlier, we leave implicit the action on the spacetime symmetries, as dictated by the group action on the supercharges of the theory.

At the level of the algebra, the symmetry of the rank $N$ E-string is $\mathfrak{e}_8 \oplus \mathfrak{su}(2)_L \oplus \mathfrak{su}(2)_R$, which reduces to $\mathfrak{e}_8  \oplus \mathfrak{su}(2)_R$ for $N=1$.
Starting from the latter case, which has the simple tensor branch configuration
\begin{align}
    [\mathfrak{e}_8] \, 1 \, ,
\end{align}
the global symmetry is encoded in the Green--Schwarz four-form:
\begin{align}\label{eq:GS-4-form_1-E-string}
    I = -c_2(F_{\mathfrak{e}_8}) + c_2(R) - \tfrac14 p_1(T) \, ,
\end{align}
where $-c_2(F_{\mathfrak{e}_8})$ is always integer since $E_8$ is simply-connected.
Following the discussion of Section \ref{sec:R-symmetry_twist}, we can consider a $\bbZ_2$-twist of the R-symmetry and tangent bundle, which indeed leads to an integer shift,
\begin{align}
    c_2(R) - \tfrac14 p_1(T) \equiv -\tfrac14 w_R^2 - \tfrac34 w_R^2 \equiv 0 \mod \bbZ \, .
\end{align}
So we conclude that the global symmetry group of the rank 1 E-string is
\begin{align}
    E_8 \times SO(3)_R \, .
\end{align}
Now, the additional $\mathfrak{su}(2)_L$ flavor symmetry of the rank $N>1$ E-string couples to all nodes of self-pairing 2 in the quiver,
\begin{align}
    [\mathfrak{e}_8] \, 1 \, \overbrace{\underbrace{\, 2 \, 2 \, \cdots \, 2 \,}_{N-1}}^{[\mathfrak{su}(2)_L]} \, ,
\end{align}
but does \emph{not} enter the topological coupling of the left-most tensor multiplet, which by itself would just be a rank one E-string.
Therefore, its topological coupling is formally identical to that in equation \eqref{eq:GS-4-form_1-E-string}, and allows a $\bbZ_2$ twisted $SU(2)_R$ bundle.
Meanwhile, the undecorated nodes of self-pairing 2 all have topological couplings of the same form \cite{Ohmori:2014kda},
\begin{align}\label{eq:GS-4-form_-2-curve}
    I^{i>1} = c_2(L) - c_2(R) \, ,
\end{align}
which, when $c_2(R)$ is fractional, also forces $c_2(L) \equiv c_2(F_{\mathfrak{su}(2)_L})$ to be fractional.
Hence, we conclude that the rank $N$ E-string has global symmetry group
\begin{align}
    E_8 \times [SU(2)_L \times SU(2)_R]/\bbZ_2 \cong E_8 \times SO(4) \, .
\end{align}

An ${\cal N}=(2,0)$ theory has a tensor branch quiver that takes the form of an ADE-type Dynkin diagram, with nodes being undecorated and having self-pairing 2.
While the SCFT has R-symmetry $\mathfrak{sp}_2 \cong \mathfrak{so}_5$, the tensor branch description sees only the $\mathfrak{so}_4 \cong \mathfrak{su}(2)_L \oplus \mathfrak{su}(2)_R$ subalgebra, where, from a $(1,0)$ perspective, the $\mathfrak{su}(2)_L$ appears as a flavor symmetry while the $\mathfrak{su}(2)_R$ is the $(1,0)$ R-symmetry.
This is analogous to the self-pairing-2 nodes of the rank $N$ E-string, including the form of the Green--Schwarz four-form in equation \eqref{eq:GS-4-form_-2-curve}, which does not couple to $p_1(T)$.
Therefore, we can naturally consider a diagonal $\bbZ_2 \subset Z(SU(2)_L \times SU(2)_R)$ twist with background field $w$, such that
\begin{align}
    c_2(L) - c_2(R) \equiv \tfrac14 w^2 - \tfrac14 w^2 \equiv 0 \mod \bbZ \, .
\end{align}
This would imply that the global symmetry of $(2,0)$ theories on the tensor branch is
\begin{align}
    SO(4) \cong (SU(2)_L \times SU(2)_R) / \bbZ_2 \, .
\end{align}
Since this is a subgroup of $SO(5)$, but not $Spin(5) \cong Sp(2)$, we predict that the $(2,0)$ SCFTs have $SO(5)$ R-symmetry group.

\section{SW-folds and Rank Two 4d \texorpdfstring{\boldmath{$\mathcal{N}=2$}}{N=2} SCFTs}\label{app:ranktwo}

In this Appendix we show that SW-folds can be used to construct nearly all of the known rank two 4d $\mathcal{N} = 2$ SCFTs.
In recent years, there has been much progress in the program of classifying low rank 4d $\mathcal{N}=2$ SCFTs by studying the complex geometry of the Coulomb branch \cite{Argyres:2015ffa,Argyres:2015gha,Argyres:2016xua,Argyres:2016xmc,Argyres:2016yzz,Caorsi:2018zsq,Caorsi:2019vex,Martone:2020nsy,Argyres:2020wmq,Argyres:2022kon}. An enumeration and analysis of the known rank two theories has appeared recently in \cite{Martone:2021ixp,Kaidi:2021tgr,Martone:2021drm,Bourget:2021csg, Kaidi:2022sng}. As has been emphasized in \cite{Martone:2021ixp}, this enumeration is by no means a classification, and there are many reasons to believe that there remain undiscovered rank two theories. Nevertheless, recent progress has unveiled an intricate structure to the rank two 4d $\mathcal{N}=2$ SCFT landscape.

Rank two SCFTs can be arranged into families that are connected via renormalization group flows. Each family possesses a collection of ``parent'' or ``top'' theories from which all other theories in the family can be obtained by mass deformation. We note that it is not necessary that each SCFT in the family comes from all top theories, only that it comes from a mass deformation of \emph{at least one} top theory. All of the ``top'' rank two theories of \cite{Martone:2021ixp} are given in Table \ref{tbl:toptwo}. In this section, we study which of the known rank two theories can be interpreted as arising from the Stiefel--Whitney twisted compactifications that have formed the topic of this paper. It is necessary only to provide an origin for the top theories, as those theories obtained via 4d mass deformation follow from the addition of continuous Wilson lines, breaking the flavor symmetry, on the $T^2$.

\begin{table}[]
    \centering
    \renewcommand{\arraystretch}{1.5}
    \begin{threeparttable}
    \begin{tabular}{cccc}
    \toprule
         Flavor Algebra & $\left\{ \Delta_u, \Delta_v\right\}$ & $(24a, 12c)$
         &  SW-fold \\\midrule
         $\left(\mathfrak{e}_8\right)_{24} \oplus \left(\mathfrak{su}_2\right)_{13}$ & $\{6, 12\}$ & $(263, 161)$
         &  \multirow{2}{*}{``Trivial SW-folds''} \\
         $\left(\mathfrak{so}_{20}\right)_{16}$ & $\{6, 8\}$ & $(202, 124)$
         &   \\\midrule
         $\left(\mathfrak{usp}_{12}\right)_{8}$ & $\left\{4, 6\right\}$ & $(130, 76)$
         & $\mathcal{T}_2^{(1)}(0,0,0,0,2)$  \\
         $\left(\mathfrak{usp}_4\right)_{7} \oplus \left(\mathfrak{usp}_8\right)_{8}$& $\left\{4, 6\right\}$ & $(128, 74)$
         & I \\
         $\left(\mathfrak{su}_2\right)_{7}^2 \oplus \left(\mathfrak{f}_4\right)_{12}$ & $\left\{6, 6\right\}$ & $(156, 90)$
         &  $\mathcal{T}_2^{(1)}(1,0,0,0,0)$ \\\midrule
         $\left(\mathfrak{su}_6\right)_{16} \oplus \left(\mathfrak{su}_2\right)_{9}$ & $\{6, 8\}$ & $(179, 101)$
         & $\mathcal{S}_2^{(1)}(0,0,0,1,0)$  \\\midrule
         $\left(\mathfrak{usp}_{14}\right)_{9}$  & $\{6, 8\}$ & $(185, 107)$
         & $\mathcal{T}_2^{(1)}(0,0,0,0,3)$ \\\midrule
         $\left(\mathfrak{su}_{5}\right)_{16}$ & $\{6, 8\}$ & $(170, 92)$
         & $\mathcal{R}_3^{(1)}(0,0,2)$  \\\midrule
         $\left(\mathfrak{usp}_8\right)_{13} \oplus \left(\mathfrak{su}_2\right)_{26}$ & $\{6, 12\}$ & $(232, 130)$
         & $\mathcal{T}_2^{(2)}(0,0,0,0,1)$  \\
         $\left(\mathfrak{su}_2\right)_{2} \oplus \left(\mathfrak{su}_2\right)_{8}$ & $\left\{3, 6\right\}$ & $(102, 54)$
         & $\mathcal{T}_4^{(1)}(1,0)$ \\\midrule
         $\left(\mathfrak{g}_2\right)_{8} \oplus \left(\mathfrak{su}_2\right)_{14}$ & $\left\{4, 6\right\}$ & $(120, 66)$
         & $\mathcal{T}_3^{(1)}(1,0,0)$ \\\midrule
         $\left(\mathfrak{su}_3\right)_{26} \oplus \mathfrak{u}(1)$ & $\{6, 12\}$ & $(219, 117)$
         & $\mathcal{S}_3^{(2)}(0,0,1)$   \\\midrule
         $\left(\mathfrak{su}_2\right)_{16} \oplus \mathfrak{u}(1)$ & $\{6, 12\}$ & $(212, 110)$
         &  $\mathcal{S}_4^{(2)}(0,1)$ \\\midrule
         $\left(\mathfrak{usp}_4\right)_{14} \oplus \left(\mathfrak{su}_2\right)_{8}$ & $\left\{4, 6\right\}$ & $(118, 64)$
         & II \\\midrule
         $\left(\mathfrak{su}_{2}\right)_{14}$ & $\left\{ \frac{12}{5}, 6\right\}$ & $(\frac{456}{5}, \frac{234}{5})$
         & $\mathcal{T}_5^{(1)}(1)$ \\\midrule
         $\left(\mathfrak{su}_{2}\right)_{14}$ & $\left\{2, 6\right\}$ & $(84, 42)$
         & $\mathcal{T}_6^{(1)}(1)$  \\\midrule
         $\left(\mathfrak{usp}_{12}\right)_{11}$ & $\{4, 10\}$ & $(188, 110)$
         & III \\\midrule
         $\varnothing$ & $\{2, 4\}$ & $(58, 28)$
         & IV \\\bottomrule
    \end{tabular}
    \end{threeparttable}
    \caption{All the ``top'' theories from \cite{Martone:2021ixp}. We list their Coulomb branch operator dimensions and their 6d origin, if known. There are four theories, labelled by I--IV, for which no SW-fold description is known. The theories marked as ``Trivial SW-folds'' are those which are obtained via compactification of a 6d $(1,0)$ SCFT on $T^2$ without turning on a Stiefel--Whitney twist.}\label{tbl:toptwo}
\end{table}

When considering a Stiefel--Whitney twisted compactification, the Coulomb branch dimension of the resulting four-dimensional theory is always at least the number of tensor multiplets, i.e., compact curves in the geometric construction, of the parent 6d theory. As such, the 4d SCFTs of low rank can only come from a highly restrictive set of 6d SCFTs. There are two ways of engineering theories with a Coulomb branch of rank two. We can consider
\begin{equation}
    \overset{\mathfrak{g}}{1}/\mathbb{Z}_\ell \,,
\end{equation}
where the $\mathbb{Z}_\ell$ Stiefel--Whitney quotient breaks the gauge algebra to $\mathfrak{g}_\text{ub}$ which is either
\begin{equation}
    \mathfrak{so}_2 \quad \text{ or } \quad \mathfrak{su}_2 \,.
\end{equation}
In either case, the methodology of \cite{Ohmori:2018ona} allows us to determine that the Coulomb branch operators, $u$ and $v$, have dimensions\footnote{We do not need to worry about the subtleties with the prescription of \cite{Ohmori:2018ona} that were highlighted in Section \ref{sec:cbe6}, as the rank two requirement on the Coulomb branch ensures that the gauge algebra after Stiefel--Whitney twist is at most rank one.}
\begin{equation}
    \left\{ \Delta_u, \Delta_v \right\} = \begin{cases}
      \left\{6, \frac{6}{\ell} + 1 \right\} \quad &\text{if} \quad \mathfrak{g}_\text{ub} = \mathfrak{so}_2 \cr
      \left\{6, 8 \right\} \quad &\text{if} \quad \mathfrak{g}_\text{ub} = \mathfrak{su}_2 \,.
      \end{cases}
\end{equation}
On the other hand, we can have
\begin{equation}
    \overset{\mathfrak{g}}{1}\overset{\mathfrak{h}}{2}/\mathbb{Z}_\ell \,,
\end{equation}
where the quotient breaks the entirety of the gauge algebra. In this case, we allow $\mathfrak{g}$ and $\mathfrak{h}$ to be trivial, and we also allow $\ell = 1$.  We can determine that
\begin{equation}
    \left\{ \Delta_u, \Delta_v \right\} = \begin{cases}
      \left\{6, \frac{12}{\ell} \right\} \quad &\text{if} \quad \mathfrak{g} = \varnothing \cr
      \left\{6, 12 \right\} \quad &\text{if} \quad \mathfrak{g} \neq \varnothing \,.
      \end{cases}
\end{equation}
Putting this together, and using that fact that $\ell = 1, \cdots, 6$ are the only valid options, we find that the possible Coulomb branch operator spectra are highly constrained. Specifically, all rank two theories obtained by Stiefel--Whitney twisted compactification have a Coulomb branch operator with scaling dimension $\Delta = 6$.

In view of these restrictions on the scaling dimensions of the Coulomb branch operators, let us consider the theories I--IV from Table \ref{tbl:toptwo}; that is, those that do not have a known SW-fold description. Theories I and II have Coulomb branch operators of dimensions $\{4, 6\}$, which are consistent with them arising from a $\mathbb{Z}_3$ Stiefel--Whitney twisted compactification of 6d SCFTs of the form
\begin{equation}
    \overset{\mathfrak{g}}{1} \,, \qquad \text{ or } \qquad 1\overset{\mathfrak{g}}{2} \,,
\end{equation}
however, it is unclear that additional 6d SCFTs of this form admitting a $\mathbb{Z}_3$ center-flavor symmetry exist. The theory labelled III has Coulomb branch operators with dimensions $\{4, 10\}$, which does not include the requisite dimension six operator for it to be able to arise from a Stiefel--Whitney twisted compactification. There are two possibilities: either theory III is not a top theory, or else it is a theory that cannot be obtained from a Stiefel--Whitney twisted torus compactification. The analysis of \cite{Cecotti:2021ouq} appears to rule out any currently unknown theory with Coulomb branch dimensions $\{6,12\}$, which would be expected for a putative top theory that mass deforms to a theory with Coulomb branch operator spectrum $\{4,10\}$, and thus we conclude that this SCFT probably does not arise from a Stiefel--Whitney twisted compactification.\footnote{We thank M. Martone for discussions on this point.} Finally, we turn to theory IV. This is the Lagrangian theory with gauge algebra $\mathfrak{sp}_2$ and a single half-hypermultiplet in the $\bm{16}$ representation, and it is also the only theory in \cite{Martone:2021ixp} that does not have any known construction in string theory. As emphasized therein, one may speculate that this SCFT sits as a descendant inside of a currently unknown family of rank two SCFTs.

At rank two it appears that almost all of the 4d $\mathcal{N}=2$ SCFTs can be obtained from torus compactifications of 6d $(1,0)$ SCFTs. Of the sixty-nine 4d SCFTs listed in \cite{Martone:2021ixp}, there are only seven for which it is not known how to obtain them in this manner. It would be interesting to understand whether other ingredients can be included in the torus compactifications to generate the complete list of rank two theories, and to determine if the preponderance of Stiefel--Whitney twists persists to higher rank 4d SCFTs.

\section{Stiefel--Whitney Twists and the Literature}\label{app:lit}

In this Appendix we present a brief survey in table format of earlier work on Stiefel--Whitney compactifications.
4d $\mathcal{N}=2$ SCFTs have been constructed from Stiefel--Whitney twisted torus compactifications of very Higgsable 6d $(1,0)$ SCFTs in previous literature \cite{Ohmori:2018ona,Apruzzi:2020pmv,Giacomelli:2020jel,Heckman:2020svr,Giacomelli:2020gee,Bourget:2020mez}. These theories form a small subset of the landscape of Stiefel--Whitney twisted theories that we discuss in the present paper, and we highlight for which values of the $E_8$-homomorphism parameters they have been studied. These particular theories are listed in Table \ref{tbl:lit}, together with the reference to where they were first explored.

\begin{table}[]
    \centering
    \small
    \begin{threeparttable}
    \begin{tabular}{ccccc}
        \toprule
        SW-fold & 6d Origin & $\mathbb{Z}_\ell$ & Alternate Name & Reference \\\midrule
        $\mathcal{T}_2^{(r)}(0,0,0,0,1)$ & $\overset{\mathfrak{su}_2}{1}\underbrace{\overset{\mathfrak{su}_2}{2}\cdots\overset{\mathfrak{su}_2}{2}}_{r-1}$ & $\mathbb{Z}_2$ & $\mathcal{S}_{E_6, 2}^{(r)}$ & \multirow{12}{*}{\cite{Giacomelli:2020jel}} \\
        $\mathcal{S}_3^{(r)}(0,0,1)$ & $\overset{\mathfrak{su}_3}{1}\underbrace{\overset{\mathfrak{su}_3}{2}\cdots\overset{\mathfrak{su}_3}{2}}_{r-1}$ & $\mathbb{Z}_3$ & $\mathcal{S}_{D_4, 3}^{(r)}$ & \\
        $\mathcal{S}_4^{(r)}(0,1)$ & $\overset{\mathfrak{su}_4}{1}\underbrace{\overset{\mathfrak{su}_4}{2}\cdots\overset{\mathfrak{su}_4}{2}}_{r-1}$ & $\mathbb{Z}_4$ & $\mathcal{S}_{A_2, 4}^{(r)}$ & \\
        $\mathcal{T}_2^{(r-1)}(1,0,0,0,0)$ & $1\underbrace{\overset{\mathfrak{su}_2}{2}\cdots\overset{\mathfrak{su}_2}{2}}_{r-1}$ & $\mathbb{Z}_2$ & $\mathcal{T}_{E_6, 2}^{(r)}$ & \\
        $\mathcal{T}_3^{(r-1)}(1,0,0)$ & $1\underbrace{\overset{\mathfrak{su}_3}{2}\cdots\overset{\mathfrak{su}_3}{2}}_{r-1}$ & $\mathbb{Z}_3$ & $\mathcal{T}_{D_4, 3}^{(r)} $ & \\
        $\mathcal{T}_4^{(r-1)}(1,0)$ & $1\underbrace{\overset{\mathfrak{su}_4}{2}\cdots\overset{\mathfrak{su}_4}{2}}_{r-1}$ & $\mathbb{Z}_4$ & $\mathcal{T}_{A_2, 4}^{(r)}$ & \\\midrule
        $\mathcal{T}_5^{(r-1)}(1)$ & $1\underbrace{\overset{\mathfrak{su}_5}{2}\cdots\overset{\mathfrak{su}_5}{2}}_{r-1}$ & $\mathbb{Z}_5$ & $\mathcal{T}_{\varnothing, 5}^{(r)}$ & \multirow{4}{*}{\cite{Giacomelli:2020gee}} \\
        $\mathcal{T}_6^{(r-1)}(1)$ & $1\underbrace{\overset{\mathfrak{su}_6}{2}\cdots\overset{\mathfrak{su}_6}{2}}_{r-1}$ & $\mathbb{Z}_6$ & $\mathcal{T}_{\varnothing, 6}^{(r)}$ & \\\midrule
        $\mathcal{S}_2^{(1)}(0,0,0,2r-1,n-2)$ & $\overset{\mathfrak{su}_{2n}}{1}\underbrace{\overset{\mathfrak{su}_{2n+8}}{2}\cdots\overset{\mathfrak{su}_{2n-8r-8}}{2}}_{r-1}$ & $\mathbb{Z}_2$ & --- & \multirow{18}{*}{\cite{Ohmori:2018ona}} \\
        $\mathcal{S}_3^{(1)}(0,0,3r-2)$ & $\overset{\mathfrak{su}_{3}}{1}\underbrace{\overset{\mathfrak{su}_{12}}{2}\cdots\overset{\mathfrak{su}_{9r-6}}{2}}_{r-1}$ & $\mathbb{Z}_3$ & --- & \\
        $\mathcal{R}_3^{(1)}(0,0,3r-1)$ & $\overset{\mathfrak{su}_{6}^\prime}{1}\underbrace{\overset{\mathfrak{su}_{15}}{2}\cdots\overset{\mathfrak{su}_{9r-3}}{2}}_{r-1}$ & $\mathbb{Z}_3$ & --- & \\
        $\mathcal{T}_3^{(1)}(0,0,3r-3)$ & $1\underbrace{\overset{\mathfrak{su}_{9}}{2}\cdots\overset{\mathfrak{su}_{9r-9}}{2}}_{r-1}$ & $\mathbb{Z}_3$ & --- & \\
        $\mathcal{S}_4^{(1)}(0,2r-1)$ & $\overset{\mathfrak{su}_{4}}{1}\underbrace{\overset{\mathfrak{su}_{12}}{2}\cdots\overset{\mathfrak{su}_{8r-4}}{2}}_{r-1}$ & $\mathbb{Z}_4$ & --- & \\
        $\mathcal{T}_4^{(1)}(0,2r-2)$ & $1\underbrace{\overset{\mathfrak{su}_{8}}{2}\cdots\overset{\mathfrak{su}_{8r-8}}{2}}_{r-1}$ & $\mathbb{Z}_4$ & --- & \\
        $\mathcal{T}_5^{(1)}(r-1)$ & $1\underbrace{\overset{\mathfrak{su}_{5}}{2}\cdots\overset{\mathfrak{su}_{5r-5}}{2}}_{r-1}$ & $\mathbb{Z}_5$ & --- & \\
        $\mathcal{T}_6^{(1)}(r-1)$ & $1\underbrace{\overset{\mathfrak{su}_{6}}{2}\cdots\overset{\mathfrak{su}_{6r-6}}{2}}_{r-1}$ & $\mathbb{Z}_6$ & --- & \\
         \bottomrule
    \end{tabular}
    \end{threeparttable}
    \caption{The Stiefel--Whitney twisted 4d $\mathcal{N}=2$ SCFTs that have appeared afore.}
    \label{tbl:lit}
\end{table}

\section{Nilpotent Orbits and Higgsing SW-folds}\label{app:nilp}

In this Appendix we track the structure of a particular class of Higgs branch flows in 6d, and their 4d descendants after a SW-twist.
Take the original 6d rank $N$ orbi-instanton SCFT of type $(\mathfrak{e}_8, \mathfrak{g)}$ which can be expressed on its partial tensor branch as
\begin{equation}
    [\mathfrak{e}_8] \overset{\mathfrak{g}}{1}  \overset{\mathfrak{g}}{2} \cdots \overset{\mathfrak{g}}{2}  [\mathfrak{g}] \,.
\end{equation}
Consider, as we did in Section \ref{sec:6d}, a Higgsing via an $E_8$-homomorphism $\Gamma_\mathfrak{g} \rightarrow E_8$ that leads to an SCFT with a non-trivial center-flavor symmetry $\mathbb{Z}_{\ell}$. Instead of immediately compactifying it on a $T^2$ with a $\bbZ_{\ell}$ Stiefel--Whitney twist, we first perform a nilpotent Higgsing in 6d of the $\mathfrak{g}$ flavor symmetry on the right of the tensor branch quiver:\footnote{Higgs branch renormalization group flows of 6d SCFTs triggered by nilpotent deformations have been studied in great detail in \cite{Heckman:2016ssk,Mekareeya:2016yal,Heckman:2018pqx,DeLuca:2018zbi,Hassler:2019eso,Baume:2021qho,Distler:2022yse}.}
\begin{equation}
    \mu_{\text{6d}}: \mathfrak{su}_2 \rightarrow \mathfrak{g} \,.
\end{equation}
Particular choices for the nilpotent orbit lead to 6d SCFT where the $\mathbb{Z}_\ell$ center-flavor symmetry is preserved. We can then take this Higgsed theory, and compactify it on a $T^2$ with a $\mathbb{Z}_\ell$ Stiefel--Whitney twist. In this way, we end up with a larger family of SW-fold theories in 4d, going beyond the scope of the theories listed in Table \ref{tbl:6dscfts}.

Nonetheless, we can still analyze these extra theories by computing their central charges and flavor central charges as in Section \ref{sec:sfolds}. We also point out that, these extra SW-fold theories can also be obtained by taking the SW-fold theories that we have obtained in the main text as in Table \ref{tbl:genSfolds} and performing the following ``induced'' nilpotent Higgsing in 4d:
\begin{equation}
    \mu_{4d}: \mathfrak{su}_2 \rightarrow \widetilde{\mathfrak{g}}
\end{equation}
where $\widetilde{\mathfrak{g}}$ is the flavor symmetry algebra in 4d that descended from the $\mathfrak{g}$ flavor symmetry on the right of the 6d SCFT. The nilpotent orbit $\mu_{4d}$ is then the nilpotent deformation in 4d which can be thought of as a ``folded'' version of the nilpotent deformation in 6d.

For the nilpotent Higgsings of the 6d SCFTs that were discussed in Section \ref{sec:e8CF}, we have $\mathfrak{g} = \mathfrak{su}_K$. Nilpotent orbits of $\mathfrak{su}_K$ are in one-to-one correspondence with integer partitions of $K$. We write
\begin{equation}\label{eqn:exppart}
    P = [1^{n_1}, 2^{n_2}, \cdots, K^{n_K}] \,,
\end{equation}
where $n_i \geq 0$ and
\begin{equation}
    \sum_{i=1}^K i n_i = K \,,
\end{equation}
to denote a partition of $K$. We are interested in the case where $K = \ell \widetilde{K}$ and where there exists a $\mathbb{Z}_\ell$ center-flavor symmetry. The RHS of the tensor branch configurations written in Table \ref{tbl:6dscfts} all have the form
\begin{equation}
    \cdots \overset{\mathfrak{su}_K}{2}\overset{\mathfrak{su}_K}{2}\cdots \overset{\mathfrak{su}_K}{2}\overset{\mathfrak{su}_K}{2} \,,
\end{equation}
and it is well-known how the tensor branch configuration is modified when Higgsing by a nilpotent orbit as in equation \eqref{eqn:exppart}. One finds\footnote{For ease of exposition, we do not focus here on the cases where the plateau is too short and the nilpotent Higgsing starts to correlate with the $E_8$-homomorphism Higgsing.}
\begin{equation}
    \cdots \underset{[\mathfrak{su}_{n_K}]}{\overset{\mathfrak{su}_{K_K}}{2}}\underset{[\mathfrak{su}_{n_{K-1}}]}{\overset{\mathfrak{su}_{K_{K-1}}}{2}}\cdots \underset{[\mathfrak{su}_{n_2}]}{\overset{\mathfrak{su}_{K_2}}{2}}\underset{[\mathfrak{su}_{n_1}]}{\overset{\mathfrak{su}_{K_1}}{2}} \,.
\end{equation}
The ranks of the flavor algebras are fixed by the exponents of the partition, and the $K_i$ are fixed from that data by anomaly cancellation. They are each required to satisfy
\begin{equation}
    2K_i - n_i - K_{i-1} - K_{i+1} = 0 \,,
\end{equation}
where we have defined $K_0 = 0$ and $K_{K+1} = K$. It is easy to see that the anomaly of the large gauge transformations of the two-forms fields, as discussed in Section \ref{sec:6d}, rules out a $\mathbb{Z}_\ell$ center-flavor symmetry unless
\begin{equation}
    n_i = \ell \widetilde{n}_i \,,
\end{equation}
for all $i$. The converse also can be shown. It is easy to see that a partition
\begin{equation}
    [1^{\ell\widetilde{n}_1}, 2^{\ell\widetilde{n}_2}, \cdots, (\ell\widetilde{K})^{\ell\widetilde{n}_{(\ell\widetilde{K})}}] \,,
\end{equation}
of $\ell \widetilde{K}$ can equivalently be written as a partition
\begin{equation}
    [1^{\widetilde{n}_1}, 2^{\widetilde{n}_2}, \cdots, \widetilde{K}^{\widetilde{n}_{\widetilde{K}}}] \,,
\end{equation}
of $\widetilde{K}$. More succinctly, $\mathbb{Z}_\ell$ center-flavor symmetry preserving nilpotent orbits of $\mathfrak{su}_K$ are in one-to-one correspondence with nilpotent orbits of $\mathfrak{su}_{\widetilde{K}}$. Physically, this reflects the fact that one can either first perform the nilpotent Higgsing by a ($\mathbb{Z}_\ell$-preserving) partition of $K$ in 6d and then compactify with Stiefel--Whitney twist to 4d, or else first perform the Stiefel--Whitney twisted compactification to 4d and then the nilpotent Higgsing by the associated partition of $\widetilde{K}$; either way, one ends up with same 4d $\mathcal{N}=2$ SCFT.

The family of theories obtained by nilpotent deformations forms a partially ordered set, capturing the network of renormalization group flows amongst the theories, that follows from the partial ordering of the nilpotent orbit inclusion: $\mu \prec \nu$ when $\text{Orbit}(\mu) \subset \overline{\text{Orbit}(\nu)}$. For $\mathfrak{su}_{K}$ nilpotent orbits, such partial ordering can be characterized by the ``dominant ordering'' of two partitions of $K$; let $\mu = [r_1, \cdots, r_{\ell_r}],\ \nu = [s_1, \cdots s_{\ell_s}]$ be weakly-decreasing partitions of $K$, then
\begin{equation}
   \mu \prec \nu\ \  \Leftrightarrow\ \  \sum_{i = 1}^j r_i \leq \sum_{i = 1}^j s_i, \quad \,\, 1 \leq j \leq \operatorname{max}(\ell_r, \ell_s) \,,
\end{equation}
where the partition with fewer elements is extended by zeroes until they are of equal length.

To be more specific, let us illustrate this construction by analyzing the network of theories obtained by starting from the SCFT in family $\mathcal{T}_3^{(N)}(p, s, 3q)$ with $p=s=q=1$. The 6d origin of this theory is given by the tensor branch description:
\begin{equation}\label{eq:6d_trivialOrbit}
   [\mathfrak{e}_8] 1 \underset{[\mathfrak{su}_3]}{\overset{\mathfrak{su}_{9}}{2}}
         \underset{[\mathfrak{su}_3]}{\overset{\mathfrak{su}_{15}}{2}}
         \underset{[\mathfrak{su}_3]}{\overset{\mathfrak{su}_{18}}{2}}
         \underbrace{\overset{\mathfrak{su}_{18}}{2}
         \cdots \overset{\mathfrak{su}_{18}}{2}}_{N-1} [\mathfrak{su}_{18}] \,.
\end{equation}
We label each theory in the nilpotent network via
\begin{equation}
    \mathcal{T}_3^{(N)}(p, s, 3q; \mu_{\text{4d}}) \,,
\end{equation}
where $\mu_{\text{4d}} = [1^{18}]$ corresponds to the $\mathcal{T}_3^{(N)}(p, s, 3q)$ SCFT discussed in Section \ref{sec:sfolds}.

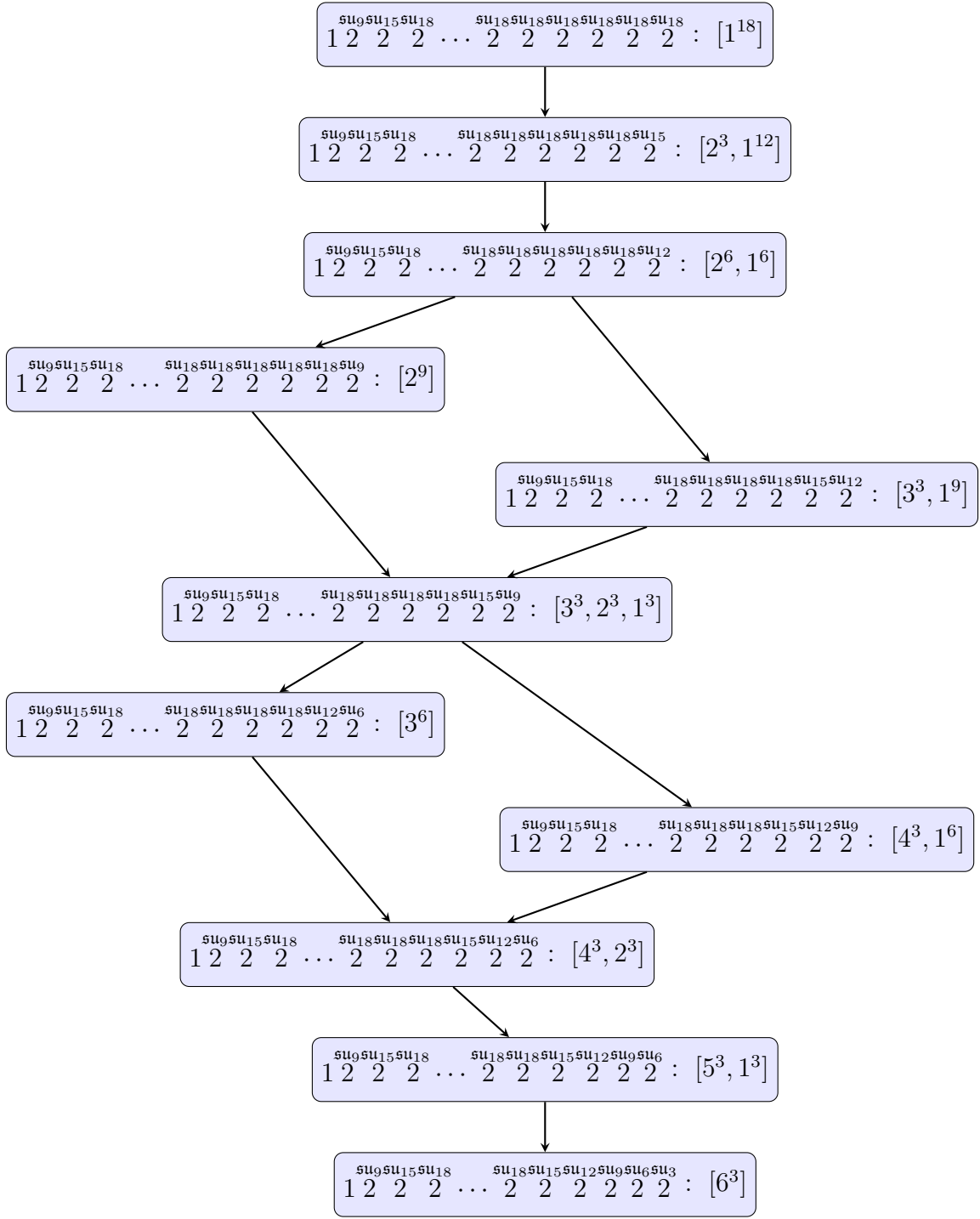
\begin{figure}
  \centering
\begin{tikzpicture}[node distance=1.8cm]

\node (0) [startstop, xshift=-1cm] {
$
\begin{gathered}
1\overset{\mathfrak{su}_{9}}{2}\overset{\mathfrak{su}_{15}}{2}\overset{\mathfrak{su}_{18}}{2}\cdots\overset{\mathfrak{su}_{18}}{2}\overset{\mathfrak{su}_{18}}{2}\overset{\mathfrak{su}_{18}}{2}\overset{\mathfrak{su}_{18}}{2}\overset{\mathfrak{su}_{18}}{2}\overset{\mathfrak{su}_{18}}{2}\end{gathered}:\ [1^{18}]
$};

\node (1) [startstop, below of=0] {
$
\begin{gathered}
1\overset{\mathfrak{su}_{9}}{2}\overset{\mathfrak{su}_{15}}{2}\overset{\mathfrak{su}_{18}}{2}\cdots\overset{\mathfrak{su}_{18}}{2}\overset{\mathfrak{su}_{18}}{2}\overset{\mathfrak{su}_{18}}{2}\overset{\mathfrak{su}_{18}}{2}\overset{\mathfrak{su}_{18}}{2}\overset{\mathfrak{su}_{15}}{2}\end{gathered}:\ [2^3, 1^{12}]
$};

\node (2) [startstop, below of=1] {
$
\begin{gathered}
1\overset{\mathfrak{su}_{9}}{2}\overset{\mathfrak{su}_{15}}{2}\overset{\mathfrak{su}_{18}}{2}\cdots\overset{\mathfrak{su}_{18}}{2}\overset{\mathfrak{su}_{18}}{2}\overset{\mathfrak{su}_{18}}{2}\overset{\mathfrak{su}_{18}}{2}\overset{\mathfrak{su}_{18}}{2}\overset{\mathfrak{su}_{12}}{2}\end{gathered}:\ [2^6, 1^6]
$};

\node (3) [startstop, below of=2, xshift=-5cm] {
$
\begin{gathered}
1\overset{\mathfrak{su}_{9}}{2}\overset{\mathfrak{su}_{15}}{2}\overset{\mathfrak{su}_{18}}{2}\cdots\overset{\mathfrak{su}_{18}}{2}\overset{\mathfrak{su}_{18}}{2}\overset{\mathfrak{su}_{18}}{2}\overset{\mathfrak{su}_{18}}{2}\overset{\mathfrak{su}_{18}}{2}\overset{\mathfrak{su}_{9}}{2}\end{gathered}:\ [2^9]
$};

\node (4) [startstop, below of=3, xshift=8cm] {
$
\begin{gathered}
1\overset{\mathfrak{su}_{9}}{2}\overset{\mathfrak{su}_{15}}{2}\overset{\mathfrak{su}_{18}}{2}\cdots\overset{\mathfrak{su}_{18}}{2}\overset{\mathfrak{su}_{18}}{2}\overset{\mathfrak{su}_{18}}{2}\overset{\mathfrak{su}_{18}}{2}\overset{\mathfrak{su}_{15}}{2}\overset{\mathfrak{su}_{12}}{2}\end{gathered}:\ [3^3, 1^9]
$};

\node (5) [startstop, below of=4, xshift=-5cm] {
$
\begin{gathered}
1\overset{\mathfrak{su}_{9}}{2}\overset{\mathfrak{su}_{15}}{2}\overset{\mathfrak{su}_{18}}{2}\cdots\overset{\mathfrak{su}_{18}}{2}\overset{\mathfrak{su}_{18}}{2}\overset{\mathfrak{su}_{18}}{2}\overset{\mathfrak{su}_{18}}{2}\overset{\mathfrak{su}_{15}}{2}\overset{\mathfrak{su}_{9}}{2}\end{gathered}:\ [3^3, 2^3, 1^3]
$};

\node (6) [startstop, below of=5, xshift=-3cm] {
$
\begin{gathered}
1\overset{\mathfrak{su}_{9}}{2}\overset{\mathfrak{su}_{15}}{2}\overset{\mathfrak{su}_{18}}{2}\cdots\overset{\mathfrak{su}_{18}}{2}\overset{\mathfrak{su}_{18}}{2}\overset{\mathfrak{su}_{18}}{2}\overset{\mathfrak{su}_{18}}{2}\overset{\mathfrak{su}_{12}}{2}\overset{\mathfrak{su}_{6}}{2}\end{gathered}:\ [3^6]
$};

\node (7) [startstop, below of=6, xshift=8cm] {
$
\begin{gathered}
1\overset{\mathfrak{su}_{9}}{2}\overset{\mathfrak{su}_{15}}{2}\overset{\mathfrak{su}_{18}}{2}\cdots\overset{\mathfrak{su}_{18}}{2}\overset{\mathfrak{su}_{18}}{2}\overset{\mathfrak{su}_{18}}{2}\overset{\mathfrak{su}_{15}}{2}\overset{\mathfrak{su}_{12}}{2}\overset{\mathfrak{su}_{9}}{2}\end{gathered}:\ [4^3, 1^6]
$};

\node (8) [startstop, below of=7, xshift=-5cm] {
$
\begin{gathered}
1\overset{\mathfrak{su}_{9}}{2}\overset{\mathfrak{su}_{15}}{2}\overset{\mathfrak{su}_{18}}{2}\cdots\overset{\mathfrak{su}_{18}}{2}\overset{\mathfrak{su}_{18}}{2}\overset{\mathfrak{su}_{18}}{2}\overset{\mathfrak{su}_{15}}{2}\overset{\mathfrak{su}_{12}}{2}\overset{\mathfrak{su}_{6}}{2}\end{gathered}:\ [4^3, 2^3]
$};

\node (9) [startstop, below of=8, xshift=2cm] {
$
\begin{gathered}
1\overset{\mathfrak{su}_{9}}{2}\overset{\mathfrak{su}_{15}}{2}\overset{\mathfrak{su}_{18}}{2}\cdots\overset{\mathfrak{su}_{18}}{2}\overset{\mathfrak{su}_{18}}{2}\overset{\mathfrak{su}_{15}}{2}\overset{\mathfrak{su}_{12}}{2}\overset{\mathfrak{su}_{9}}{2}\overset{\mathfrak{su}_{6}}{2}\end{gathered}:\ [5^3, 1^3]
$};

\node (10) [startstop, below of=9] {
$
\begin{gathered}
1\overset{\mathfrak{su}_{9}}{2}\overset{\mathfrak{su}_{15}}{2}\overset{\mathfrak{su}_{18}}{2}\cdots\overset{\mathfrak{su}_{18}}{2}\overset{\mathfrak{su}_{15}}{2}\overset{\mathfrak{su}_{12}}{2}\overset{\mathfrak{su}_{9}}{2}\overset{\mathfrak{su}_{6}}{2}\overset{\mathfrak{su}_{3}}{2}\end{gathered}:\ [6^3]
$};

\draw [arrow] (0) -- (1);
\draw [arrow] (1) -- (2);
\draw [arrow] (2) -- (3);
\draw [arrow] (2) -- (4);
\draw [arrow] (3) -- (5);
\draw [arrow] (4) -- (5);
\draw [arrow] (5) -- (6);
\draw [arrow] (5) -- (7);
\draw [arrow] (6) -- (8);
\draw [arrow] (7) -- (8);
\draw [arrow] (8) -- (9);
\draw [arrow] (9) -- (10);

\end{tikzpicture}
\caption{The subsector of the nilpotent hierarchy of the 6d SCFT in equation \eqref{eq:6d_trivialOrbit} in which each theory enjoys a $\mathbb{Z}_3$ center-flavor symmetry. We listed the quiver description of the tensor branch and the partition defining the nilpotent orbit in each case. The $\mathbb{Z}_3$ Stiefel--Whitney twisted torus compactification of the theories appearing here gives rise to the 4d $\mathcal{N}=2$ SCFTs whose nilpotent network is depicted in Figure \ref{fig:sfold_nilpotent_4d}.}
\label{fig:sfold_nilpotent}
\end{figure}

\begin{figure}
\centering
\begin{tikzpicture}[node distance=1.8cm]
\node (0) [startstop, xshift=-1cm] {
$
\mathcal{T}_3^{(N)}(1, 1, 3; \mu_{4d} = [1^6])
$};

\node (1) [startstop, below of=0] {
$
\mathcal{T}_3^{(N)}(1, 1, 3; \mu_{4d} = [2, 1^4])
$};

\node (2) [startstop, below of=1] {
$
\mathcal{T}_3^{(N)}(1, 1, 3; \mu_{4d} = [2^2, 1^2])
$};

\node (3) [startstop, below of=2, xshift=-3cm] {
$
\mathcal{T}_3^{(N)}(1, 1, 3; \mu_{4d} = [2^3])
$};

\node (4) [startstop, below of=3, xshift=6cm] {
$
\mathcal{T}_3^{(N)}(1, 1, 3; \mu_{4d} = [3, 1^3])
$};

\node (5) [startstop, below of=4, xshift=-3cm] {
$
\mathcal{T}_3^{(N)}(1, 1, 3; \mu_{4d} = [3, 2, 1])
$};

\node (6) [startstop, below of=5, xshift=-3cm] {
$
\mathcal{T}_3^{(N)}(1, 1, 3; \mu_{4d} = [3^2])
$};

\node (7) [startstop, below of=6, xshift=6cm] {
$
\mathcal{T}_3^{(N)}(1, 1, 3; \mu_{4d} = [4, 1, 1])
$};

\node (8) [startstop, below of=7, xshift=-3cm] {
$
\mathcal{T}_3^{(N)}(1, 1, 3; \mu_{4d} = [4, 2])
$};

\node (9) [startstop, below of=8, xshift=0cm] {
$
\mathcal{T}_3^{(N)}(1, 1, 3; \mu_{4d} = [5, 1])
$};

\node (10) [startstop, below of=9] {
$
\mathcal{T}_3^{(N)}(1, 1, 3; \mu_{4d} = [6])
$};

\draw [arrow] (0) -- (1);
\draw [arrow] (1) -- (2);
\draw [arrow] (2) -- (3);
\draw [arrow] (2) -- (4);
\draw [arrow] (3) -- (5);
\draw [arrow] (4) -- (5);
\draw [arrow] (5) -- (6);
\draw [arrow] (5) -- (7);
\draw [arrow] (6) -- (8);
\draw [arrow] (7) -- (8);
\draw [arrow] (8) -- (9);
\draw [arrow] (9) -- (10);

\end{tikzpicture}
\caption{The nilpotent hierarchy of 4d $\mathcal{N}=2$ SCFTs obtained from nilpotent deformations breaking the $\mathfrak{su}_6$ flavor symmetry of $\mathcal{T}_3^{(N)}(1,1,3)$. The structure of the network matches that of the 6d SCFTs in Figure \ref{fig:sfold_nilpotent}, and the $\mathbb{Z}_3$ Stiefel--Whitney twisted compactifications of each of those 6d SCFTs yields the associated 4d SCFT in this figure.}
\label{fig:sfold_nilpotent_4d}
\end{figure}
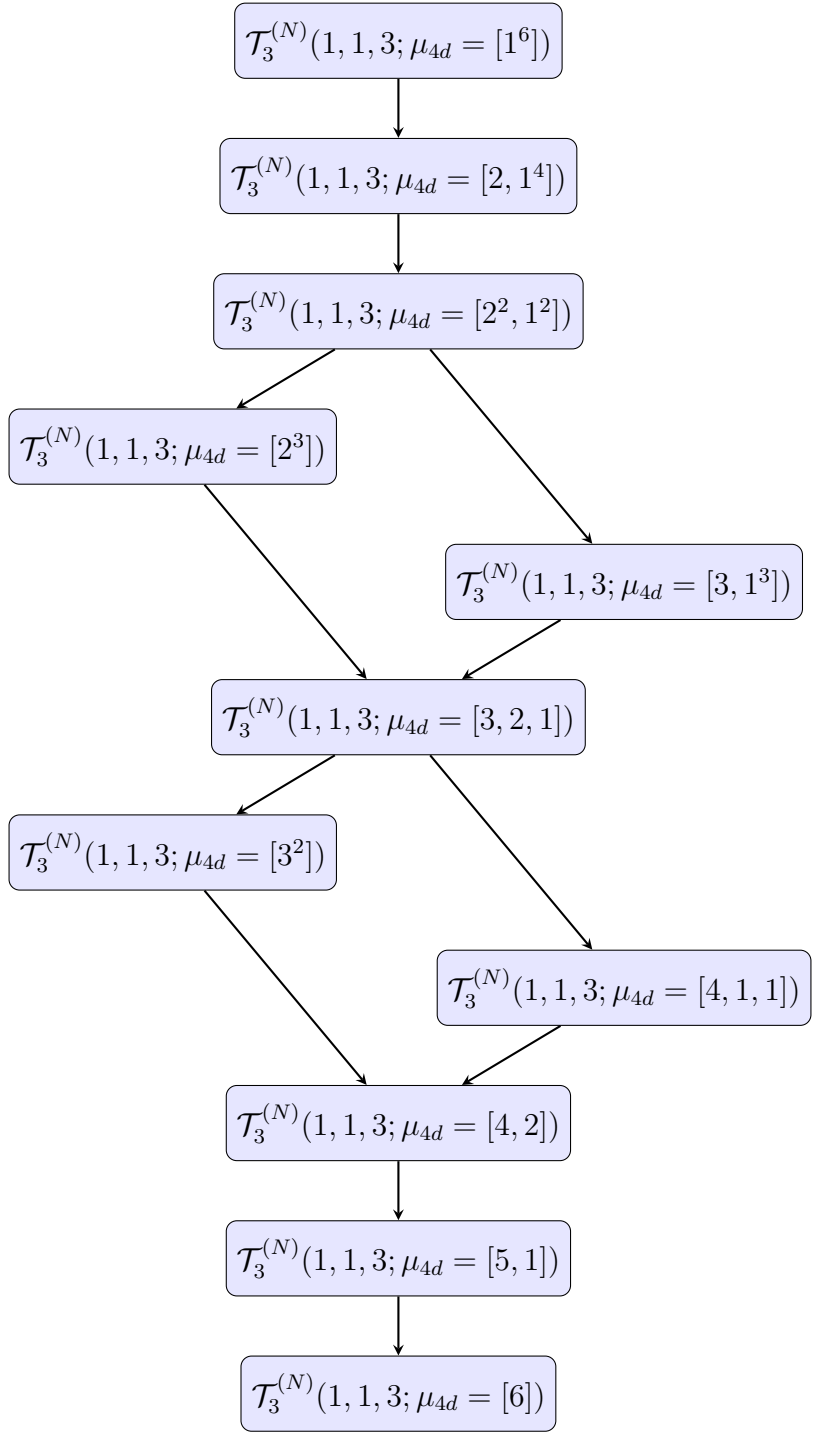

We first discuss the nilpotent network formed by the $\mathbb{Z}_3$ center-flavor preserving nilpotent deformations of the $\mathfrak{su}_{18}$ flavor symmetry. As discussed, these are specified by partitions of eighteen such that each exponent is a multiple of three. Exhaustively, there are eleven such partitions:
\begin{equation}
    [1^{18}],\, [2^3, 1^{12}],\, [2^6, 1^6],\, [2^9],\, [3^3, 1^9],\, [3^3, 2^3, 1^3],\, [3^6],\, [4^3, 1^6],\, [4^3, 2^3],\, [5^3, 1^3],\, [6^3] \,.
\end{equation}
The nilpotent network/Higgs branch flows amongst the generated 6d SCFTs is depicted in Figure \ref{fig:sfold_nilpotent}. Similarly, one can consider the network formed by performing nilpotent Higgsing of the $\mathfrak{su}_6$ flavor algebra belonging to the 4d $\mathcal{T}_3^{(N)}(p=1, s=1, 3q=3)$ SCFT. These are described by partitions of six, and the generated nilpotent hierarchy of these 4d theories is shown in Figure \ref{fig:sfold_nilpotent_4d}. The nilpotent Higgsing in 6d and 4d commutes when combined with the $\mathbb{Z}_3$ Stiefel--Whitney twisted compactification of the theories in Figure \ref{fig:sfold_nilpotent} to the theories in Figure \ref{fig:sfold_nilpotent_4d}.

\subsection{Exceptional SW-folds and Nilpotent Higgsing}

In Section \ref{sec:Esfolds}, we considered a generalization of the SW-fold theories discussed in Section \ref{sec:sfolds} to those obtained from the rank $N$ orbi-instanton theory of type $(\mathfrak{e}_8, \mathfrak{g})$, where $\mathfrak{g}$ is an algebra of type DE. In particular, in Section \ref{sec:e6sfolds}, we showed that there exist seven homomorphisms $\Gamma_{\mathfrak{e}_6} \rightarrow E_8$ such that Higgsing the $\mathfrak{e}_8$ flavor symmetry of the orbi-instanton leads to a theory with $\mathbb{Z}_3$ center-flavor symmetry.  The 6d SCFTs obtained by such Higgsing retain the $\mathfrak{e}_6$ flavor symmetry on the right of the tensor branch quiver.

\begin{table}[t]
    \centering
    \begin{threeparttable}
    \begin{tabular}{cclcc}
        \toprule
        Bala--Carter Label & Weighted Dynkin Diagram & \multicolumn{1}{c}{6d Quiver} & $\mathfrak{f}$ & $\widetilde{\mathfrak{f}}$ \\\midrule
        $0$ & $\begin{gathered}00\overset{\displaystyle 0}{0}00\end{gathered}$ & $\begin{gathered} \cdots\overset{\mathfrak{e}_6}{6}1\overset{\mathfrak{su}_3}{3}1\overset{\mathfrak{e}_6}{6}1\overset{\mathfrak{su}_3}{3}1 \end{gathered}$ & $\mathfrak{e}_6$ & $\mathfrak{g}_2$ \\
        $A_1$ &  $\begin{gathered}00\overset{\displaystyle 1}{0}00\end{gathered}$ & $\begin{gathered} \cdots\overset{\mathfrak{e}_6}{6}1\overset{\mathfrak{su}_3}{3}1\overset{\mathfrak{e}_6}{6}1\overset{\mathfrak{su}_3}{2} \end{gathered}$ & $\mathfrak{su}_6$ & $\mathfrak{su}_2$ \\
        $3A_1$ &  $\begin{gathered}00\overset{\displaystyle 0}{1}00\end{gathered}$ & $\begin{gathered} \cdots\overset{\mathfrak{e}_6}{6}1\overset{\mathfrak{su}_3}{3}1\overset{\mathfrak{e}_6}{6}12 \end{gathered}$ & $\mathfrak{su}_3 \oplus \mathfrak{su}_2$ & $\mathfrak{su}_2$ \\
        $A_2$ & $\begin{gathered}00\overset{\displaystyle 2}{0}00\end{gathered}$ & $\begin{gathered} \cdots\overset{\mathfrak{e}_6}{6}1\overset{\mathfrak{su}_3}{3}1\underset{\displaystyle 1}{\overset{\mathfrak{e}_6}{6}}1 \end{gathered}$ &  $\mathfrak{su}_3 \oplus \mathfrak{su}_3$ & $\varnothing$ \\
        $D_4$ & $\begin{gathered}00\overset{\displaystyle 2}{2}00\end{gathered}$ & $\begin{gathered} \cdots\underset{\displaystyle 1}{\overset{\mathfrak{e}_6}{6}}1\overset{\mathfrak{su}_3}{3} \end{gathered}$ & $\mathfrak{su}_3$ & $\varnothing$ \\\bottomrule
    \end{tabular}
    \end{threeparttable}
    \caption{The nilpotent orbits of $\mathfrak{e}_6$ that are consistent with a $\mathbb{Z}_3$ center-flavor symmetry. The column labelled $\mathfrak{f}$, we write the subalgebra of the $\mathfrak{e}_6$ flavor symmetry that survives the nilpotent Higgsing, and in the $\widetilde{\mathfrak{f}}$ column, we write the remnant algebra after the $\mathbb{Z}_3$ Stiefel--Whitney twisted compactification down to 4d.}
    \label{tbl:e6nilp}
\end{table}

As in the spirit of this Appendix, the $\mathfrak{e}_6$ flavor symmetry can be Higgsed by a choice of nilpotent orbit of $\mathfrak{e}_6$, and if that nilpotent orbit is compatible then the $\mathbb{Z}_3$ center-flavor symmetry can be preserved. There are only five such nilpotent orbits, which we have listed in Table \ref{tbl:e6nilp}.\footnote{We label the exceptional nilpotent orbits using the Bala--Carter notation \cite{MR417306,MR417307}; see \cite{MR1251060} for the standard reference on nilpotent orbits, and \cite{Chacaltana:2012zy} for a useful summary for the exceptional Lie algebras from the perspective of nilpotent Higgsing.} Similarly to the case where $\mathfrak{g}$ is a special unitary algebra, the $\mathfrak{e}_6$ nilpotent orbits that are compatible with the $\mathbb{Z}_3$ center-flavor symmetry are in one-to-one correspondence with the nilpotent orbits of $\mathfrak{g}_2$. Again, the 4d $\mathcal{N}=2$ SCFT obtained by the operation of nilpotent Higgsing of the $\mathfrak{e}_6$ flavor symmetry and then compactifying with $\mathbb{Z}_3$ Stiefel--Whitney twist can alternatively be obtained by first performing the $\mathbb{Z}_3$ Stiefel--Whitney twisted compactification and then Higgsing the $\mathfrak{g}_2$ flavor symmetry by the appropriate nilpotent orbit.

\begin{table}[p!]
    \centering
    \renewcommand{\arraystretch}{1.6}
    \begin{threeparttable}
    \begin{tabular}{cclc}
        \toprule
        $\mathfrak{e}_7$ Orbit & $\mathfrak{f}_4$ Orbit & \multicolumn{1}{c}{6d Quiver} & $\mathfrak{f}$, $\widetilde{\mathfrak{f}}$ \\\midrule
        $0$ & $0$ &
        $\begin{gathered} \cdots\overset{\mathfrak{e}_7}{8}1\overset{\mathfrak{su}_2}{2}\overset{\mathfrak{so}_7}{3}\overset{\mathfrak{su}_2}{2}1\overset{\mathfrak{e}_7}{8}1\overset{\mathfrak{su}_2}{2}\overset{\mathfrak{so}_7}{3}\overset{\mathfrak{su}_2}{2}1\overset{\mathfrak{e}_7}{8}1\overset{\mathfrak{su}_2}{2}\overset{\mathfrak{so}_7}{3}\overset{\mathfrak{su}_2}{2}1 \end{gathered}$ &
        $\mathfrak{e}_7 \rightarrow \mathfrak{f}_4 \oplus \cancel{\mathfrak{su}_2}$ \\

        $A_1$ & $A_1$ &
        $\begin{gathered} \cdots\overset{\mathfrak{e}_7}{8}1\overset{\mathfrak{su}_2}{2}\overset{\mathfrak{so}_7}{3}\overset{\mathfrak{su}_2}{2}1\overset{\mathfrak{e}_7}{8}1\overset{\mathfrak{su}_2}{2}\overset{\mathfrak{so}_7}{3}\overset{\mathfrak{su}_2}{2}1\overset{\mathfrak{e}_7}{8}1\overset{\mathfrak{su}_2}{2}\overset{\mathfrak{so}_7}{3}\overset{\mathfrak{su}_2}{1} \end{gathered}$ &
        $\mathfrak{so}_{12} \rightarrow \mathfrak{sp}_3 \oplus \cancel{\mathfrak{su}_2}$ \\

        $2A_1$ & $\widetilde{A}_1$ &
        $\begin{gathered} \cdots\overset{\mathfrak{e}_7}{8}1\overset{\mathfrak{su}_2}{2}\overset{\mathfrak{so}_7}{3}\overset{\mathfrak{su}_2}{2}1\overset{\mathfrak{e}_7}{8}1\overset{\mathfrak{su}_2}{2}\overset{\mathfrak{so}_7}{3}\overset{\mathfrak{su}_2}{2}1\overset{\mathfrak{e}_7}{8}1\overset{\mathfrak{su}_2}{2}\overset{\mathfrak{so}_7}{3}1 \end{gathered}$ &
        $(\mathfrak{so}_9 \rightarrow \mathfrak{su}_4 \oplus \cancel{\mathfrak{su}_2}) \oplus \cancel{\mathfrak{su}_2}$  \\

        $3A_1^\prime$ & $A_1 + \widetilde{A}_1$ &
        $\begin{gathered} \cdots\overset{\mathfrak{e}_7}{8}1\overset{\mathfrak{su}_2}{2}\overset{\mathfrak{so}_7}{3}\overset{\mathfrak{su}_2}{2}1\overset{\mathfrak{e}_7}{8}1\overset{\mathfrak{su}_2}{2}\overset{\mathfrak{so}_7}{3}\overset{\mathfrak{su}_2}{2}1\overset{\mathfrak{e}_7}{8}1\overset{\mathfrak{su}_2}{2}\overset{\mathfrak{so}_7}{2} \end{gathered}$ &
        $(\mathfrak{sp}_3\rightarrow \mathfrak{su}_2 \oplus \cancel{\mathfrak{su}_2}) \oplus \mathfrak{su}_2$ \\

        {\color{red}$A_2$} & {\color{red}$A_2$ and $\widetilde{A}_2$} &
        {\color{red}$\begin{gathered} \cdots\overset{\mathfrak{e}_7}{8}1\overset{\mathfrak{su}_2}{2}\overset{\mathfrak{so}_7}{3}\overset{\mathfrak{su}_2}{2}1\overset{\mathfrak{e}_7}{8}1\overset{\mathfrak{su}_2}{2}\overset{\mathfrak{so}_7}{3}\overset{\mathfrak{su}_2}{2}1\overset{\mathfrak{e}_7}{8}1\overset{\mathfrak{su}_2}{2}\overset{\mathfrak{su}_4}{2} \end{gathered}$} &
        {\color{red}$\mathfrak{su}_6$} \\

        $A_2+2A_1$ & $A_2 + \widetilde{A}_1$ &
        $\begin{gathered} \cdots\overset{\mathfrak{e}_7}{8}1\overset{\mathfrak{su}_2}{2}\overset{\mathfrak{so}_7}{3}\overset{\mathfrak{su}_2}{2}1\overset{\mathfrak{e}_7}{8}1\overset{\mathfrak{su}_2}{2}\overset{\mathfrak{so}_7}{3}\overset{\mathfrak{su}_2}{2}1\overset{\mathfrak{e}_7}{8}1\overset{\mathfrak{su}_2}{2}\overset{\mathfrak{su}_2}{2} \end{gathered}$ &
        $\mathfrak{su}_2 \oplus \cancel{\mathfrak{su}_2} \oplus \cancel{\mathfrak{su}_2}$ \\

        $A_3$ & $B_2$ &
        $\begin{gathered} \cdots\overset{\mathfrak{e}_7}{8}1\overset{\mathfrak{su}_2}{2}\overset{\mathfrak{so}_7}{3}\overset{\mathfrak{su}_2}{2}1\overset{\mathfrak{e}_7}{8}1\overset{\mathfrak{su}_2}{2}\overset{\mathfrak{so}_7}{3}\overset{\mathfrak{su}_2}{2}1\underset{\displaystyle 1}{\overset{\mathfrak{e}_7}{8}}1\overset{\mathfrak{su}_2}{2} \end{gathered}$ &
        $(\mathfrak{so}_7\rightarrow \mathfrak{su}_2 \oplus \mathfrak{su}_2 \oplus \cancel{\mathfrak{su}_2}) \oplus \cancel{\mathfrak{su}_2}$ \\

        $2A_2+A_1$ & $\widetilde{A}_2 + A_1$ &
        $\begin{gathered} \cdots\overset{\mathfrak{e}_7}{8}1\overset{\mathfrak{su}_2}{2}\overset{\mathfrak{so}_7}{3}\overset{\mathfrak{su}_2}{2}1\overset{\mathfrak{e}_7}{8}1\overset{\mathfrak{su}_2}{2}\overset{\mathfrak{so}_7}{3}\overset{\mathfrak{su}_2}{2}1\overset{\mathfrak{e}_7}{8}122 \end{gathered}$ &
        $\mathfrak{su}_2 \oplus \cancel{\mathfrak{su}_2}$ \\

        $(A_3+A_1)^\prime$ & $C_3(a_1)$ &
        $\begin{gathered} \cdots\overset{\mathfrak{e}_7}{8}1\overset{\mathfrak{su}_2}{2}\overset{\mathfrak{so}_7}{3}\overset{\mathfrak{su}_2}{2}1\overset{\mathfrak{e}_7}{8}1\overset{\mathfrak{su}_2}{2}\overset{\mathfrak{so}_7}{3}\overset{\mathfrak{su}_2}{2}1\underset{\displaystyle 1}{\overset{\mathfrak{e}_7}{8}}12 \end{gathered}$ &
        $\mathfrak{su}_2 \oplus \cancel{\mathfrak{su}_2} \oplus \cancel{\mathfrak{su}_2}$ \\

        $D_4(a_1)$ & $F_4(a_3)$ &
        $\begin{gathered} \cdots\overset{\mathfrak{e}_7}{8}1\overset{\mathfrak{su}_2}{2}\overset{\mathfrak{so}_7}{3}\overset{\mathfrak{su}_2}{2}1\overset{\mathfrak{e}_7}{8}1\overset{\mathfrak{su}_2}{2}\overset{\mathfrak{so}_7}{3}\overset{\mathfrak{su}_2}{2}1\overset{\displaystyle 1}{\underset{\displaystyle 1}{\overset{\mathfrak{e}_7}{8}}}1 \end{gathered}$ &
        $\cancel{\mathfrak{su}_2} \oplus \cancel{\mathfrak{su}_2} \oplus \cancel{\mathfrak{su}_2}$ \\

        $A_5^\prime$ & $B_3$ &
        $\begin{gathered} \cdots\overset{\mathfrak{e}_7}{8}1\overset{\mathfrak{su}_2}{2}\overset{\mathfrak{so}_7}{3}\overset{\mathfrak{su}_2}{2}1\overset{\mathfrak{e}_7}{8}1\overset{\mathfrak{su}_2}{2}\overset{\mathfrak{so}_7}{3}1\overset{\mathfrak{so}_9}{4} \end{gathered}$ &
        $\mathfrak{su}_2 \oplus \cancel{\mathfrak{su}_2}$ \\

        $D_4$ & $C_3$ &
        $\begin{gathered} \cdots\overset{\mathfrak{e}_7}{8}1\overset{\mathfrak{su}_2}{2}\overset{\mathfrak{so}_7}{3}\overset{\mathfrak{su}_2}{2}1\overset{\mathfrak{e}_7}{8}1\overset{\mathfrak{su}_2}{2}\overset{\mathfrak{so}_7}{3}\overset{\mathfrak{su}_2}{1}\overset{\mathfrak{so}_{12}}{4} \end{gathered}$ &
        $\mathfrak{sp}_3 \rightarrow \mathfrak{su}_2 \oplus \cancel{\mathfrak{su}_2}$ \\

        $E_6(a_3)$ & $F_4(a_2)$ &
        $\begin{gathered} \cdots\overset{\mathfrak{e}_7}{8}1\overset{\mathfrak{su}_2}{2}\overset{\mathfrak{so}_7}{3}\overset{\mathfrak{su}_2}{2}1\overset{\mathfrak{e}_7}{8}1\overset{\mathfrak{su}_2}{2}\overset{\mathfrak{so}_7}{3}1\overset{\mathfrak{so}_8}{4} \end{gathered}$ &
        $\cancel{\mathfrak{su}_2}$ \\

        $D_5$ & $F_4(a_1)$ &
        $\begin{gathered} \cdots\overset{\mathfrak{e}_7}{8}1\overset{\mathfrak{su}_2}{2}\overset{\mathfrak{so}_7}{3}\overset{\mathfrak{su}_2}{2}1\underset{\displaystyle 1}{\overset{\mathfrak{e}_7}{8}}1\overset{\mathfrak{su}_2}{2}\overset{\mathfrak{so}_7}{3} \end{gathered}$ &
        $\cancel{\mathfrak{su}_2} \oplus \cancel{\mathfrak{su}_2}$ \\

        $E_6$ & $F_4$ &
        $\begin{gathered} \cdots\underset{\displaystyle 1}{\overset{\mathfrak{e}_7}{8}}1\overset{\mathfrak{su}_2}{2}\overset{\mathfrak{so}_7}{3}\overset{\mathfrak{su}_2}{2} \end{gathered}$ &
        $\cancel{\mathfrak{su}_2}$ \\\bottomrule
    \end{tabular}
    \end{threeparttable}
    \caption{The nilpotent orbits of $\mathfrak{e}_7$ that are consistent with a $\mathbb{Z}_2$ center-flavor symmetry. In the $\mathfrak{f}$, $\widetilde{\mathfrak{f}}$ column we write the flavor symmetry in 6d, and the remnant subalgebra in 4d after the SW-twisted compactification. Scored-out algebras are removed by the SW-twist. The remnant algebra, $\widetilde{\mathfrak{f}}$ matches the flavor symmetry associated to the $\mathfrak{f}_4$ nilpotent orbit. The {\color{red} red} line is exceptional, and is discussed in the text.}
    \label{tbl:e7nilp}
\end{table}

Further, we consider the case where $\mathfrak{g} = \mathfrak{e}_7$. There are fifteen nilpotent orbits of $\mathfrak{e}_7$ that are compatible with a $\mathbb{Z}_2$ center flavor symmetry, as depicted in Table \ref{tbl:e7nilp}. The sub-Hasse diagram formed by the subset of $\mathfrak{e}_7$ nilpotent orbits appears here almost matches the Hasse diagram for $\mathfrak{f}_4$ nilpotent orbits that appears in \cite{Hanany:2017ooe}. Similarly, the flavor symmetries surviving after the Stiefel--Whitney twist and those of the $\mathfrak{f}_4$ nilpotent orbits almost always match. The one subtlety is the line denoted in {\color{red}red} in Table \ref{tbl:e7nilp}; this appears to be associated to one $\mathfrak{e}_7$ nilpotent orbit, but \emph{two} $\mathfrak{f}_4$ nilpotent orbits. This case involves the $\mathbb{Z}_2$ Stiefel--Whitney twist of a 6d theory containing an $\overset{\mathfrak{su}_4}{2}$ factor. As we discussed at the conclusion of Section \ref{sec:classs}, this leads to curious features, similar to those that occur in six dimensions when one has $\overset{\mathfrak{su}_2}{2}$ \cite{Morrison:2016djb}. We expect that a deeper understanding of these highly special configurations will lead to the resolution of this subtlety in the $\mathfrak{e}_7$ and $\mathfrak{f}_4$ nilpotent orbits, however, we leave such a study for future work.

One can use the same methods from the 6d perspective to determine the central charges, flavor symmetries, Coulomb branch operator dimensions and so forth of the 4d $\mathcal{N}=2$ SCFTs obtained from the torus compactification. A similar analysis can be carried out when $\mathfrak{g} = \mathfrak{so}_{2k}$, however we leave such an enumeration to the reader.\footnote{For $\mathfrak{g} = \mathfrak{so}_{2k}$ there is a subtlety with the fact that two distinct SCFTs, obtained from very even nilpotent Higgsing, are associated to the same tensor branch description \cite{Distler:2022yse}.}

\section{Flavor Group for Conformal Matter \& Deformations}\label{app:CM}

In Section \ref{sec:6d}, we expanded upon the proposal to determine the global structure of the global symmetry group of a 6d $(1,0)$ SCFT that was put forth in \cite{Apruzzi:2020zot}. This proposal is based on the weakly-coupled spectrum of the effective theory that lives on the generic point of the tensor branch, together with the knowledge of the Green--Schwarz couplings. In Section \ref{sec:e8CF}, we applied this prescription to the 6d SCFTs obtained from the Higgsing of the $\mathfrak{e}_8$ flavor symmetry, by a choice of $E_8$-homomorphism $\rho: \mathbb{Z}_K \rightarrow E_8$, of the rank $N$ $(\mathfrak{e}_8, \mathfrak{su}_K)$ orbi-instanton; we found that the resulting global structure of the non-Abelian part of the flavor symmetry was encoded in a simple manner in the choice of $\rho$. We highlighted an extension of this analysis to the 6d $(1,0)$ SCFTs obtained via the $\mathfrak{e}_8$ Higgsing of the rank $N$ $(\mathfrak{e}_8, \mathfrak{g})$ orbi-instanton in Section \ref{sec:Esfolds}. Furthermore, in Appendix \ref{app:nilp}, we demonstrated that, when we consider the Higgsing of the orbi-instanton by both an $E_8$-homomorphism, $\rho$, and a nilpotent orbit of $\mathfrak{g}$, $\sigma$, there is a simple prescription for the non-Abelian center-flavor symmetry of the resulting SCFT in terms of $\rho$ and $\sigma$.

Throughout this paper, we have focused on the 6d $(1,0)$ SCFTs known as the rank $N$ $(\mathfrak{e}_8, \mathfrak{g})$ orbi-instantons and the theories further obtained via Higgs branch renormalization group flows. Another broad class of 6d $(1,0)$ SCFTs are those commonly referred to as the ``Higgsable to $(2,0)$ of type $A_{N-1}$'' SCFTs \cite{Ohmori:2015pia}. These include the rank $N$ $(\mathfrak{g}, \mathfrak{g})$ conformal matter theories \cite{DelZotto:2014hpa}, corresponding to the worldvolume theory on a stack of $N$ M5-branes probing a $\mathbb{C}^2/\Gamma_\mathfrak{g}$ orbifold singularity, and the theories obtained via Higgsing of the $\mathfrak{g} \oplus \mathfrak{g}$ flavor symmetry of the conformal matter theory by a pair of nilpotent orbits $\sigma_L$ and $\sigma_R$ of $\mathfrak{g}$. The former have frequently been referred to in the literature as $\mathcal{T}_{\mathfrak{g},N}$, and the latter as $\mathcal{T}_{\mathfrak{g},N}(\sigma_L, \sigma_R)$.\footnote{We emphasize that these 6d SCFTs are distinct from the 4d SW-fold theories also labelled by $\mathcal{T}$ in Section \ref{sec:sfolds}.} In Section \ref{sec:cfanom}, we determined that the non-Abelian flavor group of rank $N$ $(\mathfrak{su}_K, \mathfrak{su}_K)$ conformal matter is
\begin{equation}
    (SU(K) \times SU(K))/\mathbb{Z}_K \,.
\end{equation}
It is straightforward to see, again from the methods presented in Section \ref{sec:6d}, that when $\mathfrak{su}_K$ is generalized to an arbitrary ADE Lie algebra $\mathfrak{g}$, the non-Abelian flavor group is
\begin{equation}
    (\widetilde{G} \times \widetilde{G})/Z(\widetilde{G}) \,,
\end{equation}
where $\widetilde{G}$ is the simply-connected group with Lie algebra $\mathfrak{g}$, and $Z(\widetilde{G})$ is the center of $\widetilde{G}$.

Now we turn to the determination of the global structure of the non-Abelian flavor symmetry after Higgsing via the pair of nilpotent orbits $(\sigma_L, \sigma_R)$. The nilpotent orbit $\sigma_L$ breaks the left $\mathfrak{g}$ flavor algebra to the semi-simple algebra $\mathfrak{h}_L$,\footnote{We ignore $\mathfrak{u}(1)$ factors in this Appendix.} and we let $\widetilde{H}_L$ denote the associated simply-connected Lie group; and similarly for the Higgsing of the right $\mathfrak{g}$ by $\sigma_R$. Both $\mathfrak{h}_L$ and $\mathfrak{h}_R$ can be read off directly from the nilpotent orbits \cite{Heckman:2016ssk}. Similarly, it is well-known how each nilpotent Higgsing modifies the tensor branch description, and thus one can use the analysis of Section \ref{sec:6d} to determine the subgroup of the center $Z(\widetilde{G})$ that is preserved after Higgsing; we refer to these subgroups as $Z_L(\widetilde{G})$ and $Z_R(\widetilde{G})$ for $\sigma_L$ and $\sigma_R$, respectively. For $\widetilde{G} = SU(K)$, $E_6$, and $E_7$ these subgroups have been discussed in Appendix \ref{app:nilp}.

Putting all this together, the global structure of the non-Abelian flavor group of the Higgsed conformal matter theory, $\mathcal{T}_{\mathfrak{g},N}(\sigma_L, \sigma_R)$, can be shown to be
\begin{equation}\label{eqn:NCMstr}
    (\widetilde{H}_L \times \widetilde{H}_R)/(Z_L(\widetilde{G})\cap Z_R(\widetilde{G})) \,.
\end{equation}
Here the quotient is by the common subgroup of $Z_L(\widetilde{G})$ and $Z_R(\widetilde{G})$ inside $Z(\widetilde{G})$. In particular, if we consider $\widetilde{G} \neq Spin(4K)$, then we have
\begin{equation}
    Z(\widetilde{G}) = \mathbb{Z}_K \,, \qquad Z_L(\widetilde{G}) = \mathbb{Z}_{K_L} \,, \qquad Z_R(\widetilde{G}) = \mathbb{Z}_{K_R} \,,
\end{equation}
for some $K$, and $K_L$, $K_R$ divisors of $K$. The quotient is then by
\begin{equation}
    (Z_L(\widetilde{G})\cap Z_R(\widetilde{G})) = \mathbb{Z}_{\gcd(K_L, K_R)} \,.
\end{equation}
For $\widetilde{G} = Spin(4K)$ it is a little more technical due to the product structure of the center.

We now make this explicit in one example. Consider the rank $N$ $(\mathfrak{su}_{18}, \mathfrak{su}_{18})$ conformal matter theory.\footnote{We assume that $N > 7$, as this guarantees that the specific nilpotent deformations that we turn on, on the left and on the right, do not start to cross-correlate.} As is by now familiar, the tensor branch description is 
\begin{equation}
    \overbrace{\,
    \underset{\displaystyle [\mathfrak{su}_{18}]}{\overset{\mathfrak{su}_{18}}{2}}
    \overset{\mathfrak{su}_{18}}{2}\,
    \cdots\,
    \overset{\mathfrak{su}_{18}}{2}
    \underset{\displaystyle [\mathfrak{su}_{18}]}{\overset{\mathfrak{su}_{18}}{2}}
    \,}^{N - 1 \text{ $(-2)$-curves}} \,.
\end{equation}
We consider the Higgs branch deformations triggered by turning on vacuum expectation values associated to the nilpotent orbits 
\begin{equation}
    \sigma_L = [1^6, 4^3] \,, \qquad \sigma_R = [1^6, 2^6] \,,
\end{equation}
of the $\mathfrak{su}_{18}$ flavor algebras on the left and right. It is clear from the analysis in Appendix \ref{app:nilp} that $\sigma_L$ preserves a $\mathbb{Z}_3$ center-flavor subgroup, as the exponents of the partition are all multiples of three, and similarly, $\sigma_R$ preserves a $\mathbb{Z}_6$ center-flavor subgroup:
\begin{equation}
    Z_L(SU(18)) = \mathbb{Z}_3 \,, \qquad Z_R(SU(18)) = \mathbb{Z}_6 \,.
\end{equation}
After performing the nilpotent Higgsing, the renormalization group flow ends at an interacting 6d $(1,0)$ SCFT with tensor branch description
\begin{equation}
    \overbrace{\,
    \underset{\displaystyle [\mathfrak{su}_6]}{\overset{\mathfrak{su}_9}{2}}\,
    \overset{\mathfrak{su}_{12}}{2}\,
    \overset{\mathfrak{su}_{15}}{2}\,
    \underset{\displaystyle [\mathfrak{su}_3]}{\overset{\mathfrak{su}_{18}}{2}}\,
    \overset{\mathfrak{su}_{18}}{2}\,
    \cdots\,
    \overset{\mathfrak{su}_{18}}{2}\,
    \underset{\displaystyle [\mathfrak{su}_6]}{\overset{\mathfrak{su}_{18}}{2}}\,
    \underset{\displaystyle [\mathfrak{su}_6]}{\overset{\mathfrak{su}_{12}}{2}}
    \,}^{N - 1 \text{ $(-2)$-curves}} \,.
\end{equation}
To determine the global structure of the non-Abelian flavor symmetry, we can apply the procedure described in Section \ref{sec:6d}. From that perspective, we determine that the non-Abelian flavor group is 
\begin{equation}
    (SU(6) \times SU(3) \times SU(6) \times SU(6))/\mathbb{Z}_3 \,,
\end{equation}
as expected from equation \eqref{eqn:NCMstr}.

\bibliographystyle{utphys}
\bibliography{CenterFolds}

\end{document}